\documentclass[utf8]{FrontiersinHarvard} 

\usepackage{url,hyperref,lineno,microtype,subcaption}
\usepackage[onehalfspacing]{setspace}

\def\keyFont{\fontsize{8}{11}\helveticabold }
\def\firstAuthorLast{L. Tortorelli \& A. Mercurio} 
\def\Authors{Luca Tortorelli\,$^{1,2,*}$ and Amata Mercurio\,$^{3,4}$}
\def\Address{$^{1}$ Faculty of Physics,University Observatory, Ludwig-Maximilians-Universit\"at M\"unchen,  Munich, Germany \\
$^{2}$ Institute for Particle Physics and Astrophysics,  ETH Z\"urich, Z\"urich,  Switzerland \\
$^{3}$ Dipartimento di Fisica “E.R. Caianiello”,  Universit\`a Degli Studi di Salerno,  Fisciano (Salerno),  Italy \\
$^{4}$ INAF-Osservatorio Astronomico di Capodimonte,  Napoli, Italy}
\def\corrAuthor{L.  Tortorelli}

\def\corrEmail{Luca.Tortorelli@physik.lmu.de}

\begin{document}
\onecolumn
\firstpage{1}

\title[\textsc{morphofit}]{\textsc{morphofit}: An automated galaxy structural parameters fitting package} 

\author[\firstAuthorLast ]{\Authors} 
\address{} 
\correspondance{} 

\extraAuth{}

\maketitle

\begin{abstract}

In today's modern wide-field galaxy surveys, there is the necessity for parametric surface brightness decomposition codes characterised by accuracy,  small degree of user intervention,  and high degree of parallelisation.  We try to address this necessity by introducing \textsc{morphofit},  a highly parallelisable \textsc{python} package for the estimate of galaxy structural parameters.  The package makes use of wide-spread and reliable codes,  namely \textsc{sextractor} and \textsc{galfit}.  It has been optimised and tested in both low-density and crowded environments, where blending and diffuse light makes the structural parameters estimate particularly challenging.  \textsc{morphofit} allows the user to fit multiple surface brightness components to each individual galaxy,  among those currently implemented in the code.  Using simulated images of single S\'ersic and bulge plus disk galaxy light profiles with different bulge-to-total luminosity ($\mathrm{B/T}$) ratios, we show that \textsc{morphofit} is able to recover the input structural parameters of the simulated galaxies with good accuracy.  We also compare its estimates against existing literature studies, finding consistency within the errors.  We use the package in \citealt{Tortorelli2023} to measure the structural parameters of cluster galaxies in order to study the wavelength dependence of the Kormendy relation of early-type galaxies. The package is available on github\footnote{\href{https://github.com/torluca/morphofit}{https://github.com/torluca/morphofit}} and on the Pypi server \footnote{\href{https://pypi.org/project/morphofit/}{https://pypi.org/project/morphofit/}}.

\tiny
 \keyFont{ \section{Keywords:} galaxies: morphology, methods: data analysis, techniques: photometric, galaxies: photometry, galaxies: fundamental parameters,  galaxies: clusters: individual: Abell S1063, galaxies: clusters: individual: MACS J0416.1-2403,  galaxies: clusters: individual: MACS J1149.5+2223} 
\end{abstract}

\section{Introduction}
\label{sec:intro}

The accurate measurement of galaxy structural parameters (e.g.  total magnitudes,  sizes,  S\'ersic indices,  position angles,  ellipticities) provides extremely useful insights into galaxy formation and evolution studies.  The photometric structural analysis of galaxies allows us to perform their morphological classification. This sheds light on the presence of common processes taking place during galaxy formation and evolution and provides qualitative information that supports galaxy classification schemes, which also helps to constrain the dynamical modelling of galaxies (e.g.  \citealt{Binney1990,vanderMarel1991,Cappellari2013}). Furthermore,  magnitudes and sizes can be used to measure empirical scaling relations (e.g.  the effective brightness - effective radius (or Kormendy) relation \citealt{Kormendy1977}), while ellipticities and position angles can be used for shear and intrinsic alignment measurements (e.g. \citealt{Lee2018,Kannawadi2019}).  

The measurement of galaxy structural parameters is usually conducted with 1D or 2D methods.  The 1D methods,  e.g.  isophotal analysis,  work well in low signal-to-noise conditions, since they azimuthally average the surface brightness along the elliptical isophotes.  2D methods are routinely performed by creating parametrised models for the 2D galaxy surface brightness.  They have the advantage of allowing us to disentangle among different galaxy components, as well as taking into account the point-spread-function (PSF) of the observations through its convolution with the galaxy light profile (see \citealt{Mendez-Abreu2008,Peng2010,Bonfini2014,Gao2017} and references therein for differences between 1D and 2D techniques).  The most common parametrisation for the surface brightness distribution is the S\'ersic model \citep{Sersic1963}. It consists of a smooth surface brightness distribution with seven degrees of freedom,  involving the profile centroid,  the total magnitude,  the effective radius,  the S\'ersic index, the axis ratio,  and the position angle.  The model owns his success not only to the fact that it is a good approximation of the surface brightness distribution of galaxies,  but also to the fact that it can be used to disentangle between classical and pseudo-bulges (\citealt{Fisher2008}, but see \citealt{Costantin2018} for a different approach), and, for $n=1$, it approximates the exponential behaviour of galaxy disks.  The S\'ersic profile can also be used to discover hidden substructures.  The latter may consist of spiral arms, rings, bars,  that deviate from a smooth surface brightness distribution and,  therefore,  require additional modelling components with respect to the simple S\'ersic profile \citep{Sonnenfeld2022}.  The morphological description of galaxies can be conducted also with non-parametric techniques. However,  their use is not advised in low signal-to-noise conditions: non-parametric techniques mostly rely on being able to measure a gradient in galaxies and that is possible only for high signal-to-noise conditions, especially in the case of shallow profile gradients. Additionally,  the required background subtraction might cause a sky over-estimation when dealing with low signal-to-noise shallow gradients\footnote{See \href{https://users.obs.carnegiescience.edu/peng/work/galfit/TFAQ.html}{https://users.obs.carnegiescience.edu/peng/work/galfit/TFAQ.html} for a detailed discussion.}.

Modern photometric and spectroscopic surveys have enabled the measurement of galaxy structural parameters for millions of galaxies.  This allows astronomers to characterise galaxy properties and build relations among them with high statistical precision (e.g.  \citealt{LaBarbera2010,Simard2011,Kelvin2012,Kawinwanichakij2021}).  This rapidly growing sample of galaxies with quality imaging has sparkled renovated interests in the development of codes for the measurement of galaxy structural parameters and galaxy morphological classification that are accurate,  time efficient,  and require very few input from the user. These codes involve the use of non-parametric fitting (\textsc{cas} \citealt{Conselice2003},  \textsc{gini} \citealt{Lotz2004},  \textsc{morfometryka} \citealt{Ferrari2015}),  parametric fitting (\textsc{gim2d} \citealt{Simard2002},  \textsc{budda} \citealt{DeSouza2004}, \textsc{gasp2d} \citealt{Mendez-Abreu2008},  \textsc{pymorph} \citealt{Vikram2010},  \textsc{galapagos} \citealt{Haussler2011,Haussler2013},  \textsc{imfit} \citealt{Erwin2015},  \textsc{galight} \citealt{Ding2021},  \textsc{galapagos-2} \citealt{Haussler2022}) or machine learning methodologies (\textsc{deeplegato} \citealt{Tuccillo2018},  \textsc{gamornet} \citealt{Ghosh2020},  \textsc{galnets} \citealt{Li2022}).

In this work,  we present a package called \textsc{morphofit}\footnote{\href{https://github.com/torluca/morphofit}{https://github.com/torluca/morphofit}}.  The package seeks to address the necessity of modern galaxy surveys for parametric surface brightness decomposition codes that are accurate,  parallelisable,  and automatic,  meaning that they require a small degree of human inspection and intervention. The package is written in \textsc{python} and relies upon the well-tested and widespread codes \textsc{galfit} \citep{Peng2010,Peng2011} and \textsc{sextractor} \citep{Bertin1996}.  \textsc{morphofit} allows the user to measure galaxy structural parameters in stamps around individual objects or by simultaneously fitting multiple galaxies in the image.  It has been optimised and widely tested in low and high-density environments, the latter being characterised by the presence of blending and diffuse light effects.  It contains routines to create PSF images from the knowledge of star positions in the images, as well as the ability to run \textsc{sextractor} in forced photometry mode to use its parameters as either individual galaxy properties estimates or starting point for the \textsc{galfit} fit.  

The package is written in a modular nature,  such that each individual module can be run independently depending on the user necessities.  The modules can be chained in a pipeline that constitutes a novel method to fit galaxies in a cluster environment.   In particular, this method involves the surface brightness profile fit of galaxies in images of increasing sizes. This approach has already been adopted in \citealt{Tortorelli2018} (hereafter,  LT18) to measure the sample dependence of the Kormendy relation.  In this work,  we improve the methodology and we successfully apply it in \citealt{Tortorelli2023} (hereafter,  LT23) to measure the wavelength dependence of the Kormendy relation.  It differs from the usual approach adopted in the literature, where the authors limit their surface brightness fit to stamps around individual galaxies.  It has been shown (see figure 5 of LT18) that this approach negatively impacts the estimated galaxy properties.  By fitting the surface brightness profiles in stamps around galaxies, the minimisation algorithm converges to biased structural parameters since the local sky background is dominated, in the case of galaxy clusters, by the intracluster light and not by the actual image background.  We refer the reader to section 3.3 of LT18 for a more detailed analysis of this aspect. The cluster environment is particularly challenging for the fit in stamps because of the intracluster light and the high galaxy density.  The latter requires a simultaneous fit of multiple sources to take into account light contamination among neighbouring objects.  The former requires a careful modelling of its contribution, usually by adding an additional large S\'ersic light profile to the brightest cluster galaxy (BCG) \citep{Kluge2021},  as well as a careful estimate of the image background.  The methodology we propose addresses these two aspects.  In order to deal with nearest neighbours,  we use an iterative approach where we analyse images of increasing size.  In order to deal with the intracluster light contamination in flux, instead,  we average the structural parameters measured with different background estimation methods and we add a secondary S\'ersic light profile \citep{Kluge2021} to the BCG fit.  This methodology is, therefore,  especially useful in the case of galaxy clusters and compact groups. 

In addition,  since as mentioned earlier,  sometimes galaxies are best fit by a multi-component surface brightness distribution,  the package is designed such that each individual galaxy can be fit with multiple components among those currently implemented in the code (i.e.  S\'ersic,  deVaucouleurs,  exponential disk).  However,  the multi-component capabilities of \textsc{morphofit} have been tested up to two light profile components,  therefore the user must use care when fitting a larger number of components for its own specific scientific case.  Although galaxies may have a lot of structures (like rings, pseudorings, lenses, bars, envelops, and spiral arms), a parametric bulge-disk decomposition is one of the first possible quantitative steps to tag a galaxy as an early or late-type system. 

This work is structured as follows.  In section \ref{morphofit_features}, we present the main features of \textsc{morphofit}.  In section \ref{simulations}, we test the package against simulated data generated as both single S\'ersic (section \ref{single_sersic_sims}) and  bulge plus disk profiles with a variable $\mathrm{B/T}$ ratio (section \ref{bulge_disk_sims}). Section \ref{section:comparison_with_catalogues} contains the comparison between the results obtained with this automated methodology and those from already published catalogues.  We provide the main conclusions in section \ref{section:conclusions}.

\begin{figure}
\centering
\includegraphics[scale=0.4]{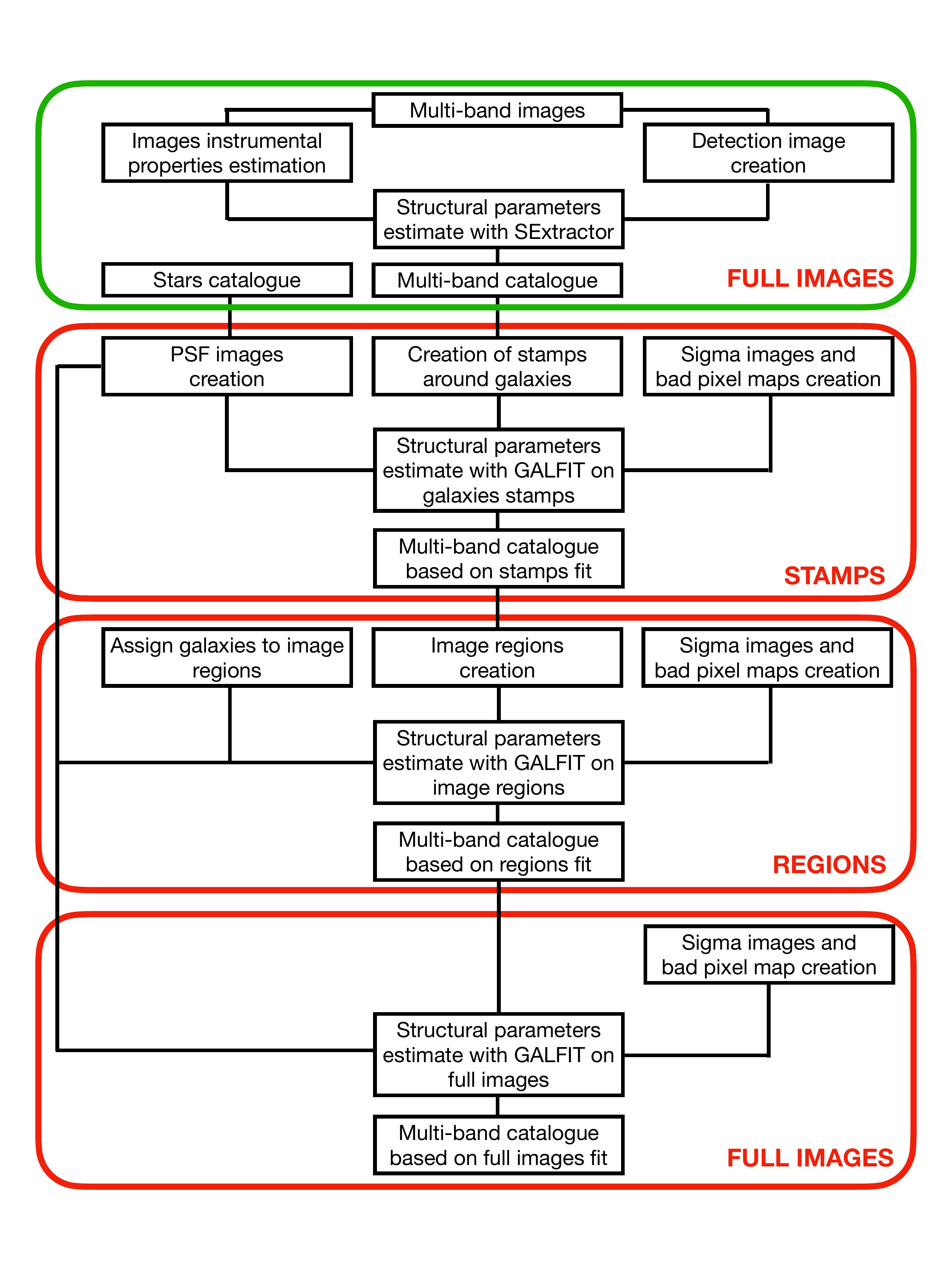}
\caption{Flowchart describing the structural parameters measurement steps of the pipeline adopted in LT18 and LT23.  The red boxes refer to estimates using \textsc{galfit}, while the green box using \textsc{sextractor}. }
\label{flowchart}
\end{figure}

\section{\textsc{morphofit} main features}
\label{morphofit_features}

\textsc{morphofit} has a modular nature.  Each module can be run independently with respect to the others.  For the sake of clarity,  we describe the different modules that constitute \textsc{morphofit} following the pipeline for the morphological analysis of galaxies first introduced in LT18,  improved in this work, and further used in LT23 (Figure \ref{flowchart}).  This pipeline has been designed to mitigate the challenging aspects of measuring the structural parameters in galaxy clusters,  e.g.   intracluster light and blending. It allows for the measurement of the structural parameters via an iterative approach by fitting images of increasing size (to deal with nearest neighbours), with different PSF image estimates,  different sigma (noise) images and on multiple background estimations (to deal with the ICL contamination in flux).  The various steps of the pipeline correspond to the different modules \textsc{morphofit} is built upon: \textsc{sextractor} run in forced photometry mode to get an initial estimate of the galaxy structural parameters; creation of PSF images for the surface brightness profile convolution; \textsc{galfit} run on stamps using the \textsc{sextractor} properties as input; \textsc{galfit} run on regions and full images,  using the fit on stamps and regions,  respectively,  as initial guesses; creation of multi-wavelength catalogues of best-fitting structural parameters.  We design the pipeline to be especially useful for fitting galaxy clusters and compact groups.

Each \textsc{morphofit} module is constituted by two scripts.  The first one creates an \textsc{hdf5} table that stores file paths and variable values.  The second script reads the \textsc{hdf5} table and performs the measurement tasks.  The scripts are run via \textsc{esub-epipe}\footnote{\href{https://cosmo-gitlab.phys.ethz.ch/cosmo\_public/esub-epipe}{https://cosmo-gitlab.phys.ethz.ch/cosmo\_public/esub-epipe}} \citep{Zuercher2021,Zuercher2022},  a \textsc{python} package that takes a single properly formatted \textsc{python} executable file and runs it as either single core job or multi-core job,  on both local machines and computer clusters via job schedulers. 

\subsection{\textsc{SExtractor} structural parameters estimate}
\label{sextractor_params_estimate}

The initial estimate of the structural parameters is performed using \textsc{sextractor}.  \textsc{morphofit} first stores the science (\textit{sci}),  the root mean square (\textit{rms}) and the exposure time (\textit{exp}) images paths, the photometric parameters for \textsc{sextractor} and the paths to \textsc{sextractor} executable and required files into an \textsc{hdf5} file.  Then,  in the main script,  it derives the non-photometric parameters needed by \textsc{sextractor} ($\mathrm{GAIN}$, $\mathrm{SATURATION}$, $\mathrm{EXPOSURE\_TIME}$, $\mathrm{MAGNITUDE\_ZEROPOINT}$),  the image background value and the PSF full width at half maximum (FWHM).  The non-photometric parameters are obtained by reading them from the image headers.  If they are not present in the image header,  they can be estimated via survey-specific functions the users might need to implement in \textsc{morphofit} (e.g.  saturation by finding the highest value pixel in the image).  The image background value is estimated by sigma clipping the image,  where each source is masked according to a segmentation map.  We obtain the latter by running \textsc{sextractor} on the image in single image mode.  The resulting background amplitude and rms are given by the median and the standard deviation of the sigma clipped masked image.  The PSF FWHM of the image is estimated by fitting stars with a two-dimensional circular Moffat profile.  The stars positions are provided by the user by means of an external catalogue (see Section  \ref{psfmeasurement}).  The different estimates of the FWHMs from the stars in the image are then median combined as described in Section  \ref{psfmeasurement}.

In order to exploit the full potential of having multi-wavelength data,  \textsc{morphofit} runs \textsc{sextractor} in forced photometry mode.  To do that, it needs a detection image.  The latter is created by following the prescription in \citealt{Coe2006} and references therein.  This prescription involves stacking the \textit{sci} images in the available wavelength bands by normalising them according to the rms of the background noise in each waveband. The detection image has the average PSF of all the stacked images. \textsc{sextractor} estimates the isophotal apertures and sizes from this detection image. The inclusion of all the wavelengths allows \textsc{morphofit} to obtain a detection image where objects of different colours are present and detected by \textsc{sextractor}.  If provided,  the \textit{rms} images are used as weight images for \textsc{sextractor}. \textsc{morphofit} creates the \textit{rms} image for the detection image by doing the root sum squared of the \textit{rms} images for each waveband normalised by the rms of the background noise in each waveband.

The final catalogue contains the measurement for all the detected objects in the available wavebands.  \textsc{morphofit} then excludes stars by removing sources which have the $\tt{CLASS\_STAR}$ parameter greater than $\tt{CLASS\_STAR} \ge 0.95$ in all wavebands. It also excludes sources that have $\tt{FLUX\_RADIUS} \le 0$ and $\tt{MAG\_AUTO} > 35$ ABmag.  \textsc{morphofit} also defines for each waveband an aperture and PSF corrected magnitude labelled as $\tt{MAG\_ISO\_CORR}$ following the prescription in \citealt{Coe2006}. 

\begin{figure}
\centering
\includegraphics[width=4.3cm]{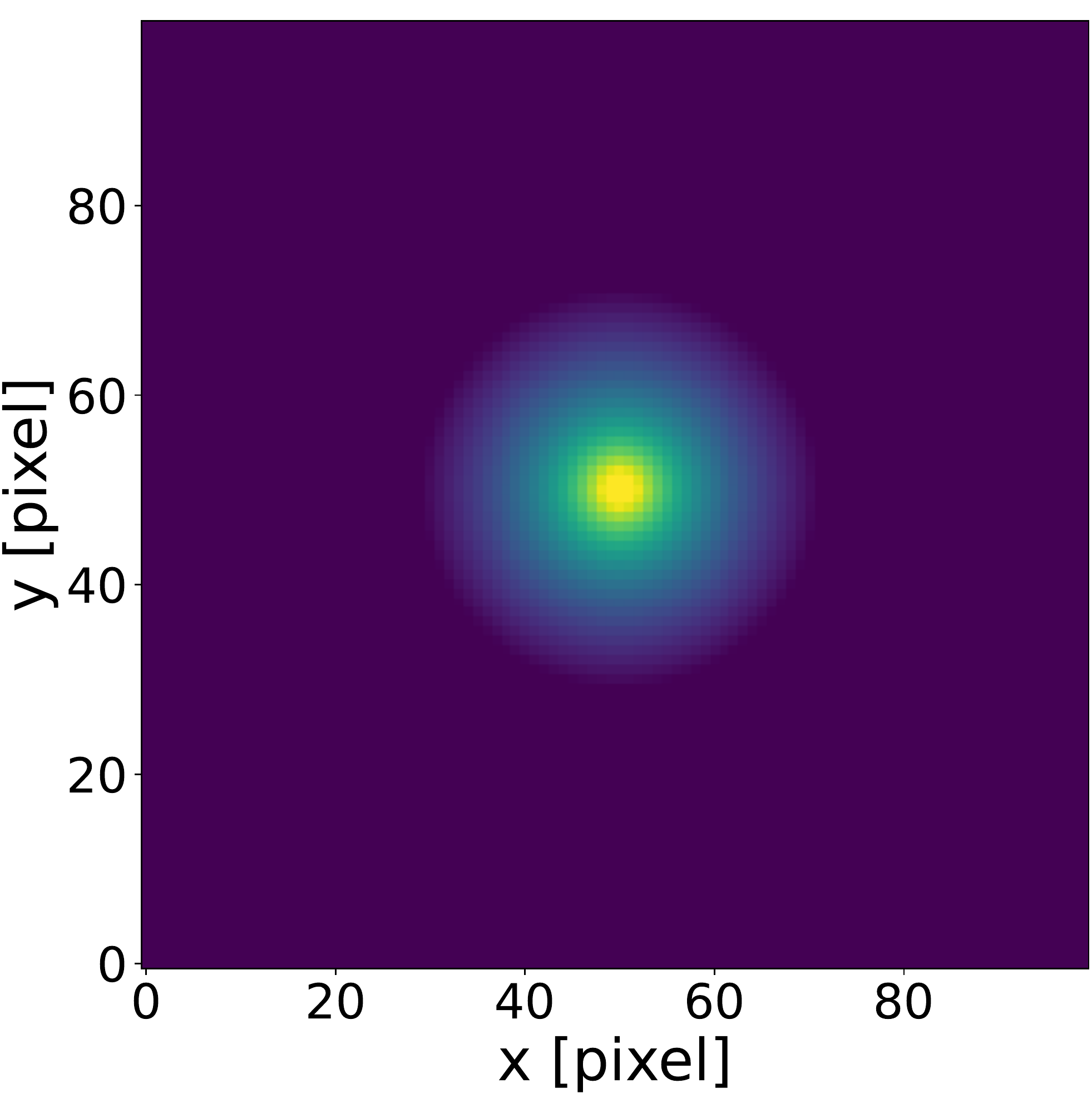}
\includegraphics[width=4.3cm]{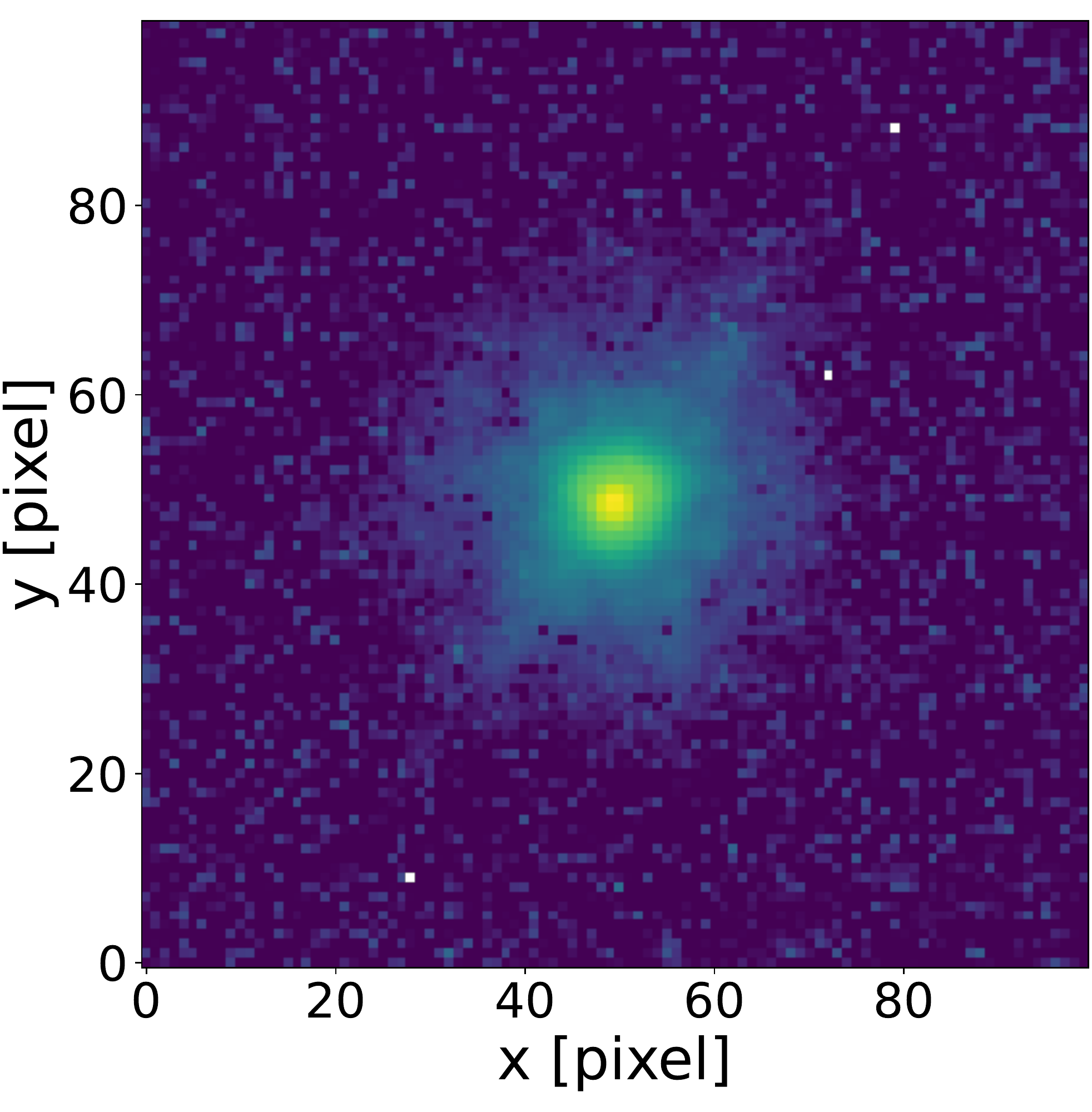}
\includegraphics[width=4.3cm]{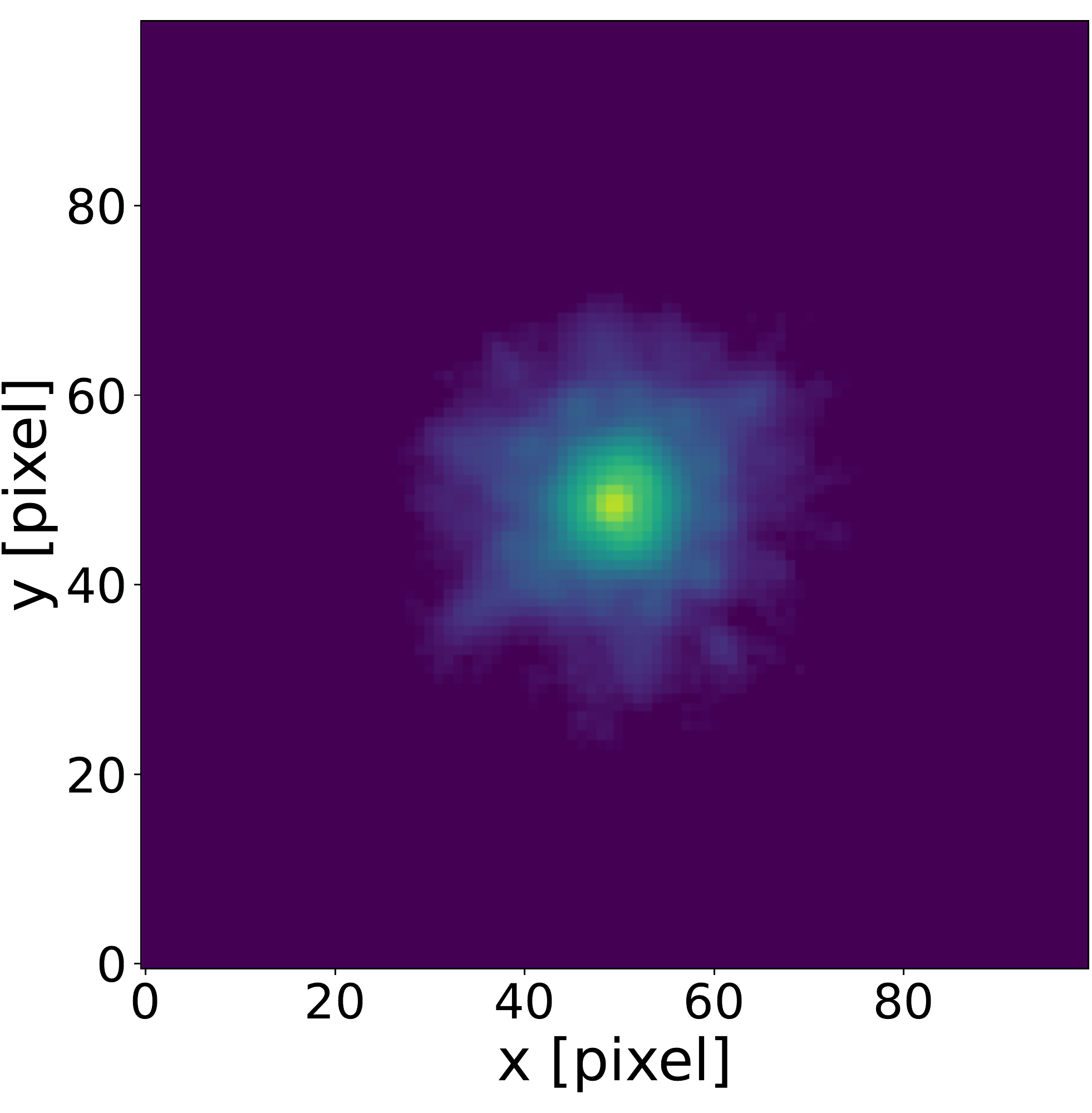}
\includegraphics[width=4.52cm]{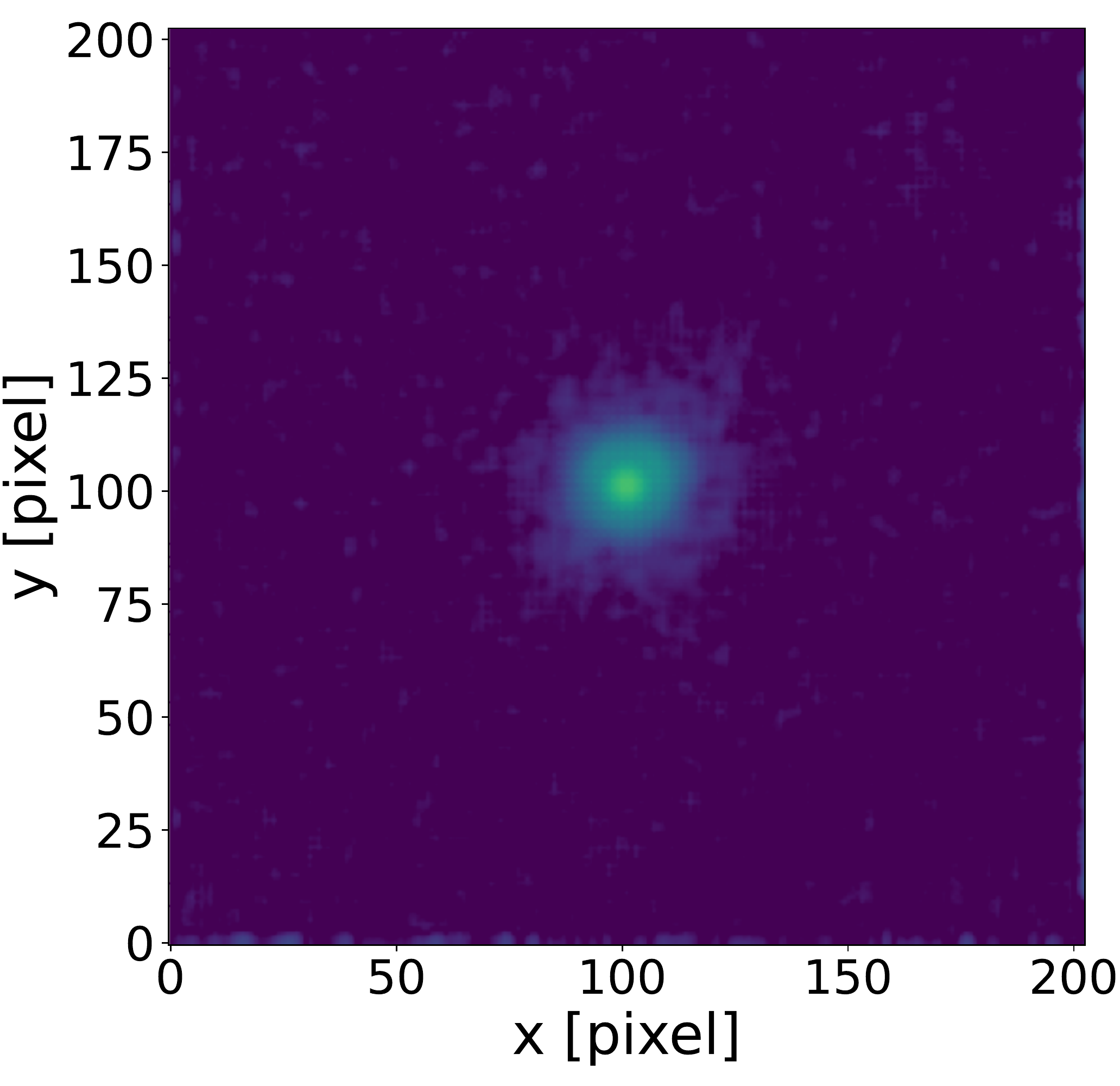}
\caption{The panels show a comparison among the PSF images obtained with the methods described in Section \ref{psfmeasurement}. They refer to those of the Abell S1063 cluster in the \textit{F814W} waveband from the LT23 analysis. From left to right, the panels show the `Moffat PSF', the `Observed PSF',  the `PCA PSF' and the `Effective PSF'.  The first three PSF images are $100 \times 100$ pixels wide, while the`Effective PSF' is $200 \times 200$ pixels since the minimum oversampling factor of the `Effective PSF' relative to the input stars along each axis is 2 (see \textsc{photutils} manual for a detailed explanation).}
\label{psfs_comparison}
\end{figure}

\subsection{PSF images creation}
\label{psfmeasurement}

Accurate surface brightness fitting requires convolution of the galaxy light profile with the image PSF.  \textsc{morphofit} contains a module that creates PSF images using four different methods (more methods can be easily added by the user).  The PSF images can be created for each individual \textit{sci} image (or parts of it) and wavelength.  To create the PSF images,  the package requires an input star catalogue for each waveband covering the same sky portion of the original image.  The star catalogue can be either provided by a third-party source (e.g.  GAIA catalogue, \citealt{Gaia2021}) or generated by the user.  In LT23,  we generate the star catalogue by running \textsc{sextractor} in single image mode on the \textit{sci} image.  Since the measurement of magnitudes and sizes are independent of the seeing estimate given to \textsc{sextractor}, we set the $\tt{CLASS\_STAR}$ keyword to the nominal ACS value of $0.1\ arcsec$. Then, we select as stars those belonging to the locus of fixed size in the magnitude-size plane. Once selected, we visually inspect them to ensure that they are not saturated and they are sufficiently isolated.  Using the stars positions,  \textsc{morphofit} creates $M \times N$ pixels wide PSF images for every waveband,  where $M,\ N$ are user defined.

The PSF modelling is a critical aspect of the structural parameters derivation.  A wrong choice of the PSF model could severely affect the measurements.  Each image might have peculiarities that make the choice of the PSF used for fitting not universal.  Since \textsc{morphofit} can be used on different types of images, either ground-based or space-based,  in order to make it as generalisable as possible,  we implement four different methods to create a PSF image.  The different PSF creation routines are mutuated from methods that are widely used in the literature for different scientific applications.  

Each method has as initial step the creation of $M \times N$ pixels wide stamps around observed stars from the target image.  \textsc{morphofit} then subtracts the background value in each stamps to obtain a background subtracted observed star image. The background value estimate in each stamp is obtained by sigma-clipping the image and taking its median as background estimate.  The star stamps are normalised to the maximum pixel value. The four PSF estimation methods implemented in \textsc{morphofit} are:
\begin{itemize}
\item `Moffat PSF': The implementation of a `Moffat PSF' profile was motivated by a number of general properties it possesses.  The `Moffat PSF' is numerically well-behaved when modelling narrow PSFs in Hubble Space Telescope (HST) images thanks to its polynomial structure \citep{Trujillo2001}.  Additionally,  it constitutes also an analytical approximation to the PSF predicted from atmospheric turbulence theory when $b \sim 4.765$, allowing the modelling of the PSF wings present in real images obtained from ground-based telescopes.  However, as detailed in \citealt{Trujillo2001}, it is not sufficient to consider a circular Moffat PSF profile to model the effects of seeing on the surface brightness distribution when the ratio of the effective radius to the FWHM is less than 2.5.  \citealt{Trujillo2001} provides a prescription to correct for this effect.  To create a `Moffat PSF', \textsc{morphofit} models the star light profile with a two-dimensional circular Moffat function:
\begin{equation}
\phi (x, y) = \mathrm{bkg} +  \frac{\mathrm{amplt}}{\left[ 1 + \frac{(x - x_0)^2 + (y - y_0)^2}{a^2} \right]^{b}} \  ,
\end{equation}
where $\mathrm{amplt}$ is the model amplitude,  $(x_0, y_0)$ are the positions of the Moffat model maximum,  $a = \mathrm{FWHM} / \left(2 \sqrt{2^{1/b}-1} \right)$ is the core width of the Moffat model,  and $b$ is the Moffat model power index.  We also add a background term,  $\mathrm{bkg}$,  to account for any residual contamination after the background subtraction in the $M \times N$ star images.  Using stars in both simulated and real images (see section \ref{section:comparison_with_catalogues}), we check that the best-fitting $\mathrm{bkg}$ values are consistent with zero within the errors.  The best-fitting $a$, $b$, $\mathrm{amplt}$,  and $\mathrm{bkg}$ of each star are then stored and the mean values of the best-fitting parameters are computed.  \textsc{morphofit} then constructs the PSF image by using the analytical profile of the circular Moffat function with parameters given by the mean values aforementioned.

\item `Observed PSF': The `Observed PSF' is a fast method to approximate the PSF seen on the images that already includes the different seeing and detector effects.  The final PSF image is given by the median of the normalised background subtracted star stamps.

\item `PCA PSF': The `PCA PSF' is a method that is commonly used in planetary science where the detailed star modelling is crucial \citep{Amara2012}. We first compute a mean star stamp using all the background-subtracted normalised star stamps. We subtract the mean star stamp from the latter.  \textsc{morphofit} flattens the mean subtracted star stamps and stores them into an array of size $\mathrm{s = [number\ of\ stars, stamp\ size^2]}$. It then performs a Principal Component analysis (PCA) on this array.  The final PSF image is the linear combination of an user-defined number of Principal components resulting from the PCA.  In principle, by having larger statistics, one can complexify the resulting PSF including the effects of skewness and asymmetries by adding higher order principal components \citep{Herbel2018}.

\item `Effective PSF': The `Effective PSF' is an empirical oversampled model of the HST PSF.  The method relies on the Anderson empirical PSF library \citep{Anderson2000, Anderson2016} that was obtained from observations of star associations with HST.   \textsc{morphofit} builds the PSF image using the \textsc{EPSFBuilder} class of the \textsc{photutils} \textsc{python} package \citep{photutils}. This package relies upon the method in \citealt{Anderson2000, Anderson2016} to build the final PSF image.  Its use is recommended only in the case of HST images.
\end{itemize}

In Figure \ref{psfs_comparison}, we show a comparison among the four PSF types from the LT23 analysis created using stars from the HST \textit{F814W} images of the Abell S1063 cluster. The `Moffat PSF' appears smooth and regular given its analytical nature.  The `Observed PSF' and the `PCA PSF' reproduce details from the actual stars in the HST images,  including the wings of the PSF profile. The `Effective PSF' presents a smoother PSF with less pronounced wings. The difference among the last three methods is mostly due to the small sample of stars used to construct them, an effect that is less pronounced for the `Observed' and the `PCA' PSF.  The latter is able to handle the small sample of stars by controlling the number of principal components used to construct it. The former provides a good result even with just one object by definition of observed star.  In LT23,  we do not use Tiny Tim \citep{Krist2011} to model the HST PSF.  Tiny Tim model PSFs are designed to be a good approximation of the single exposure ".flt" images and not of the multidrizzled images\footnote{See \href{https://www.stsci.edu/hst/instrumentation/focus-and- pointing/focus/tiny-tim-hst-psf-modeling}{https://www.stsci.edu/hst/instrumentation/focus-and- pointing/focus/tiny-tim-hst-psf-modeling} for more details.}.

\subsection{\textsc{GALFIT} run on stamps}
\label{galfit_on_stamps}

In the flowchart shown in Figure \ref{flowchart},  the parameter values estimated with \textsc{sextractor} constitute the starting point for the \textsc{galfit} fit on stamps module.  Since the run on stamps requires as input a catalogue,  in principle any other external one can be used,  provided that the appropriate properties are included. 

For the fit on stamps,  \textsc{morphofit} requires the user to provide a target and a source catalogue.  The target catalogue contains the initial parameter values of the galaxies we are interested in.  The source catalogue contains the properties of all the remaining sources in the image.  This allows the user to simultaneously fit neighbouring objects falling in the stamps cut around the target galaxies.

Both the target and the source catalogues need to contain the properties that are required as initial values for the surface brightness fit with \textsc{galfit}.  The surface brightness components implemented in \textsc{morphofit} are the S\'ersic,  the deVaucouleurs,  and the exponential disk profiles. These require the galaxy centroids,  the magnitudes,  the effective radii,  the S\'ersic indices,  the axis ratios,  and the position angles.  \textsc{morphofit} implements the possibility of fitting multiple components to the same galaxy (among the three previously mentioned),  therefore the target and source catalogues additionally need information about which profile is being fit,  the component number and whether to keep the values fixed or let them free to vary in the fit.  The column names corresponding to the needed properties are provided by the user via command-line when running the code.

The first step of this module involves the creation of an \textsc{hdf5} file that contains the \textit{sci},  \textit{rms},  \textit{exp} and segmentation map ($\mathrm{seg}$) images paths.  Additionally, the table includes the target and source catalogues path,  as well as a list of all the possible fit combinations for the different kind of PSF images,  sigma images, background estimations,  and wavebands for all the target galaxies.

In the main script,  the first step is to create cutouts around each target galaxy.  The cutouts are centred on the galaxy centroid.  Following \citealt{Vikram2010}, the $x_{\mathrm{size}}$ and $y_{\mathrm{size}}$ pixel sizes of the stamps are given by
\begin{eqnarray}
x_{\mathrm{size}} &=& r_{\mathrm{e}} \gamma \left( |\cos{\theta}| + q |\sin{\theta}| \right) \ , \\
y_{\mathrm{size}} &=& r_{\mathrm{e}} \gamma \left( |\sin{\theta}| + q |\cos{\theta}| \right) \  ,
\label{size_equation}
\end{eqnarray}
where $r_{\mathrm{e}}$ is the effective radius of the target galaxy in pixels, $\gamma$ is an user-defined enlarging factor,  $\theta$ is the angle between the galaxy major axis and the image x-axis,  and $q$ is the galaxy axis ratio.  Since the estimates might be different for different wavebands,  the user can provide a reference band, whose values are used to create the cutout around the target galaxy. This ensures that the cutouts have the same pixel size for every waveband analysed. In LT23,  these parameters are set to values of the \textsc{sextractor} estimates in the HST \textit{F814W} waveband \footnote{See SExtractor's user-manual for a detailed explanation of the keywords \href{https://sextractor.readthedocs.io/en/latest/Param.html}{https://sextractor.readthedocs.io/en/latest/Param.html}}:
\begin{eqnarray}
x &=& \tt{XWIN\_IMAGE} \ , \\
y &=& \tt{YWIN\_IMAGE} \ , \\
r_{\mathrm{e}} &=& \tt{FLUX\_RADIUS} \ , \\
\gamma &=& 20 \ , \\
\theta &=& \tt{THETAWIN\_SKY} * \pi / 180 \ , \\
q &=& \tt{BWIN\_IMAGE / AWIN\_IMAGE} \ .
\end{eqnarray}

Rather than considering all the galaxies that fall inside the stamp,  \textsc{morphofit} simultaneously fit only those neighbouring objects that have distance $d$ in arcseconds from the target galaxy that is less than
\begin{equation}
d < d_{\mathrm{max}} = \sqrt{(x_{\mathrm{size}} / 2)^2 + (y_{\mathrm{size}} / 2)^2} \times \mathrm{pixel\ scale} \  ,
\end{equation}
where $x_{\mathrm{size}}$ and $y_{\mathrm{size}}$ are given by Eq. \ref{size_equation} with $\gamma = 10$ and $\mathrm{pixel\ scale}$ is the image pixel scale in $arcsec / pixel$.

\textsc{galfit} requires a `bad pixel image' to know which pixels of the image need to be fit.  \textsc{morphofit} creates this `bad pixel image' by assigning zeros to pixels belonging to the target and neighbouring galaxies,  and greater than zero values to pixels belonging to sources not included in the fit.  The latter might be galaxies having distances greater than $d$ from the target galaxy or stars which are masked during the fitting procedure.  In LT23,  the `bad pixel image' is created using the \textsc{sextractor} output segmentation map. 

The sigma image used in \textsc{galfit} can be provided by the user, generated by \textsc{morphofit} or internally generated by \textsc{galfit} itself.  We refer to them as `external sigma image', `custom sigma image',  and `internal sigma image', respectively.  \textsc{morphofit} creates the last two, assuming the \textit{sci} images are provided in units of $\mathrm{electrons/s}$, according to the following prescriptions:
\begin{itemize}
\item `custom sigma image': \textsc{morphofit} creates the sigma image $\mathrm{\sigma_{img}}$ following \textsc{galfit}'s user manual\footnote{\href{users.obs.carnegiescience.edu/peng/work/galfit/README.pdf}{users.obs.carnegiescience.edu/peng/work/galfit/README.pdf}}
\begin{equation}
\mathrm{\sigma_{img}} = \sqrt{\mathrm{(sci - bkg\_amp) / exp + rms^2}} \ , 
\end{equation}
where $\mathrm{sci}$ is the science image,  $\mathrm{bkg\_amp}$ is the amplitude of the science image background in the same units as the \textit{sci} image,  $\mathrm{exp}$ is either a fixed exposure time value or an exposure time image,  and $\mathrm{rms}$ is the root mean square image.  The negative values of the \textit{sci} image are set to zero in creating the sigma image.  The sigma image is further smoothed with a Gaussian kernel having a width value of $w = 1.2$ pixels and its pixel values are set to $\mathrm{nan}$ where the \textit{rms} image pixel values exceeds $10^5$. To avoid systematic errors and have a correct value of the reduced $\chi ^2$,  the units of the images and their header keywords need to be adjusted such that $\mathrm{GAIN \times sci \times NCOMBINE = electrons}$, where $\mathrm{NCOMBINE}$ is the number of images used to create the \textit{sci} image.  In LT23,  we work with HST images in units of $\mathrm{electrons/s}$. Therefore, we set the following drizzled \textit{sci} image keywords to:
\begin{eqnarray}
\mathrm{EXPTIME} &=& 1 \ , \\
\mathrm{GAIN} &=& \mathrm{exptime} \ , \\
\mathrm{NCOMBINE} &=& 1 \ ,
\end{eqnarray}
where $\mathrm{exptime}$ is the drizzled image total exposure time in seconds,  $\mathrm{EXPTIME}$ is the image header keywords detailing the \textit{sci} image exposure time,  $\mathrm{GAIN}$ details the detector instrumental gain,  and $\mathrm{NCOMBINE}$ is the number of images combined to produce the \textit{sci} image. 

\item `Internal sigma image': in order to let \textsc{galfit} correctly create the sigma image,  the \textit{sci} image needs to be converted from $\mathrm{electrons/s}$ units to $\mathrm{ADU}$. To do that,  \textsc{morphofit} divides the \textit{sci} image by the instrumental gain and multiplies it by the \textit{exp} image or, if not provided,  by the exposure time.  A similar operation is performed on the background amplitude value.  The package further sets the \textit{sci} image header keyword $\mathrm{BUNIT} = \mathrm{ADU}$, the $\mathrm{GAIN}$ keyword to its instrumental value and the image zeropoint is updated such that $\mathrm{zpt}^{'} = \mathrm{zpt} - 2.5 \log_{10}{(\mathrm{GAIN})}$.
\end{itemize}

\textsc{galfit} requires also an estimate of the sky background of an image.  \textsc{morphofit} provides the possibility to either pass an user-defined value or to let \textsc{GALFIT} freely fit the background amplitude and gradient in the x and y directions.  In LT23,  the initial estimate of the sky background is obtained from the sigma clipped masked image (see Section \ref{sextractor_params_estimate}).  \textsc{morphofit} includes also the possibility to provide \textsc{galfit} with a parameter constraints file.

\begin{figure*}
\centering
\includegraphics[width=5.8cm]{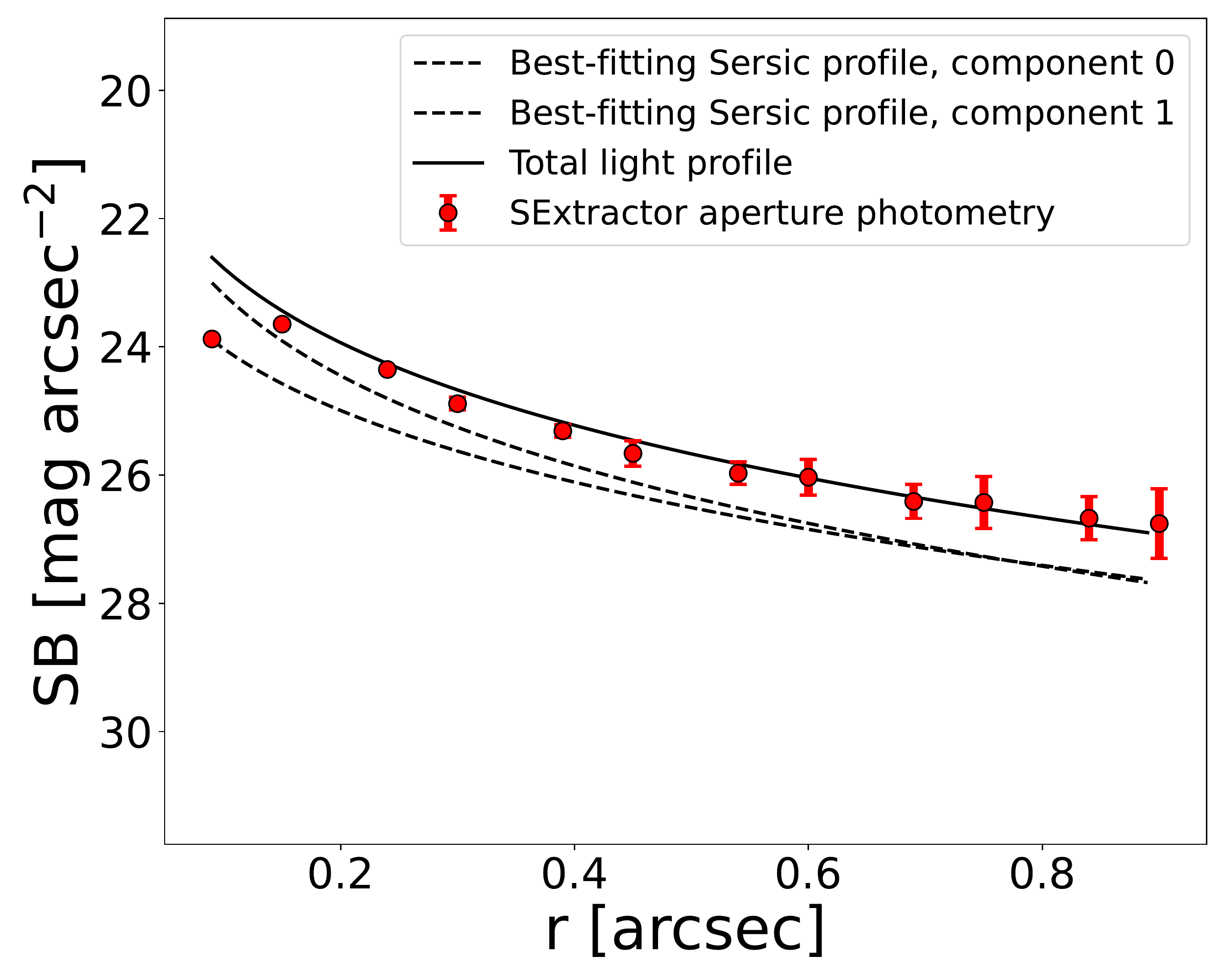}
\includegraphics[width=5.8cm]{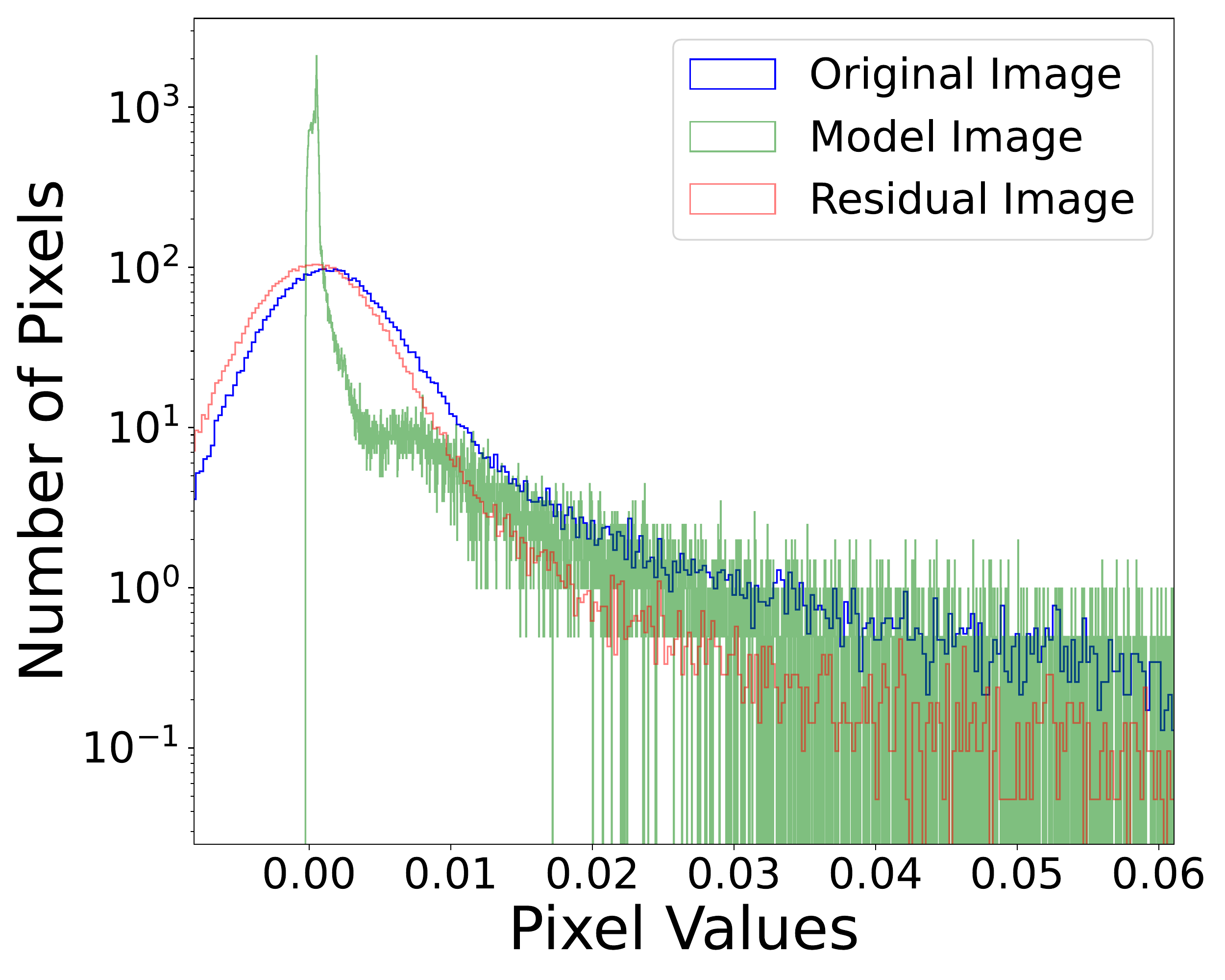}
\includegraphics[width=5.8cm]{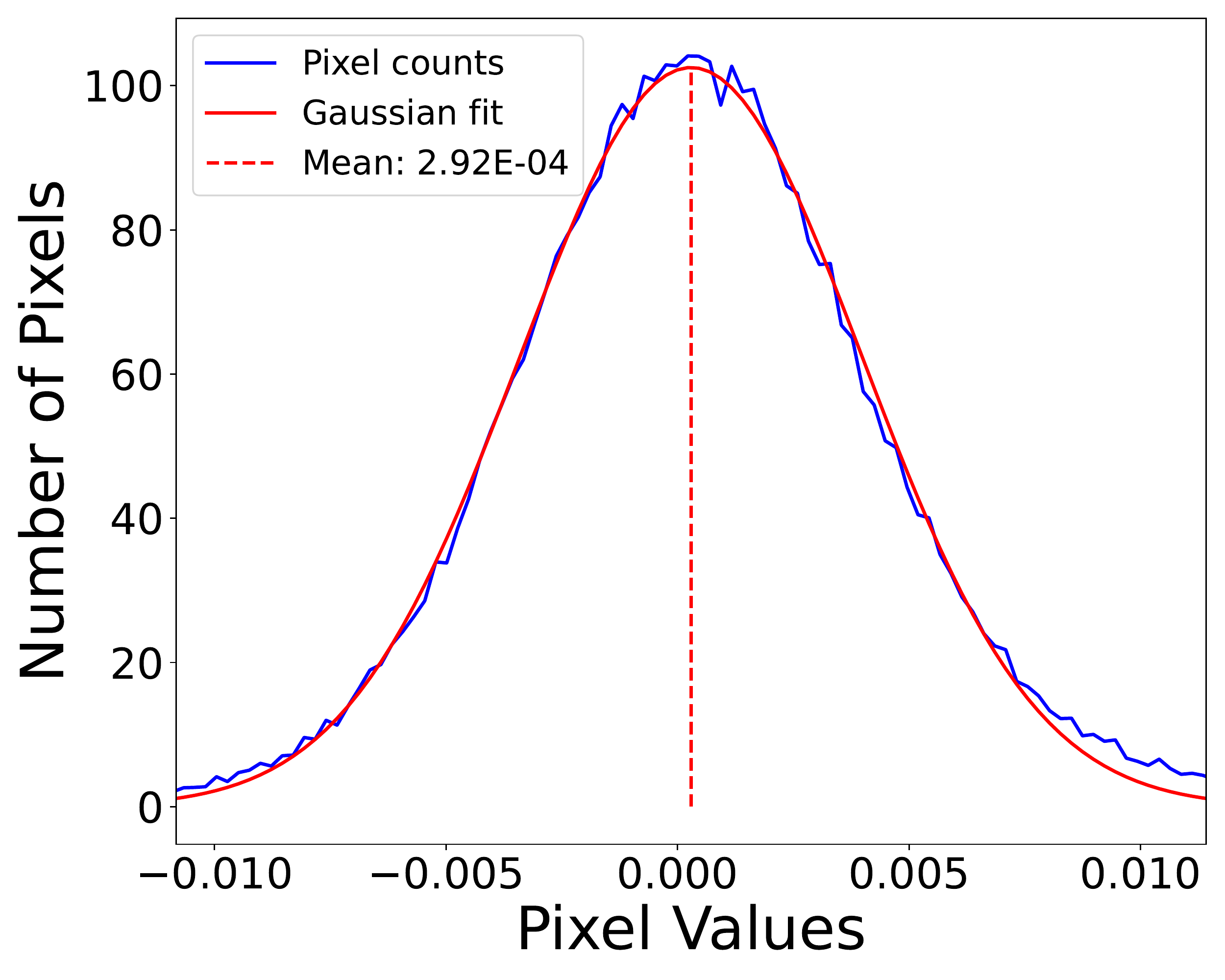}
\includegraphics[width=5.8cm]{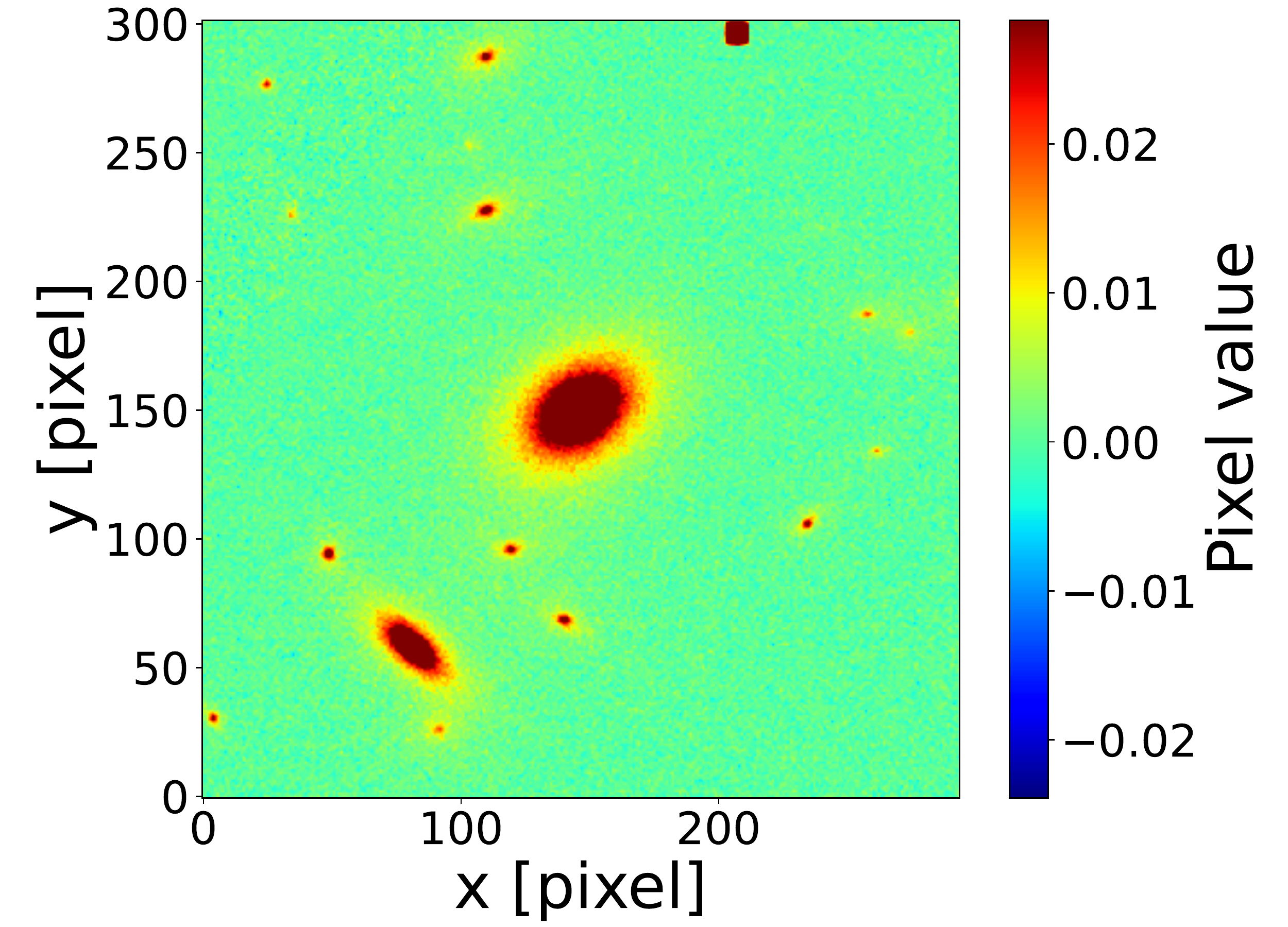}
\includegraphics[width=5.8cm]{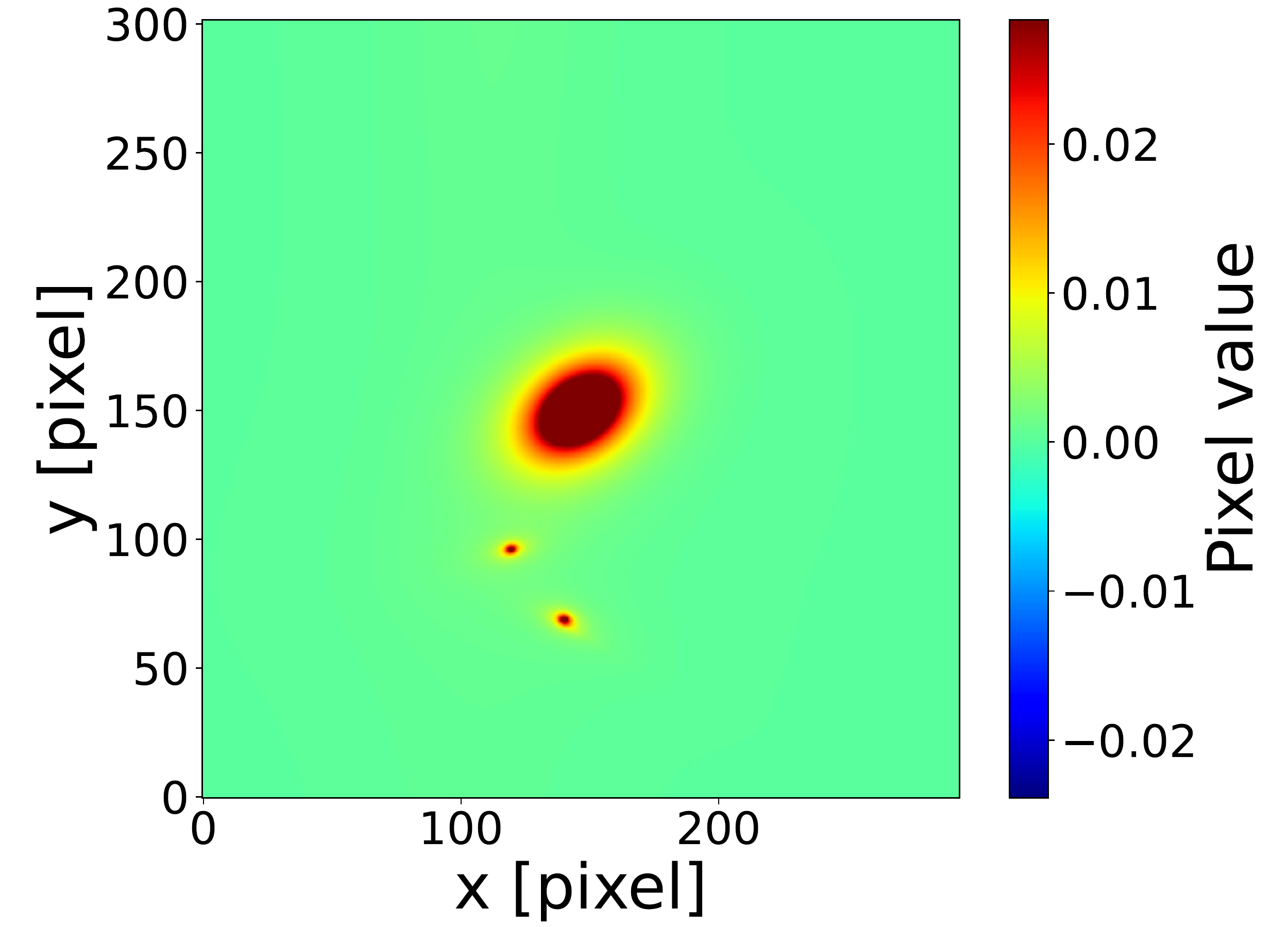}
\includegraphics[width=5.8cm]{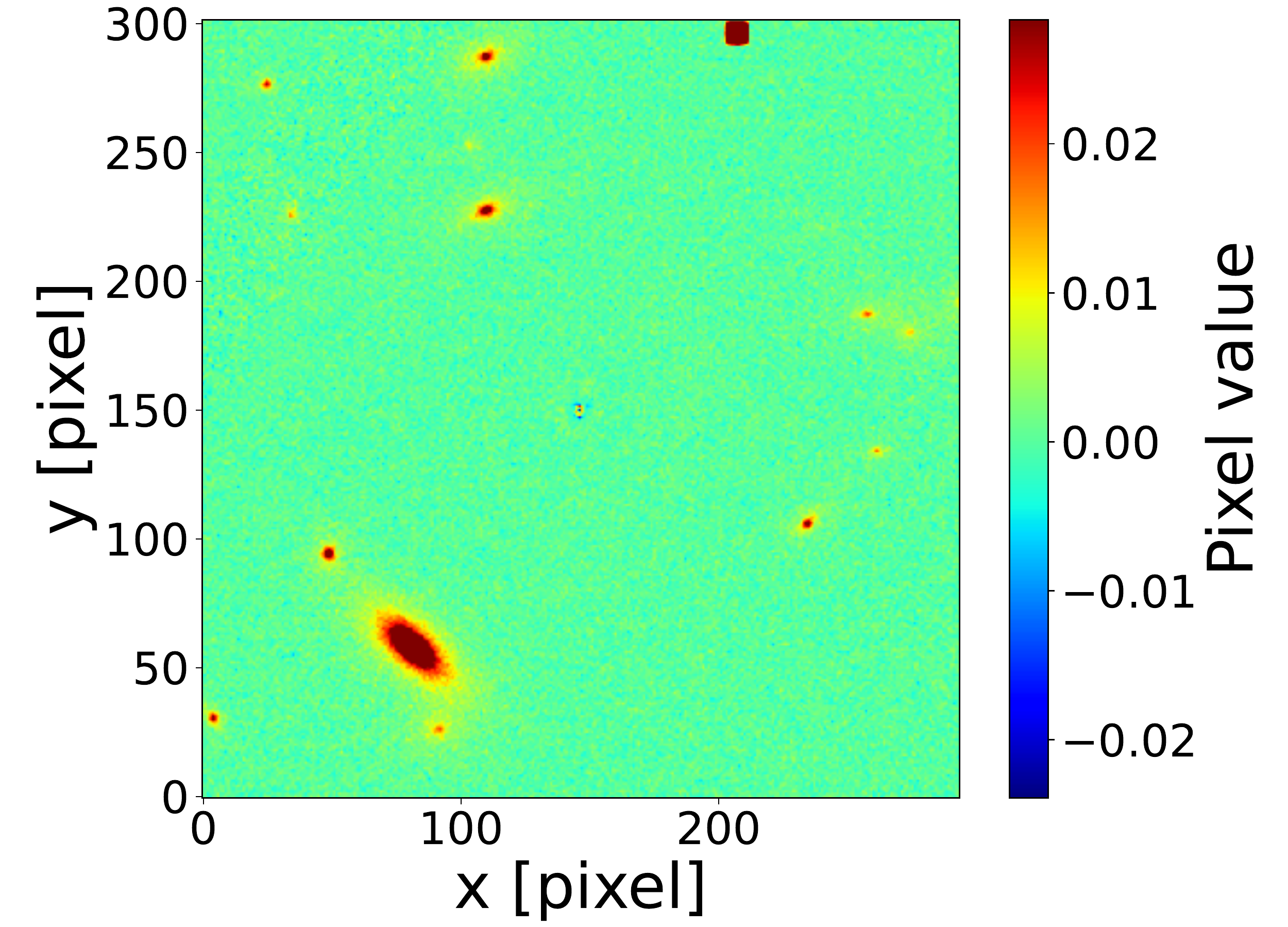}
\caption{This figure shows the diagnostic plots automatically created by \textsc{morphofit}.  Upper left panel shows the comparison between the two component surface brightness fit and the \textsc{sextractor} aperture photometry for the target galaxy.  The upper central panel shows the pixel value histograms of the original,  model and residual images generated by \textsc{galfit}. The upper right plot shows a Gaussian fit to the pixel values distribution of the residuals image.  The bottom panels show the original,  model and residuals images after the fit with \textsc{galfit}.}
\label{diagnostic_plots}
\end{figure*}

\textsc{morphofit} creates the input file for \textsc{galfit},  runs it on the individual stamps,  reads and saves in a \textsc{FITS} table the best-fitting parameters from the \textsc{galfit} output files.  It also creates diagnostic plots for each fitted galaxy.  These diagnostic plots are:
\begin{itemize}
\item original,  model,  and residuals images from the \textsc{galfit} fit;
\item pixel count histograms of original,  model,  and residuals images;
\item Gaussian fit of the residuals image pixel count histogram;
\item comparison between the best-fitting surface brightness profile and the azimutally averaged aperture fluxes. 
\end{itemize}
Figure \ref{diagnostic_plots} shows an example of the diagnostic plots for the fit on a stamp around one of the simulated galaxies created in Section \ref{simulations} in the \textit{F140W} HST waveband.  The galaxy is simulated as a bulge plus disk component.  \textsc{morphofit} is able to correctly fit the double component as shown by the model and the residual images,  and by the comparison between the best-fitting total light profile and the aperture photometry obtained with \textsc{sextractor}.

\subsection{Image regions creation}
\label{regions_creation}

The pipeline,  as illustrated in Figure \ref{flowchart}, involves the use of the results from the fit on stamps for estimating the galaxy properties on increased size images.  These are obtained by creating image regions.  \textsc{morphofit} contains a module that crops or divides into a number $N_{\mathrm{reg}}$ of equally sized regions the \textit{sci}, \textit{rms},  and \textit{exp} images. After creating the \textsc{hdf5} file containing the images paths and the $N_{\mathrm{reg}}$ value,  the main script crops the image (if the appropriate keyword is set to `True') and then divides it into $N_{\mathrm{reg}}$ regions.  The crop routine is further sub-divided into two kinds: a size-based crop,  where the user specifies the x and y pixel ranges,  and a catalogue-based crop,  where the initial image is restricted to the area containing sources inside an user-provided catalogue.

\subsection{\textsc{GALFIT} run on regions}
\label{galfit_on_regions}

This module performs the surface brightness fitting of sources inside the regions created in Section \ref{regions_creation}.  The \textsc{hdf5} file contains the \textit{sci}, \textit{rms},  \textit{exp} and $\mathrm{seg}$ images paths, the to-be-fitted galaxy catalogue path and the list of possible fit combinations for the different kind of PSF images,  sigma images,  backgroud estimations and wavebands.  In the main script, \textsc{morphofit} performs similar operations as the ones described in Section \ref{galfit_on_stamps}. The only difference is that,  in place of creating stamps,  it finds the sources from the user-provided catalogue that fall into the specific region and use their parameter values as initial points of the surface brightness fit.  In LT23,  the fit on regions uses as initial values of the structural parameters those obtained from the \textsc{galfit} run on stamps,  but any user-provided catalogue can be used as input, provided that the appropriate properties are included.  The typical number of simultaneously fitted sources in LT23 is $\sim 20$.

\subsection{\textsc{GALFIT} run on full images}
\label{galfit_on_fullimages}

\begin{figure*}
\centering
\includegraphics[scale=0.09]{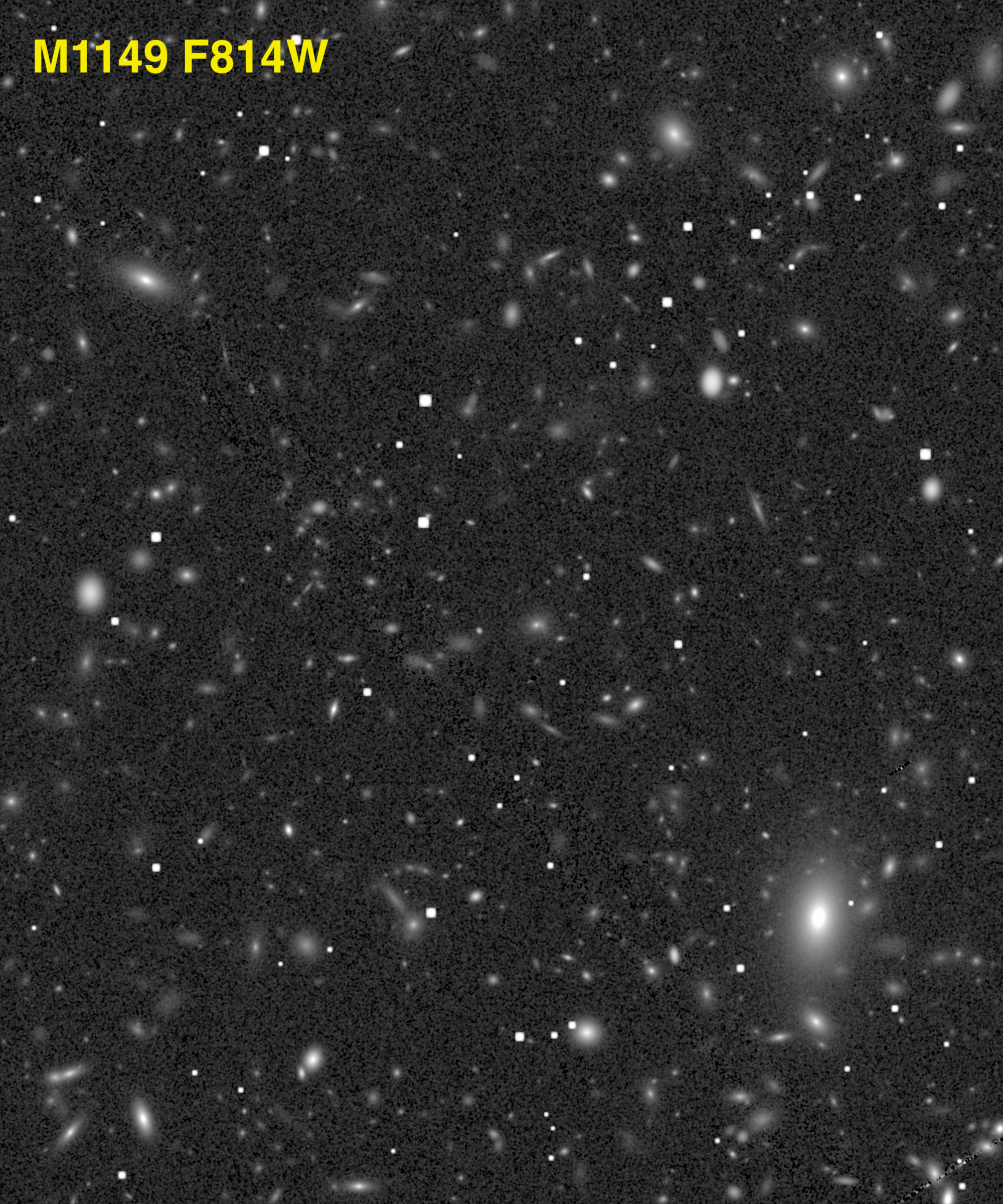}
\includegraphics[scale=0.09]{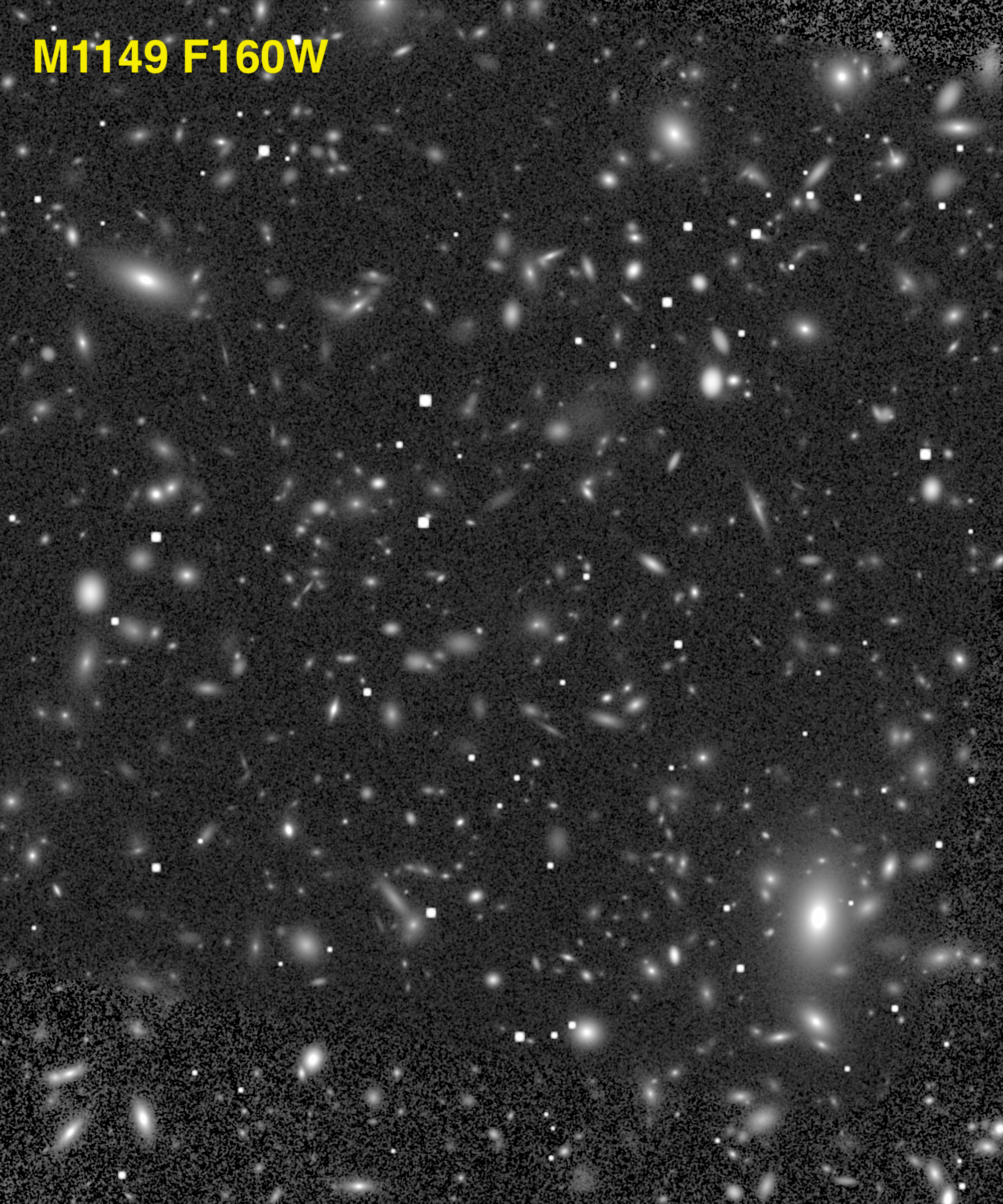}

\vspace{0.2cm}

\includegraphics[scale=0.09]{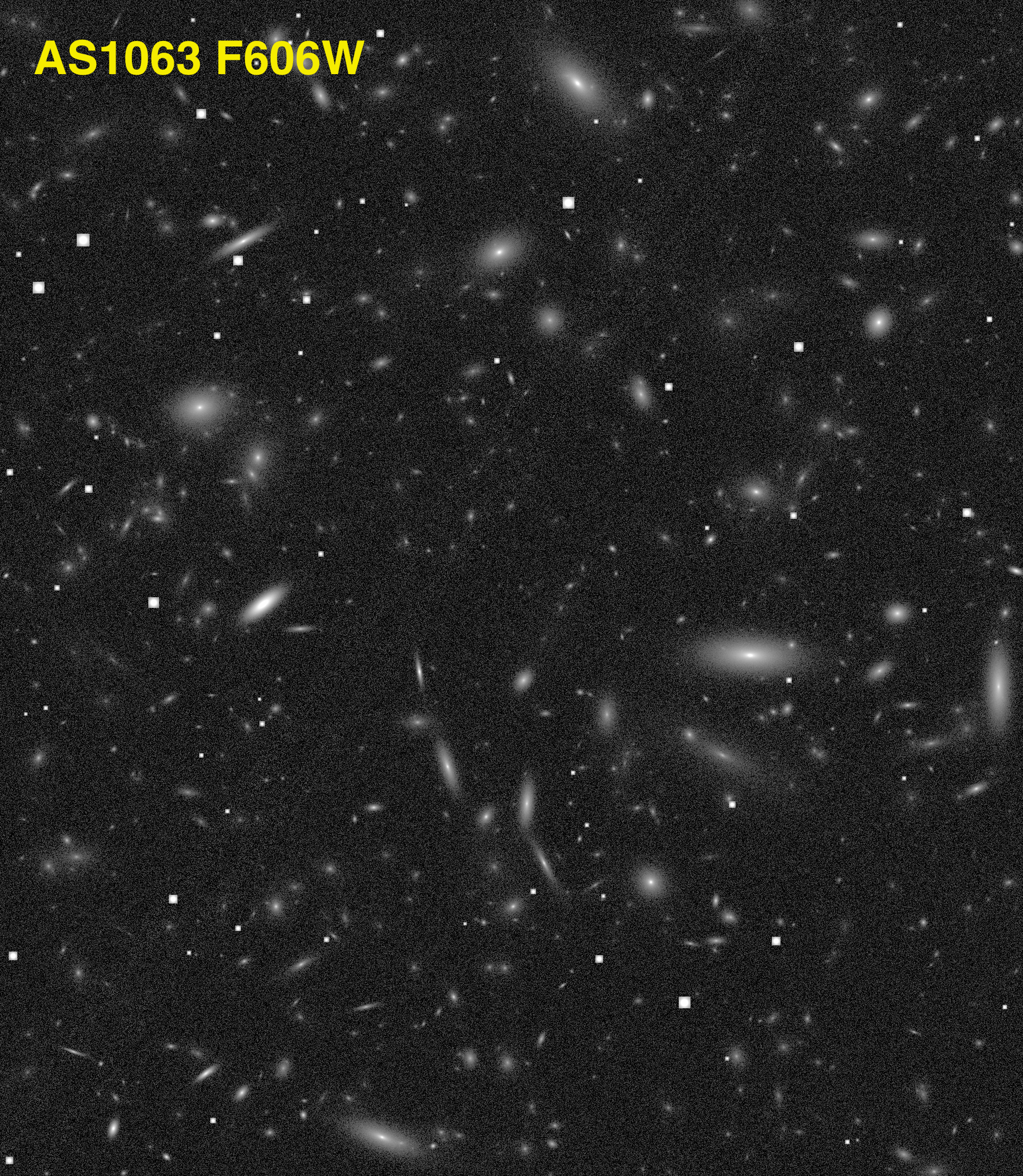}
\includegraphics[scale=0.09]{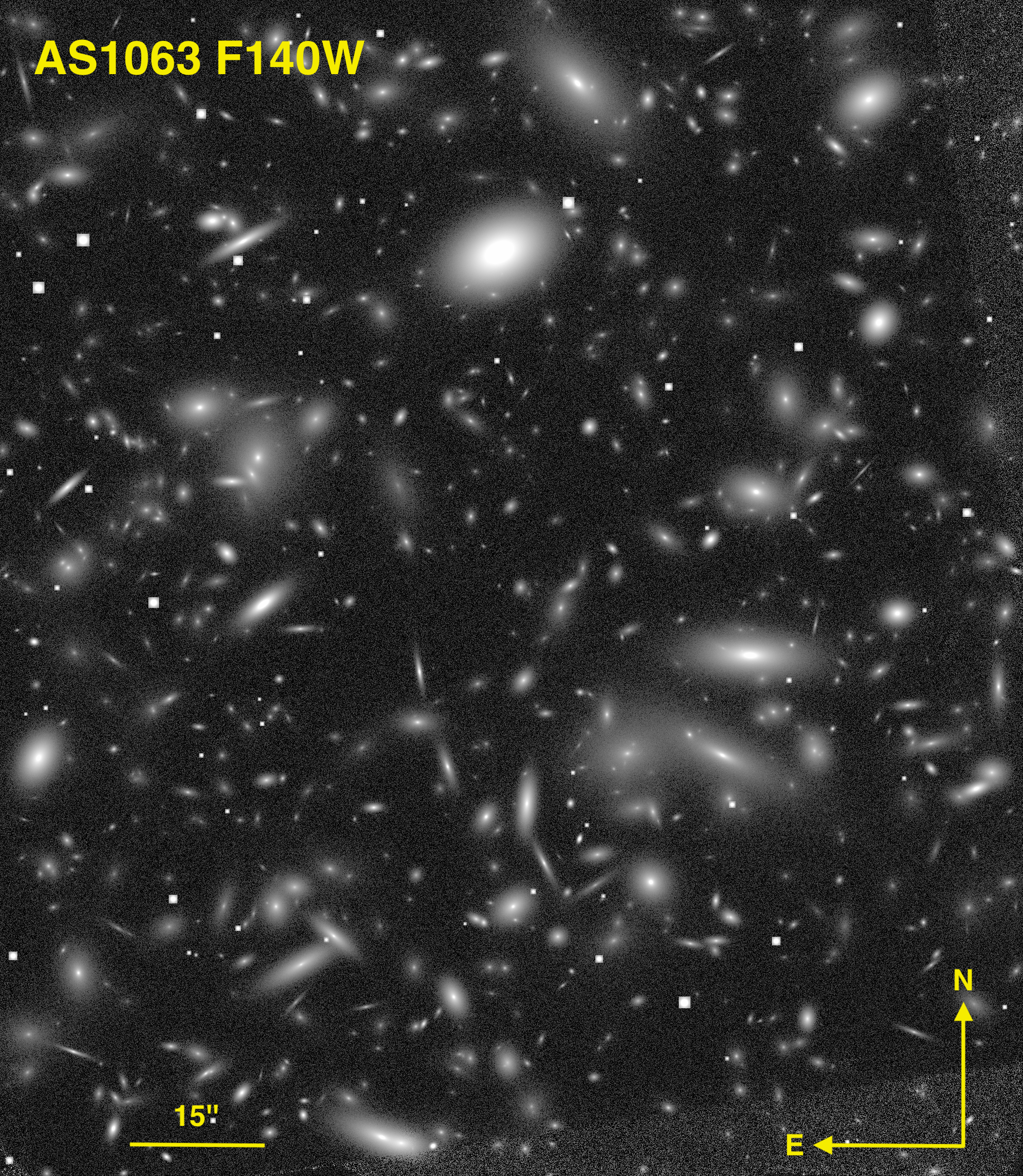}
\caption{This figure shows the simulated HST Frontier Fields images ($2000 \times 2300$ pixels) for the clusters M1149 (upper panels) and AS1063 (lower panels).  The M1149 simulations are shown in the \textit{F814W} (upper left panel) and \textit{F160W} wavebands (upper right panel). Galaxies in this cluster are rendered as single S\'ersic profiles.  The AS1063 simulations are shown in the \textit{F606W} (bottom left panel) and \textit{F140W} wavebands (bottom right panel). Galaxies in this cluster are rendered as bulge plus disk components with different $\mathrm{B/T}$ ratios.  The simulated images have been created matching the background noise and the PSF FWHM of the real Frontier Fields observations.  The matching noise features coming from the real \textit{rms} images are visible for instance in the bottom and upper right corner of the lower right image. The images orientation and scale are the same,  and are highlighted in the lower right figure.}
\label{sim_images}
\end{figure*}

The next step of the pipeline in Figure \ref{flowchart} is the fit of galaxies using the full image.  This \textsc{morphofit} module performs the surface brightness fitting of user-specified sources inside the full image.  This is the preferred choice if the user does not want to follow the procedure highlighted in LT18,LT23 and is not interested in cutting stamps around individual galaxies, but rather prefers to fit them simultaneously.  The module works similarly to those in Sections \ref{galfit_on_stamps}, \ref{galfit_on_regions}.  The main difference is that it does not cut stamps around galaxies and it does not determine which sources fall into the image,  but uses all the sources included in the user-provided catalogue. The main factor limiting the amount of components the user can simultaneously fit with \textsc{galfit} is the Random Access Memory (RAM).  The amount of RAM we could allocate per core was our limiting factor as well.  In this work,  we use the high-performance computing facility `Euler'\footnote{\href{https://scicomp.ethz.ch/wiki/Euler}{https://scicomp.ethz.ch/wiki/Euler}} at ETH Z\"urich. The maximum amount of RAM we could request per core to avoid entering an unreasonable long queue waiting time was 32 gigabyte (Gb). This amount of RAM prevented us to simultaneously fit all the sources visible in Figure \ref{sim_images}.  For the application of this methodology in LT23,  the amount of sources does not constitute a limitation since the aim is to fit only the spectroscopically confirmed cluster member galaxies ($\sim 90$).

\subsection{Best-fitting catalogues creation}
\label{best_fit_cat}

At the end of the \textsc{galfit} on stamps,  regions and full images modules,  a number of surface brightness profile fits with different combinations of PSFs,  sigma images,  backgrounds,  and wavebands are available for each target galaxy.  In LT23,  we use the seven wavebands of the HST Frontier Fields data \citep{Lotz2017},  all four PSF estimation methods described in section \ref{psfmeasurement},  custom and internally generated sigma images, and both fixed and free to vary backgrounds. \textsc{morphofit} provides modules to combine these measurements into a single estimate of the structural parameters per individual galaxy via a weighted sum.  This is crucial to penalise the combinations of PSFs,  sigma images,  and backgrounds that produce incorrect surface brightness fits,  thereby providing a more robust estimate of the structural parameters with realistic errors.  Indeed, it is known that the formal errors computed by \textsc{galfit} account only for statistical uncertainties in the flux \citep{Peng2011}.

The estimates $X_{\mathrm{est}}$ of the structural parameters for each individual galaxy per waveband are obtained via a weighted average of all the measurements $X_i$ obtained with different combinations of PSFs,  sigma images,  and backgrounds:
\begin{eqnarray}
\label{weighted_average_formula}
X_{\mathrm{est}} &=& \frac{ \sum_{i=1}^{N_{\mathrm{comb}}} X_i \ w_i}{\sum_{i=1}^{N_{\mathrm{comb}}} w_i} \ ,  \\
w_i &=& \frac{1}{X_{i,\mathrm{err}}^2} \ ,
\end{eqnarray}
where $w_i$ are the weights, $X_{i,\mathrm{err}}$ the errors quoted by \textsc{galfit} for each fit and `est',  `comb' stand for estimate and combinations,  respectively.

The error on the estimates $\sigma_{\mathrm{est}}$ is computed as the square root of the unbiased weighted estimator sample variance:
\begin{equation}
\sigma_{\mathrm{est}}^2 = \frac{\sum_{i=1}^{N_{\mathrm{comb}}} w_i}{\left( \sum_{i=1}^{N_{\mathrm{comb}}} w_i \right)^2 - \sum_{i=1}^{N_{\mathrm{comb}}} w_i^2} \sum_{i=1}^{N_{\mathrm{comb}}} w_i \left( X_i - X_{\mathrm{est}} \right)^2 \ ,
\end{equation}
where the terms have the same meaning as in Equation \ref{weighted_average_formula}.

\section{Test on simulated HST images}
\label{simulations}

In order to test \textsc{morphofit} and the performance of the pipeline,  we create simulated HST images matching the characteristics of the data used in LT23.  The images are simulated using the non-public code `Ultra Fast image generator'\footnote{\href{https://cosmology.ethz.ch/research/software-lab/ufig.html}{https://cosmology.ethz.ch/research/software-lab/ufig.html}} (\textsc{ufig},  \citealt{Berge2013,Bruderer2016,Herbel2017,Tortorelli2018b,Fagioli2020,Tortorelli2020,Tortorelli2021}).  \textsc{ufig} is a fast software written in \textsc{python} wrapped around \textsc{c++} code that simulates astronomical images in different optical filter bands.  From internally generated or user-provided catalogues, \textsc{ufig} renders galaxies pixelated light profiles, including observational and instrumental effects, such as noise, PSF and pixel saturation.  In the standard implementation,  galaxy properties are drawn from a simple yet realistic galaxy population model that has been calibrated in \citealt{Tortorelli2020,Tortorelli2021}.  Galaxies are rendered on the image with random positions according to S\'ersic light profiles.  

We generate simulations matching the instrumental and observational effects of real HST Frontier Fields survey images for the clusters Abell S1063 (AS1063) and MACS J1149.5+2223 (M1149).  We use the Frontier Fields \textit{rms} image to generate realistic background noise, we set the $\mathrm{GAIN}$, $\mathrm{SATURATION}$, $\mathrm{PIXEL\ SCALE}$ and $\mathrm{MAGNITUDE\ ZEROPOINT}$ of the simulated images to the same values as the real HST images.  Furthermore, we set the FWHM of the image to the value measured on stars on the real image as described in Section \ref{psfmeasurement}.  

We create two kinds of simulations with \textsc{ufig}: galaxies rendered as single S\'ersic profiles following the S\'ersic indices distribution used in \citealt{Tortorelli2020,Tortorelli2021} and galaxies rendered as bulge plus disk with B/T values in the range $0.1 \le \mathrm{B/T} \le 0.9$ and values of the S\'ersic indices fixed to $n=4$ and $n=1$, respectively.  Figure \ref{sim_images} shows an example of the quality of the simulated images we generate.  The upper panels show single S\'ersic profile simulations in the \textit{F814W} and \textit{F160W} wavebands with instrumental parameters matching those from the real M1149 Frontier Fields images.  The bottom panels show the bulge plus disk simulations in the \textit{F606W} and \textit{F140W} wavebands with instrumental parameters matching those from the real AS1063 Frontier Fields images.  The simulation of a realistic noise is particularly visible in the \textit{F140W} and \textit{F160W} wavebands, where noise features coming from the real \textit{rms} images are visible. The differences with respect to the observed Frontier Fields images reside in the simplistic treatment of the stars that are rendered as circular Moffat profiles in the simulations, in the absence of galaxy clustering and in the more complicated morphology of real galaxies. However, these differences do not constitute a problem, since our aim is to test whether our methodology is able to recover the input magnitudes, sizes and S\'ersic indices of the simulated objects, and whether the structural parameters estimated on real data are consistent with already existing literature values.

\subsection{Single S\'ersic profile simulated galaxies}
\label{single_sersic_sims}

\begin{figure*}
\centering
\includegraphics[scale=0.22]{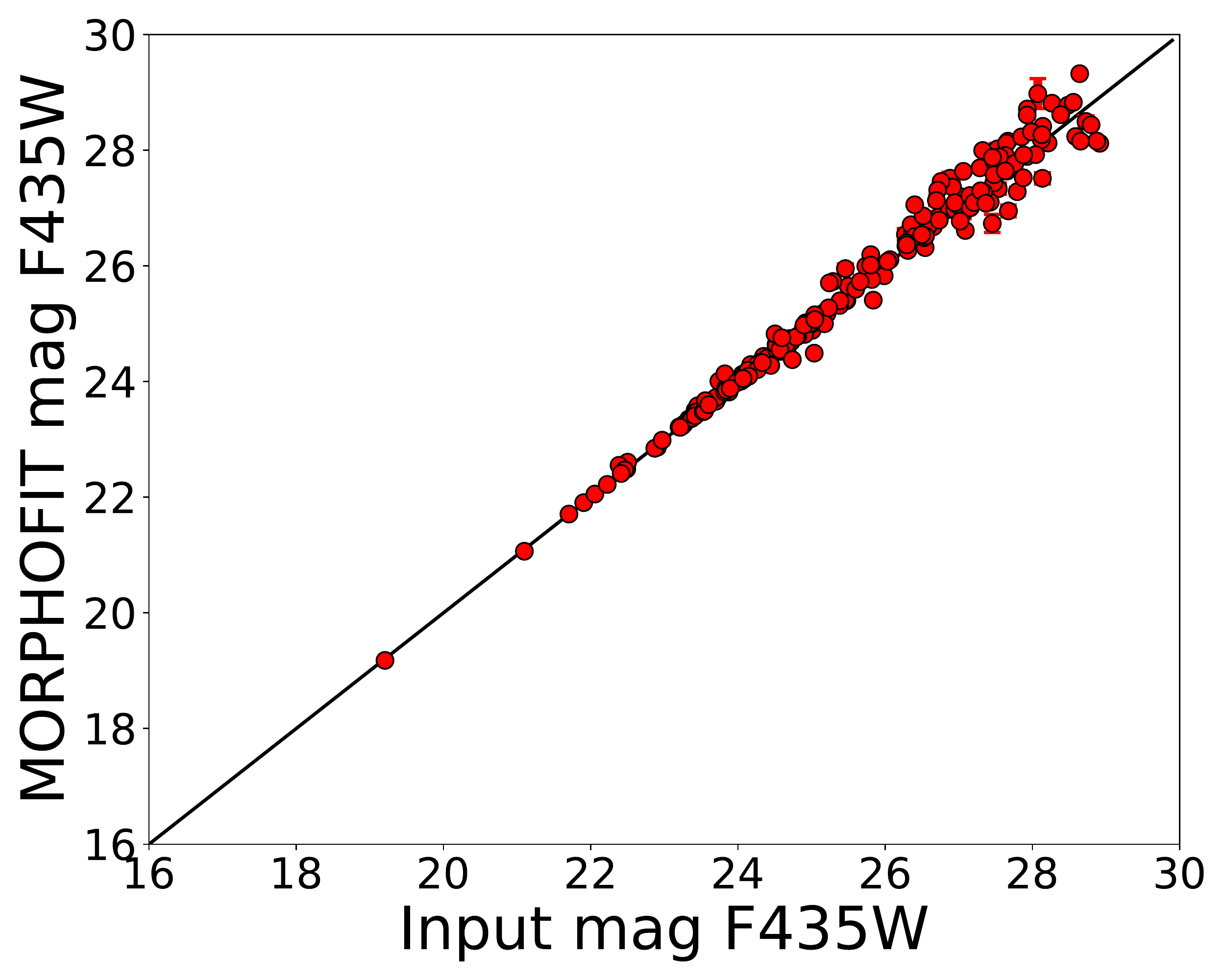}
\includegraphics[scale=0.22]{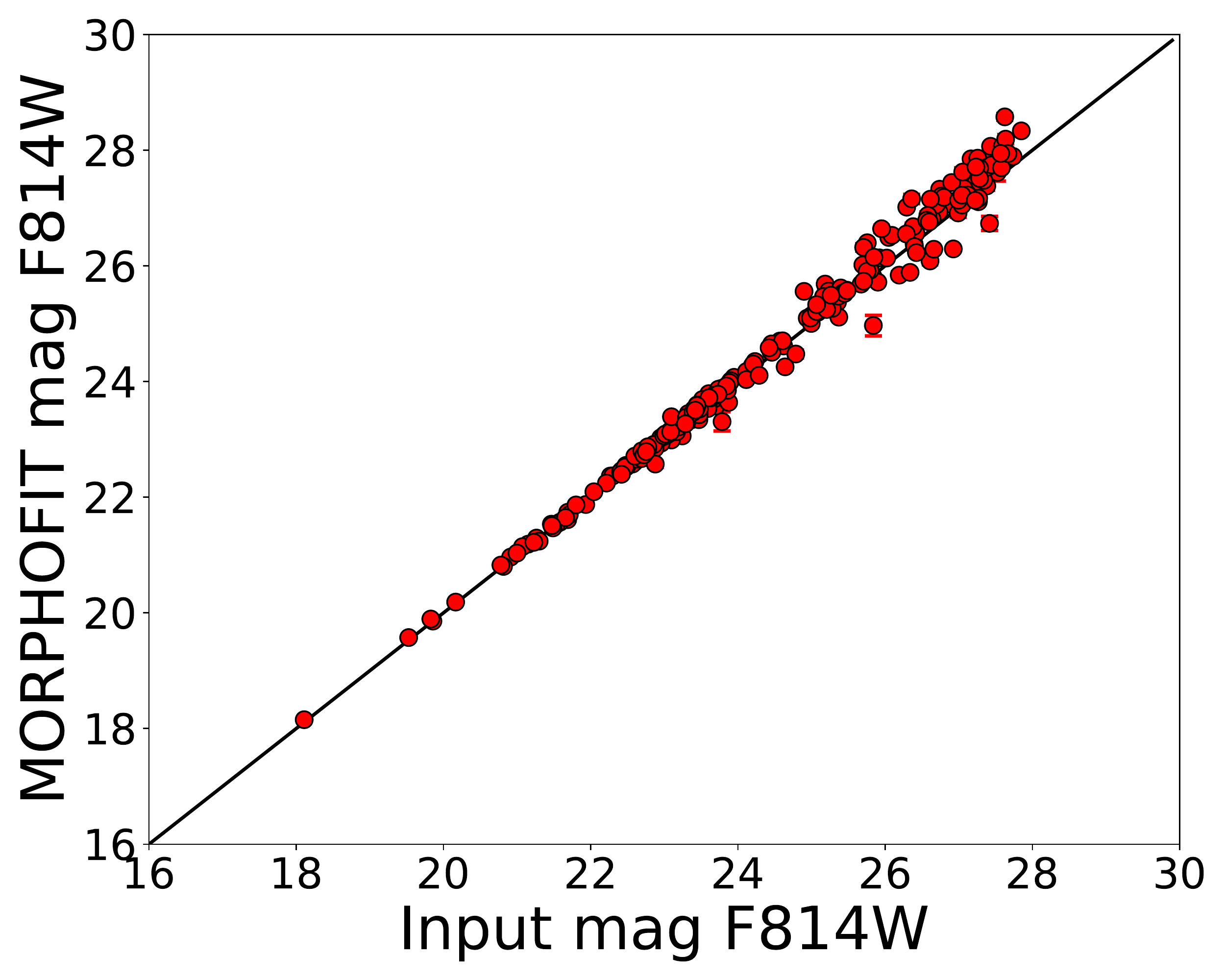}
\includegraphics[scale=0.22]{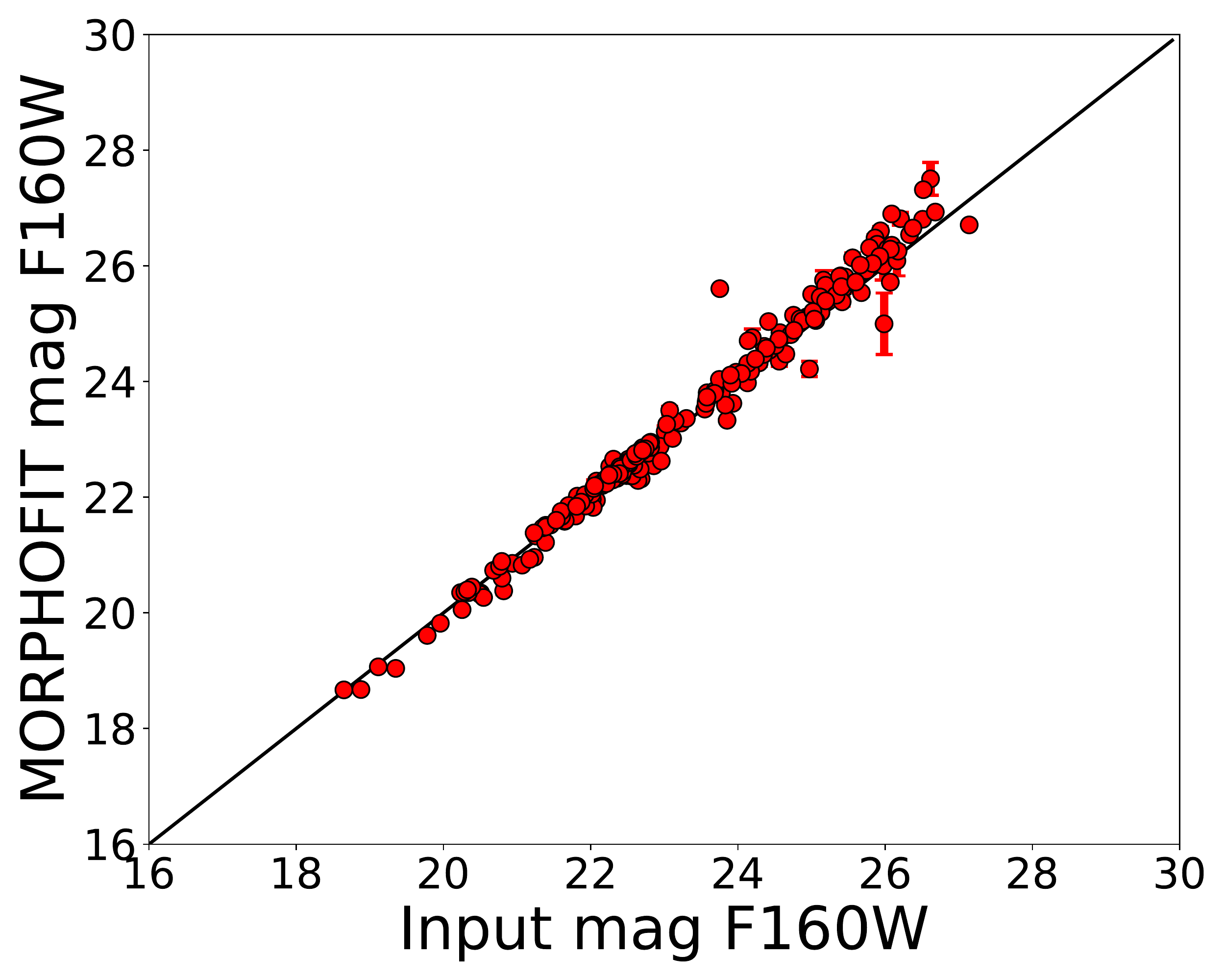}
\includegraphics[scale=0.22]{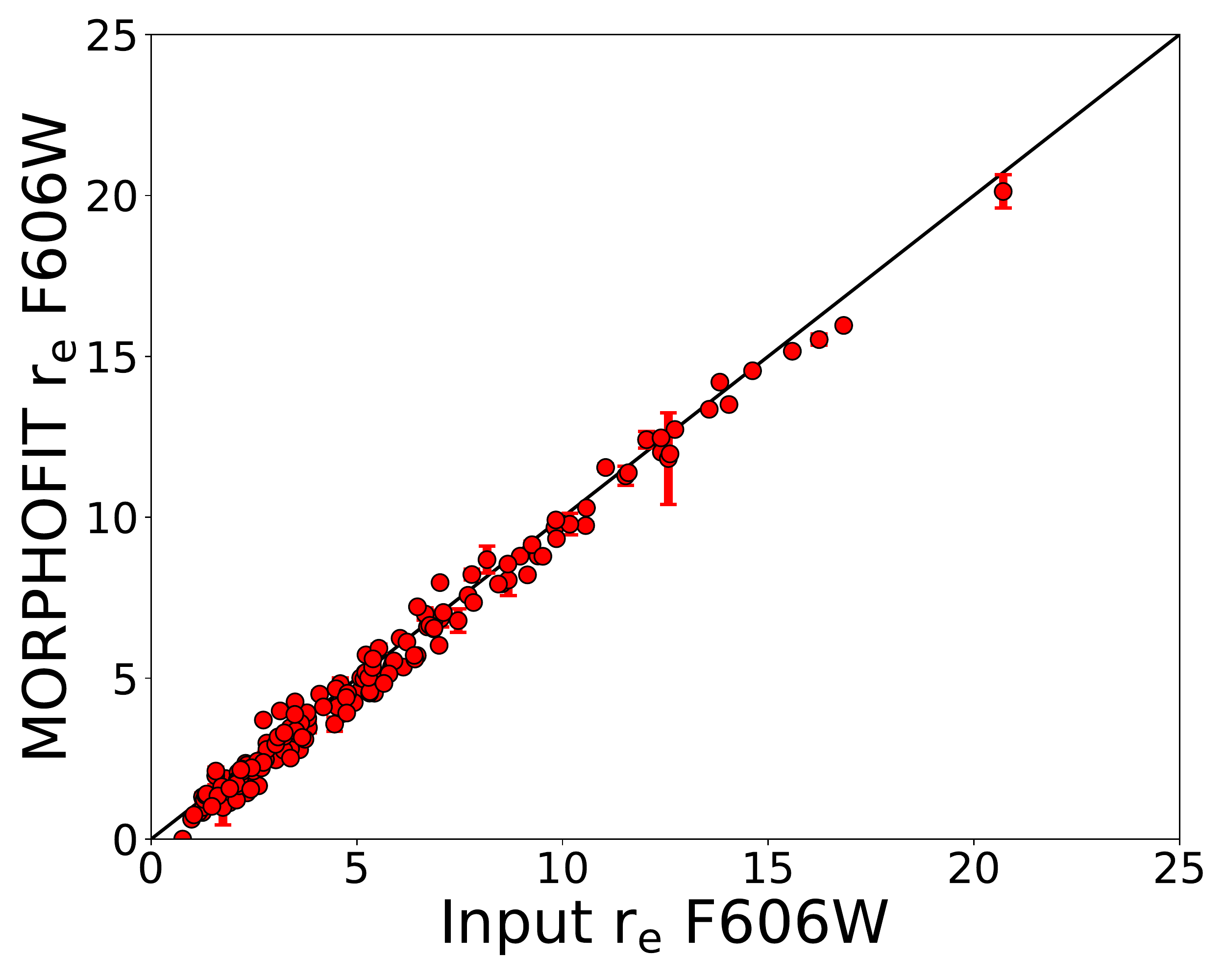}
\includegraphics[scale=0.22]{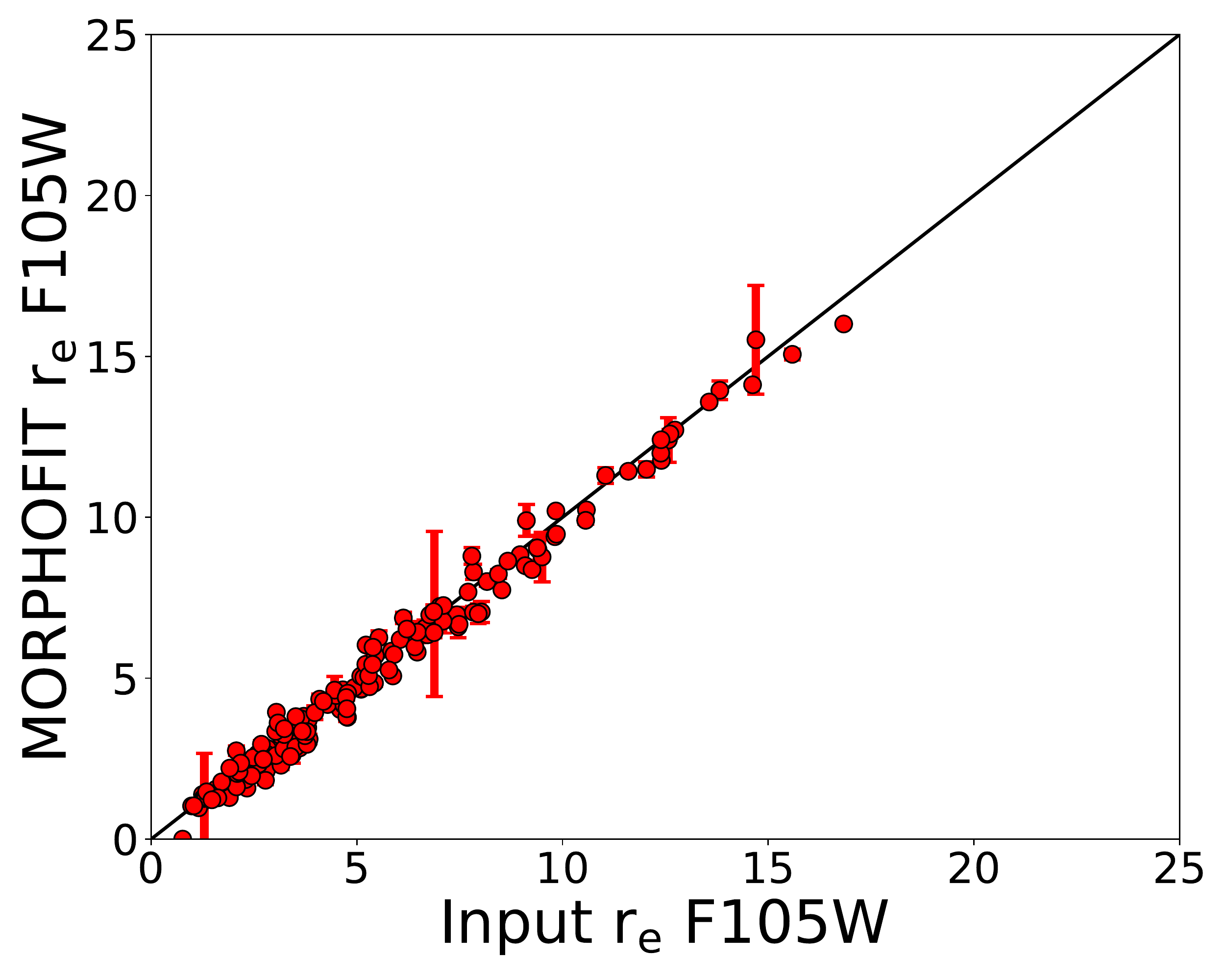}
\includegraphics[scale=0.22]{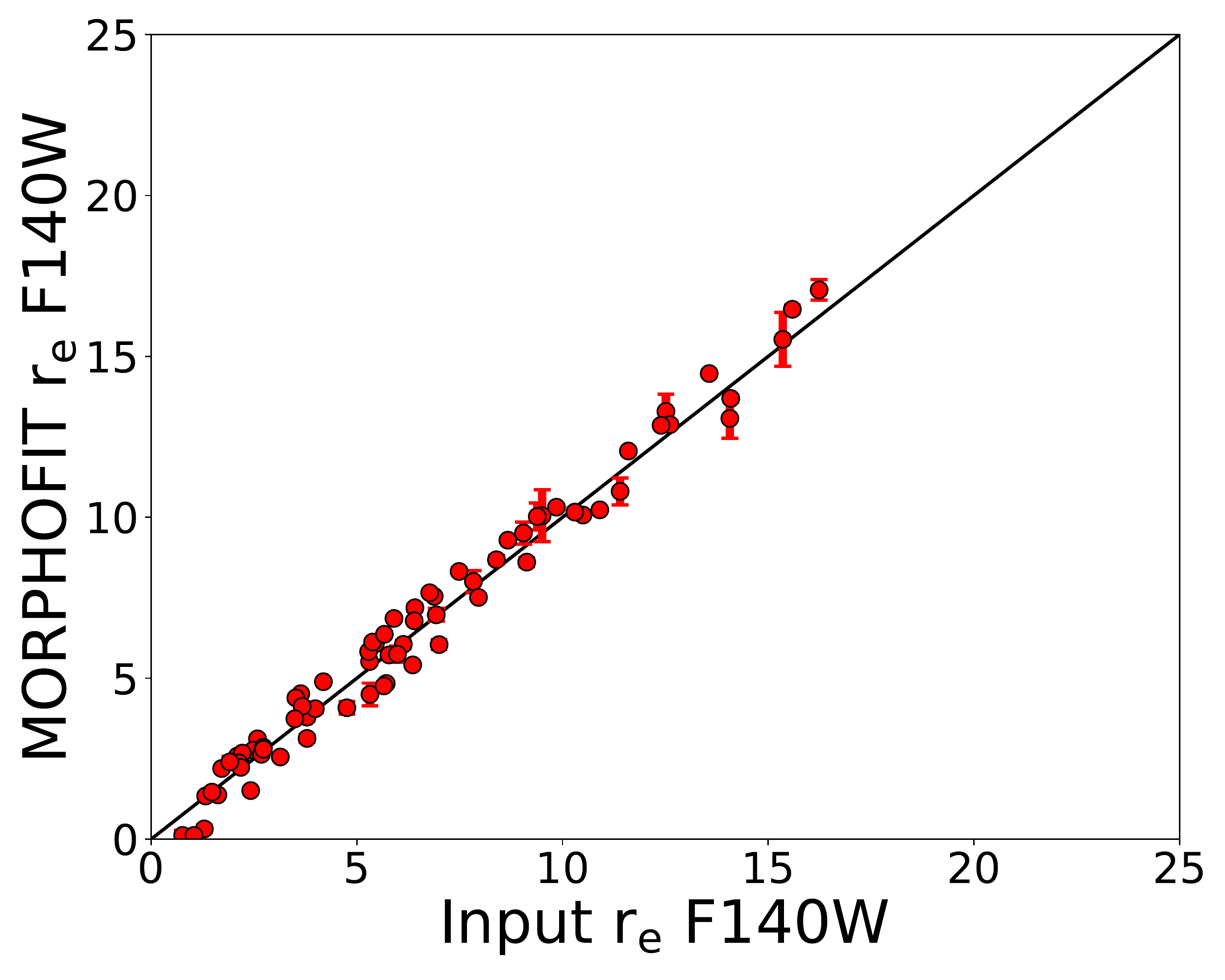}
\includegraphics[scale=0.22]{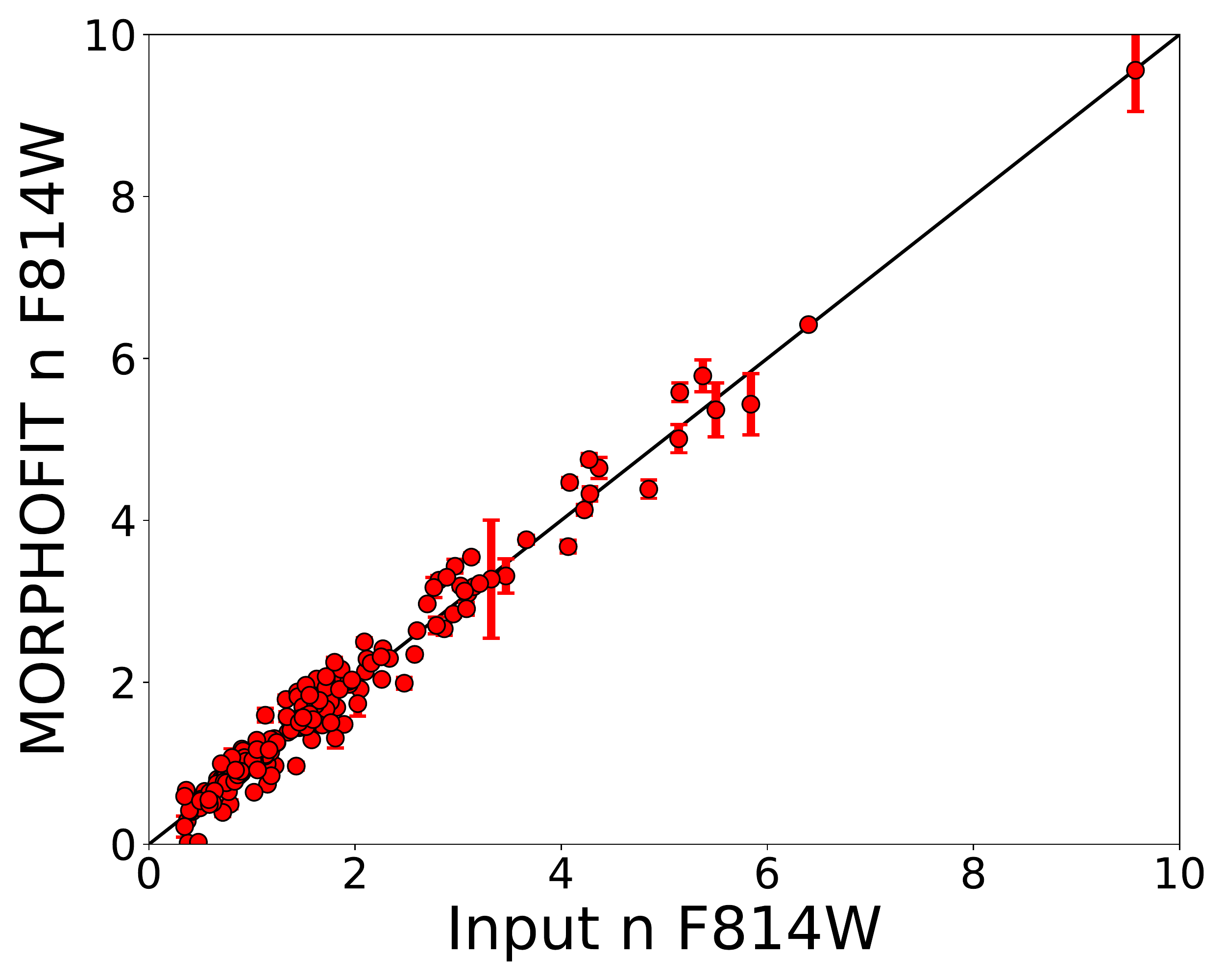}
\includegraphics[scale=0.22]{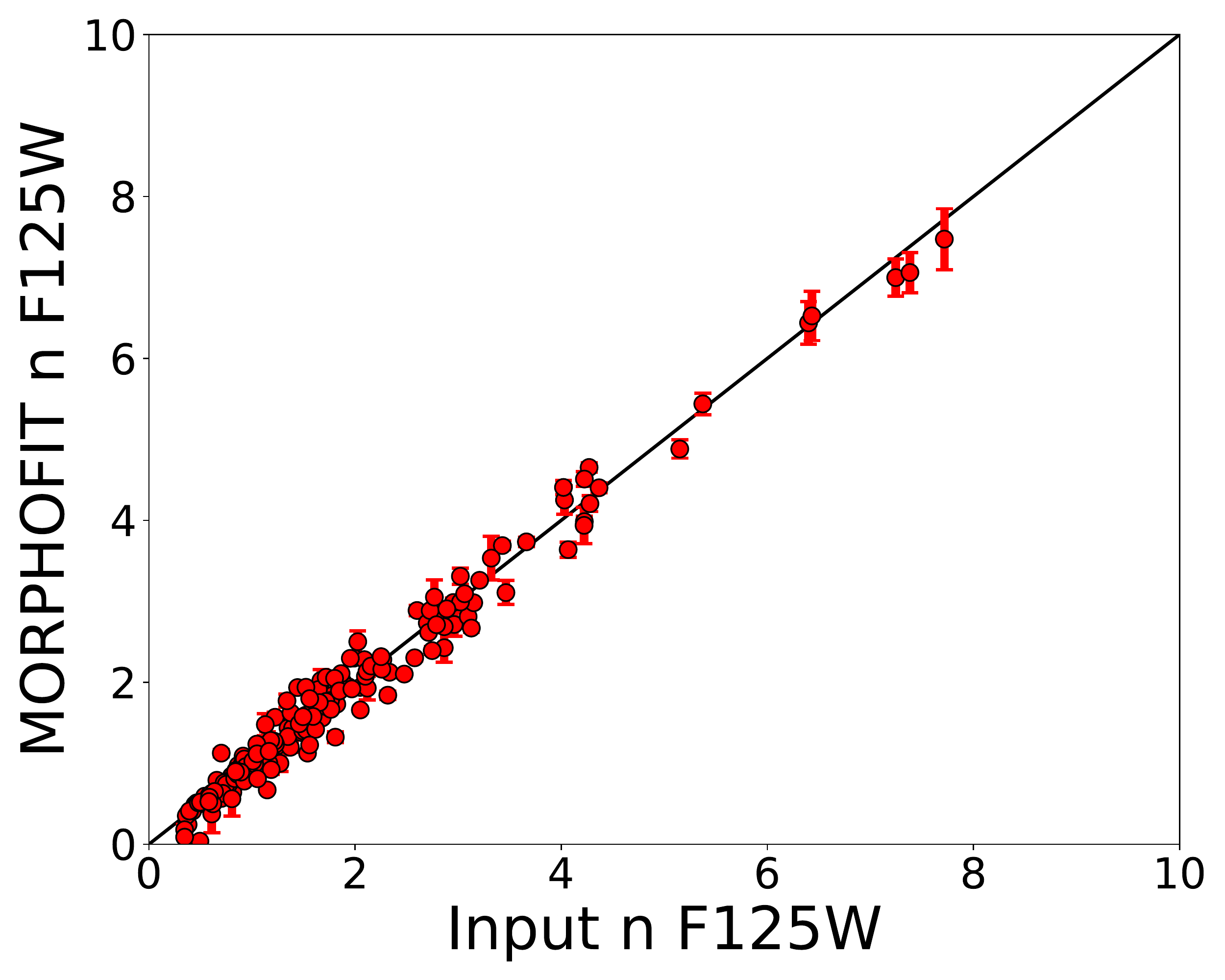}
\includegraphics[scale=0.22]{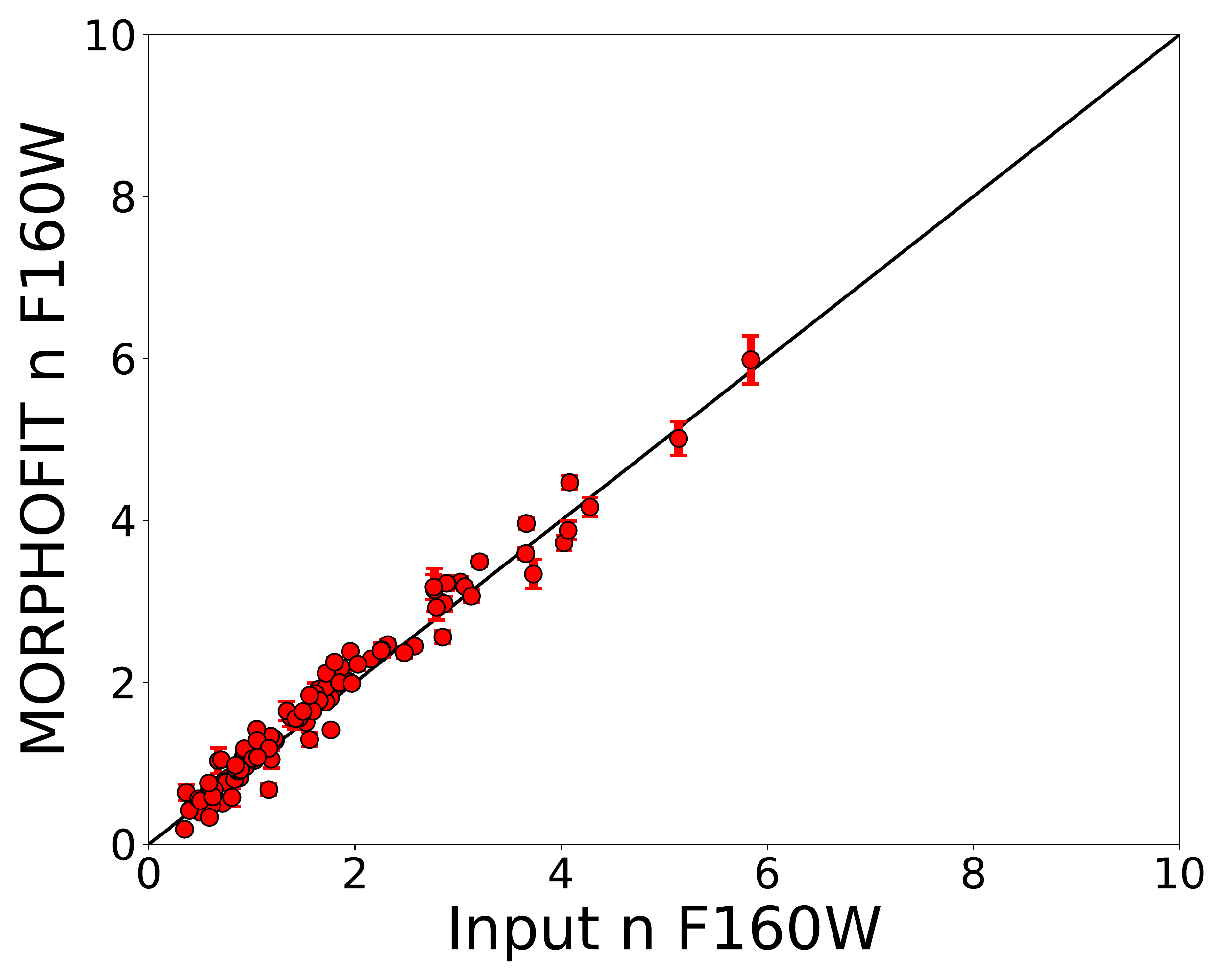}
\caption{The top, middle and bottom panels show the comparison between the input simulated galaxy magnitudes, the galaxy effective radii, the galaxy S\'ersic indices and the best-fitting magnitudes,  effective radii, S\'ersic indices from \textsc{morphofit}, respectively.  Magnitudes are in the AB system,  while the effective radii are expressed in units of pixels.  $\mathrm{mag}$, $\mathrm{r_{\mathrm{e}}}$ and $\mathrm{n}$ refer to the total magnitude,  the effective radius and the S\'ersic index of the input and fitted S\'ersic profile, respectively.  For each row,  we only select a sub-sample of the seven wavebands for plot clarity. Red points refer to the simulated target galaxies and their neighbouring objects. The black solid line represents the one-to-one relation between the input and the best-fitting quantities.}
\label{input_galfit_single_comparison}
\end{figure*}

In the single S\'ersic profile simulations, galaxy intrinsic properties are drawn from the empirical galaxy population model calibrated in \citealt{Tortorelli2020,Tortorelli2021}. \textsc{ufig} uses these properties to render galaxies on a pixelated grid.  To perform the surface brightness fitting of galaxies in the single S\'ersic profile simulations,  we follow the same pipeline adopted for the LT23 work.  We run \textsc{sextractor} on the simulated images and we use the resulting structural parameters as initial values for the surface brightness fitting with \textsc{galfit} on stamps.  We select as target galaxies that have magnitude $m_{\mathrm{F814W}} \le 22.5$ (measured with \textsc{sextractor} $\tt{MAG\_AUTO}$),  which is the completeness limit of the images in LT23.  These are 54 for AS1063 and 70 for M1149. We use the seven wavebands of the Frontier Fields survey (\textit{F435W, F606W, F814W, F105W, F125W,F140W, F160W}),  two methods for the PSF image creation (`Moffat' and `observed'),  `custom' and `internal' generated sigma images and we let the background both fixed and free to vary.  This leads to a total of 3024 combinations to fit for AS1063 and 3920 for M1149.  For the fit on regions,  the number of combinations is 336 for both clusters.  For the fit on the full images, the number is 56 per cluster.

Thanks to the high degree of parallelization of the package,  these combinations can all be fit simultaneously, provided that the appropriate number of cores is available to the user.  We use the high-performance computing facility `Euler' at ETH Z\"urich for all the surface brightness fit carried out in this work and for the image simulations.  The seven bands simulated images for both clusters are created using two cores and 8 Gb of RAM in roughly 10 minutes.  The fit on stamps uses 6944 cores,  4 Gb of RAM per core and a run-time that ranges from minutes to a couple of hours depending on how large the stamps are and, therefore, how many sources are simultaneously fit.  The fit on regions uses 672 cores,  8 Gb of RAM per core and a run-time that ranges from less than an hour to almost 24 hours depending on the number of sources per region.  The fit on the full images requires more computational resources and is more time intensive.  It uses less cores,  112,  due to the smaller number of combinations,  but 32 Gb of RAM per core and a run-time that ranges from 12 hours to roughly 3-4 days. The latter is especially true for the bulge-disk component fit (see next section) that doubles the amount of simultaneous components to fit. 

We use \textsc{morphofit} to fit the surface brightness distributions with single S\'ersic profiles with the following 7 free parameters:
\begin{itemize}
\item $x$, $y$ centroid: we set the initial value of the centroid pixel coordinates by converting the \textsc{sextractor} $\tt{ALPHAWIN\_J2000}$ and $\tt{DELTAWIN\_J2000}$ sky coordinates in the \textit{F814W} waveband into the corresponding pixel coordinates for each individual stamp.
\item $mag$: we set the initial value of the total galaxy magnitude using the parameter $\tt{MAG\_AUTO}$ in the waveband of the galaxies we want to fit.
\item $r_\mathrm{e}$: we set the initial value of the effective radius in pixel using the \textsc{sextractor} parameter $\tt{FLUX\_RADIUS}$ in the waveband of the galaxies we want to fit.
\item $n$: we set the initial value of the S\'ersic index to $n = 2.5$.
\item $q$: we set the initial value of the axis ratio using the \textsc{sextractor} parameters $\tt{BWIN\_IMAGE}$ and $\tt{AWIN\_IMAGE}$ to $q = \tt{ BWIN\_IMAGE / AWIN\_IMAGE}$ in the waveband of the galaxies we want to fit.
\item $P.A.$: we set the initial value of the position angle using the \textsc{sextractor} parameter $\tt{THETAWIN\_SKY}$ in the waveband of the galaxies we want to fit.
\end{itemize}

We then create the catalogue containing the estimates of structural parameters for every single galaxy as described in Section \ref{best_fit_cat}.  These are then used as initial values for the fit on regions and the results are,  in turn,  used for the fit on the full images. We use windowed positional parameters as initial values following the suggestions in the \textsc{sextractor} manual\footnote{\href{http://astroa.physics.metu.edu.tr/MANUALS/sextractor/sextractor.pdf}{http://astroa.physics.metu.edu.tr/MANUALS/sextractor/sextractor.pdf}}.  The manual recommends the use of windowed parameters rather than its isophotal equivalents because, despite being more CPU intensive,  they provide a less noisy estimate.

We show in Figure \ref{input_galfit_single_comparison} the comparison between the structural parameters measured with \textsc{morphofit} and the input properties of the simulated images.  We compare the magnitudes,  the effective radii  and the S\'ersic indices of the target galaxies and their neighbours in the seven Frontier Fields wavebands.  The magnitude comparison shows that bright galaxies ($m_{\mathrm{F814W}} \le 22.5$) at all wavebands are all consistent within the errors with the input values.  The neighbouring objects that have magnitudes greater than $m_{\mathrm{F814W}} > 22.5$ lie close to the one-to-one relation as well,  but they are consistent with the input values only at the $3 \sigma$ level.  The scatter of the effective radii and of the S\'ersic indices is also remarkably small.  Roughly $80\%$ of the galaxies are consistent within the errors with the input values, while all of them become consistent with the input when we consider the $3 \sigma$ level. 

\subsection{Bulge plus disk simulated galaxies}
\label{bulge_disk_sims}

\begin{figure*}
\centering
\includegraphics[scale=0.2]{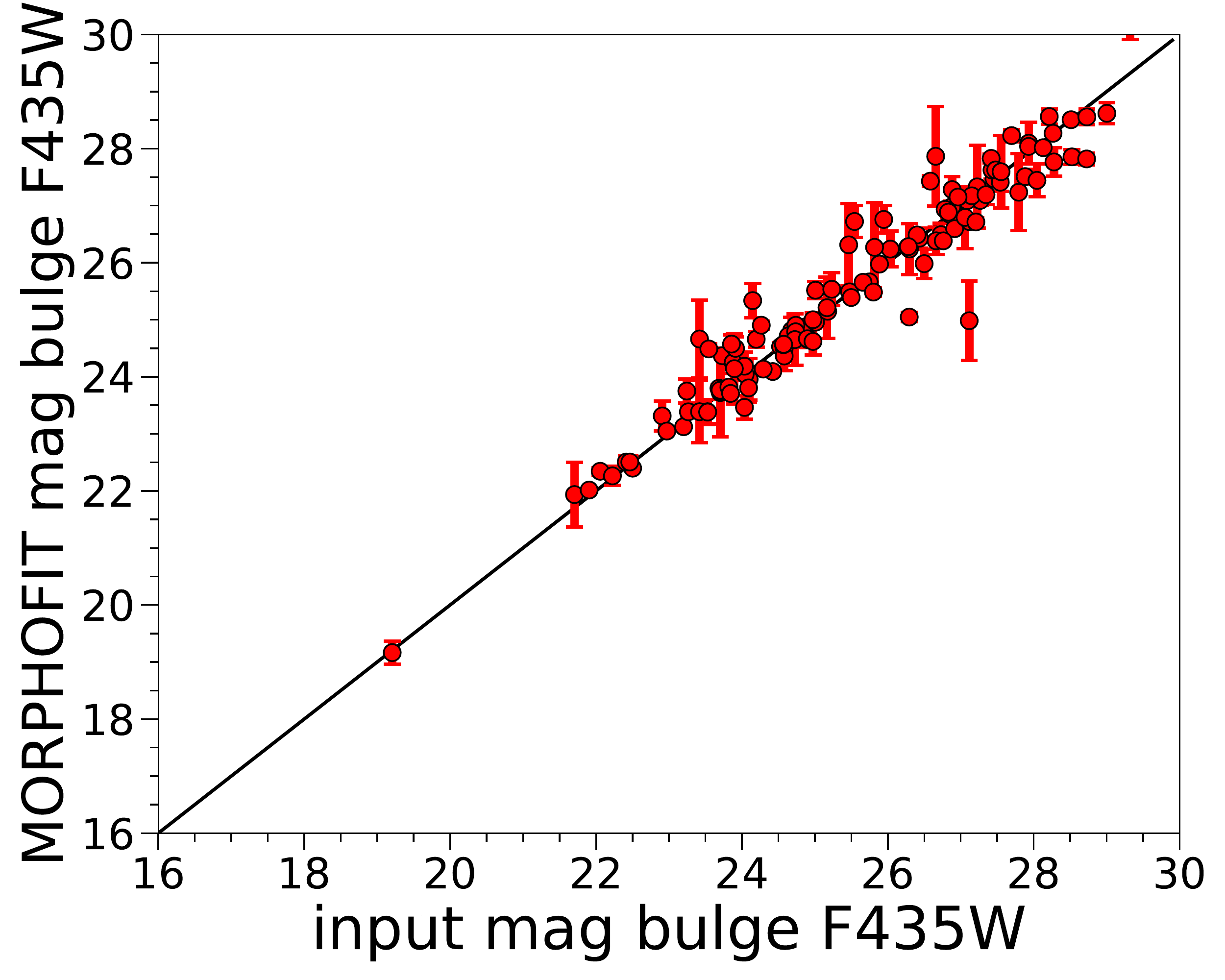}
\includegraphics[scale=0.2]{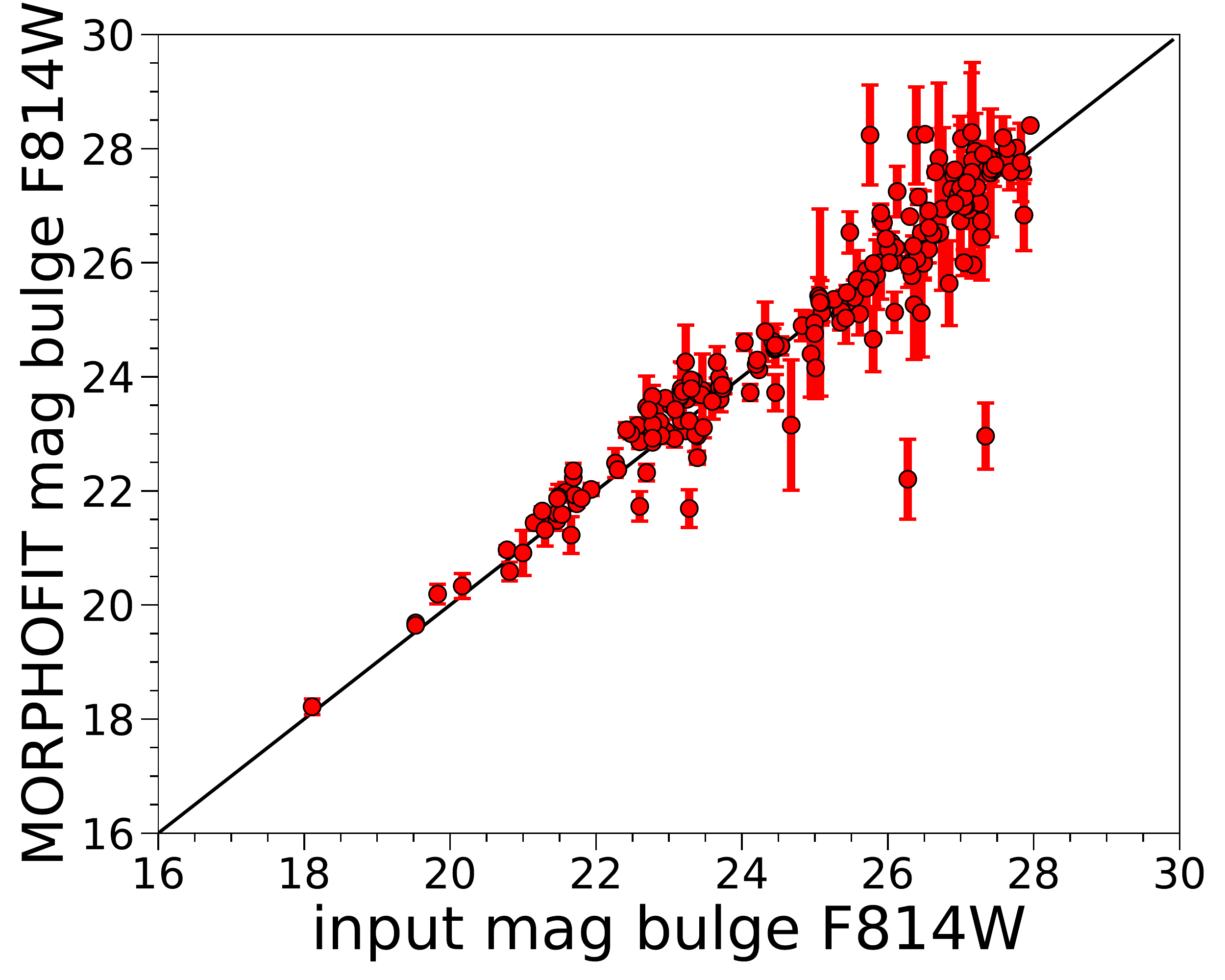}
\includegraphics[scale=0.2]{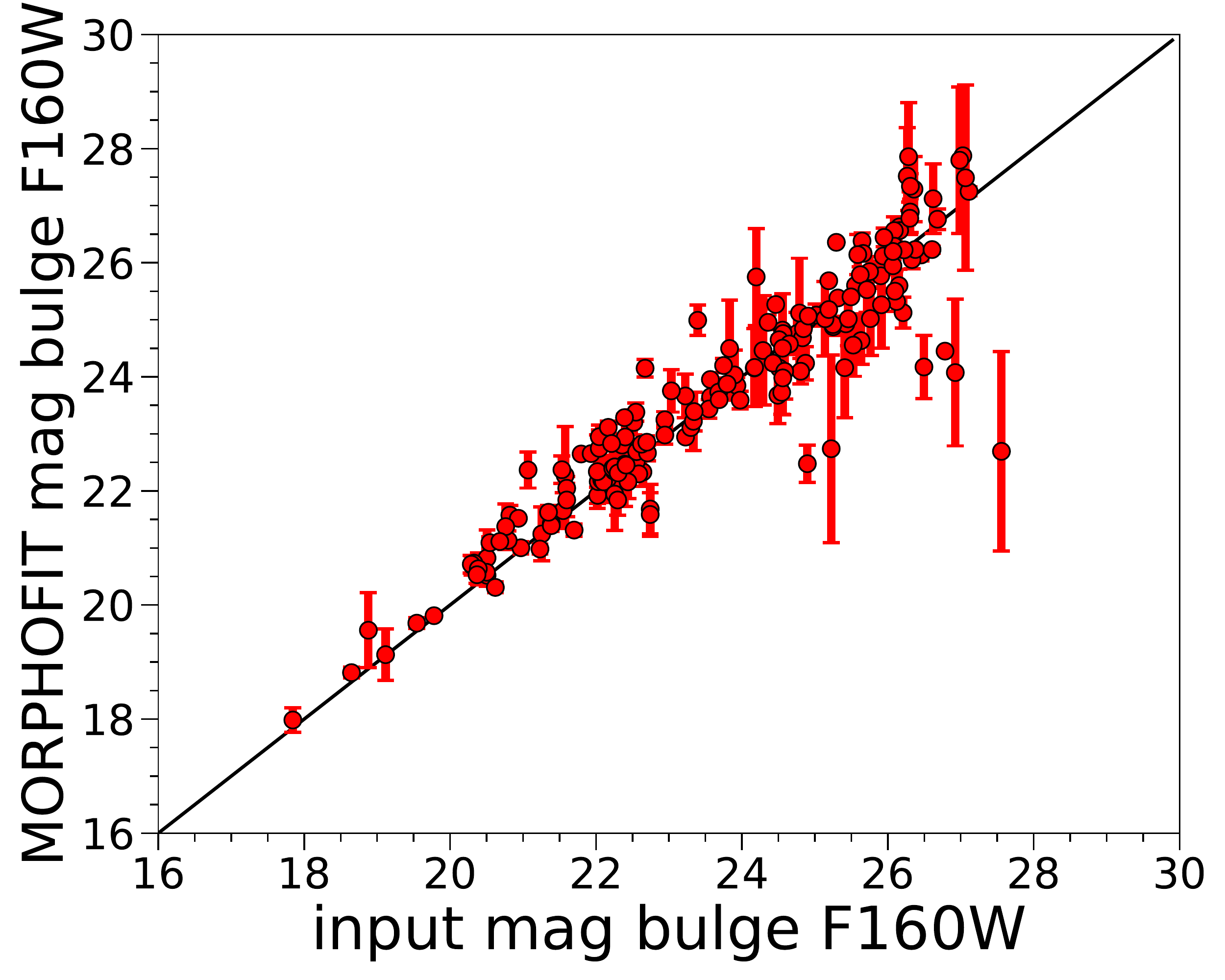}
\includegraphics[scale=0.2]{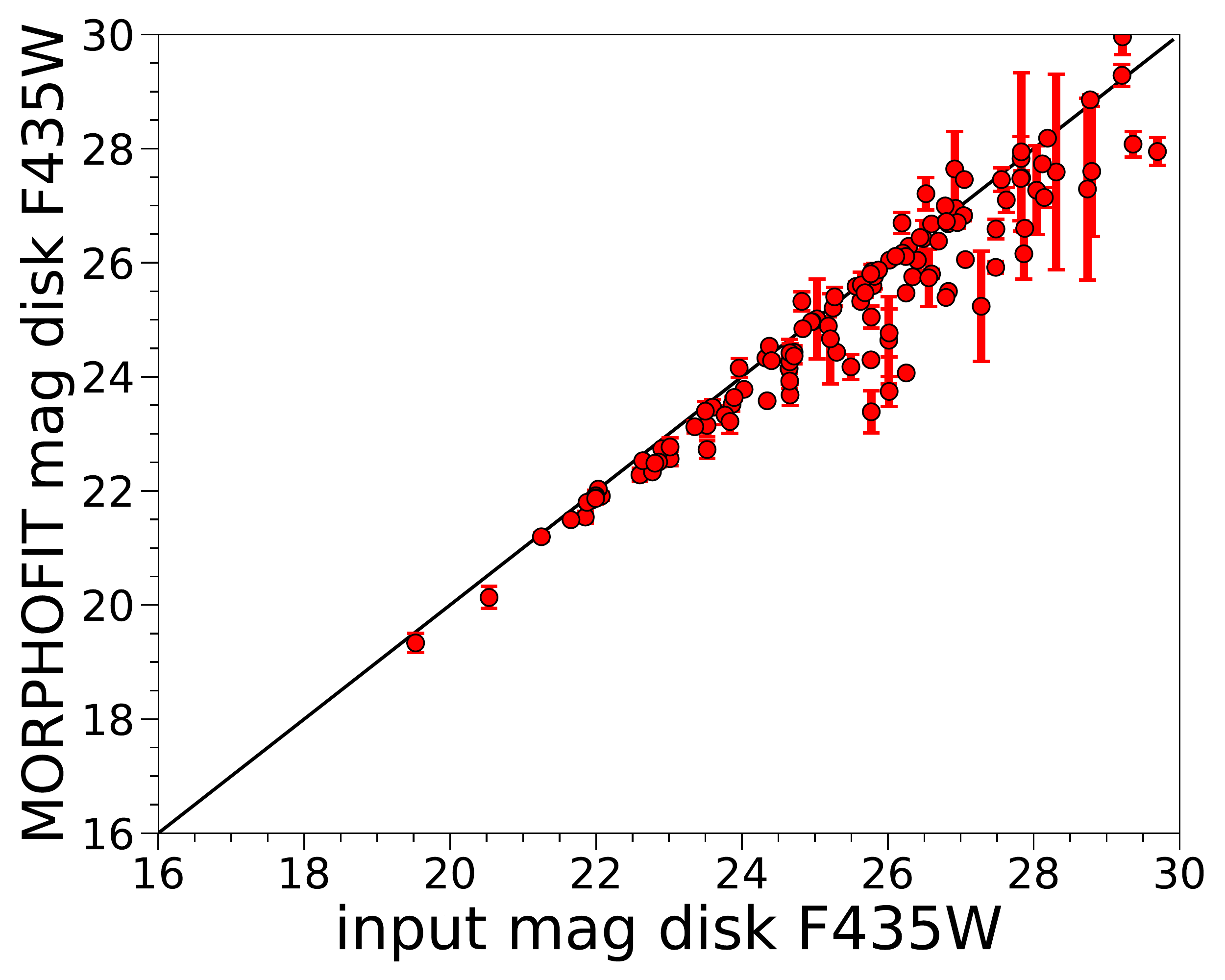}
\includegraphics[scale=0.2]{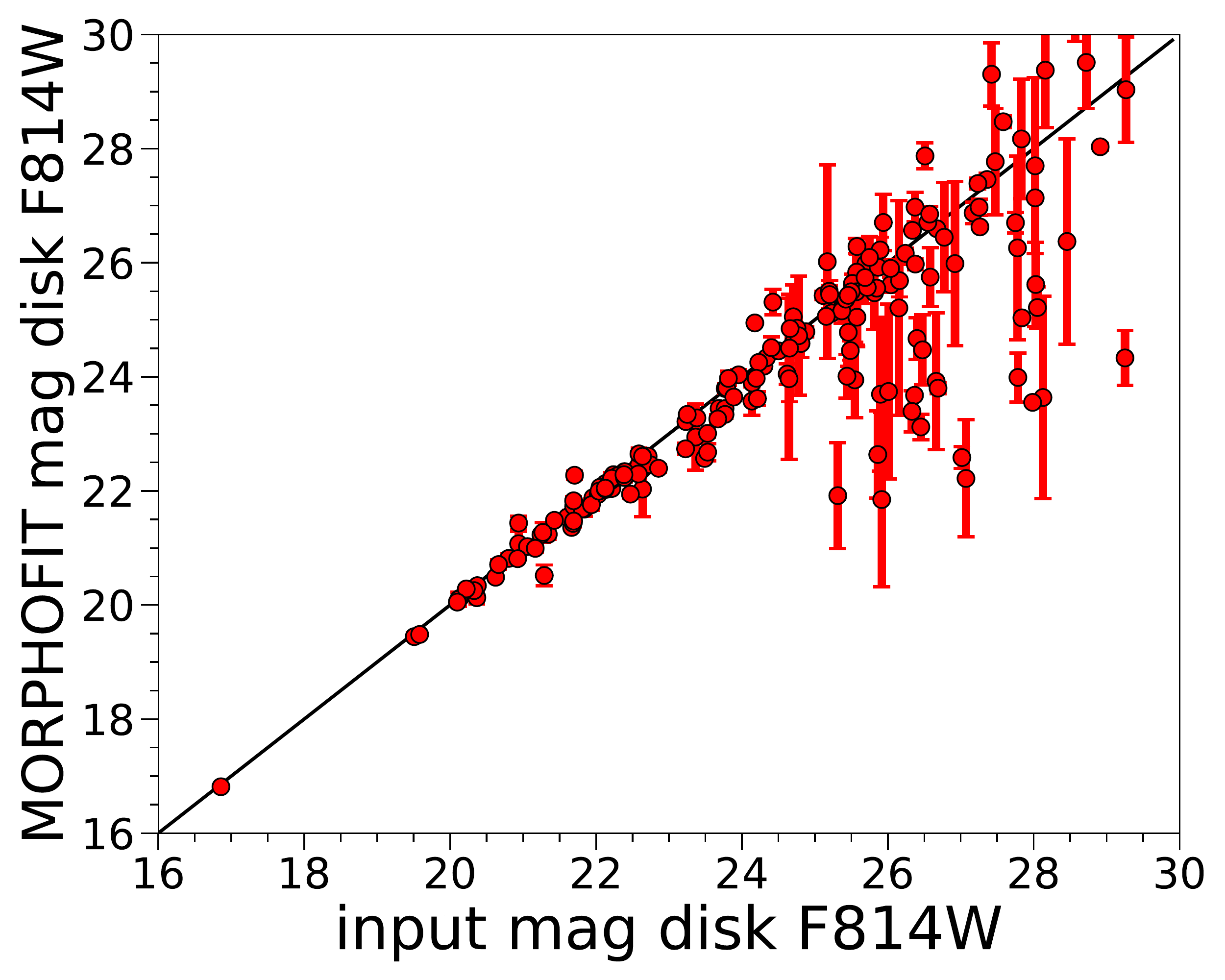}
\includegraphics[scale=0.2]{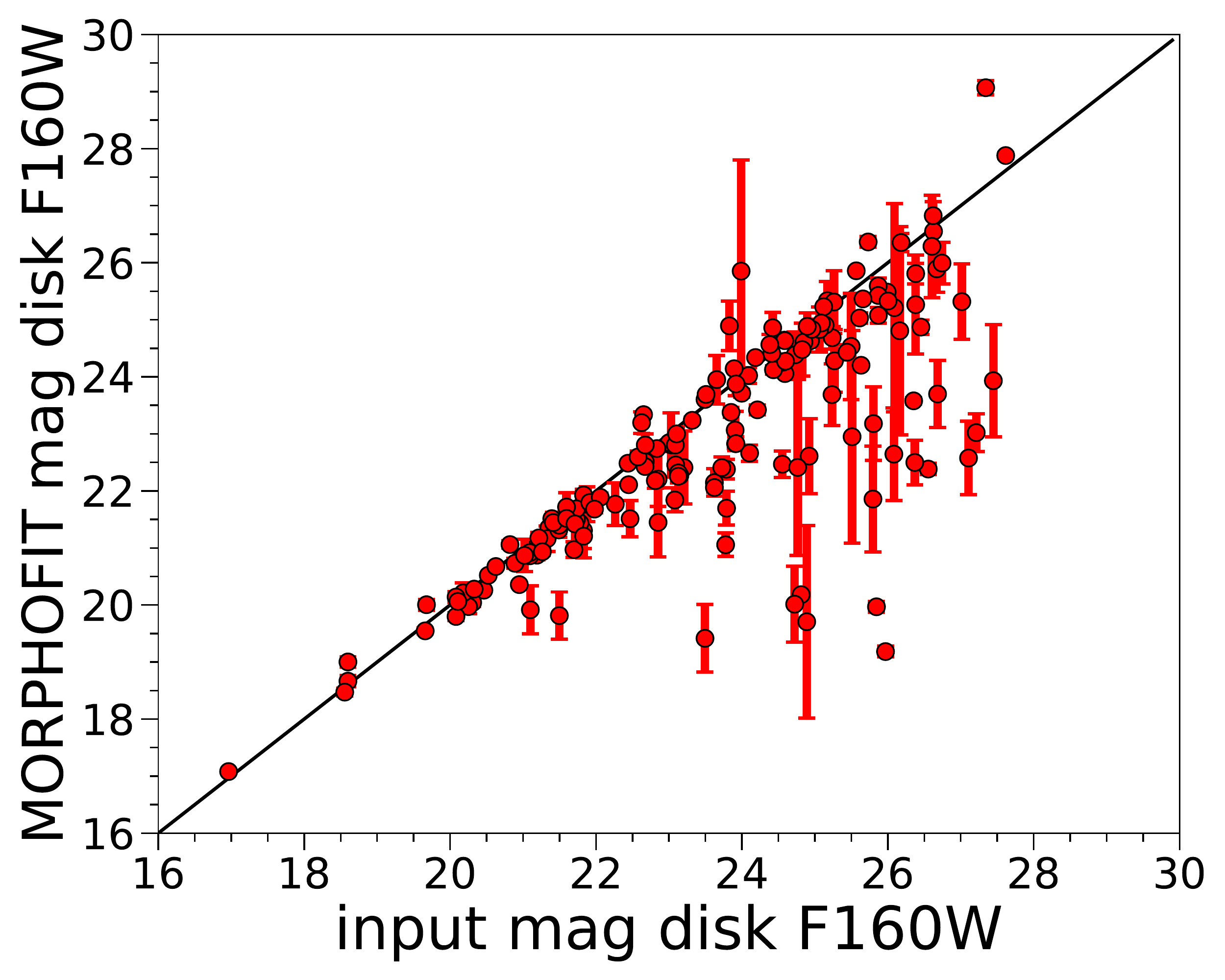}
\includegraphics[scale=0.2]{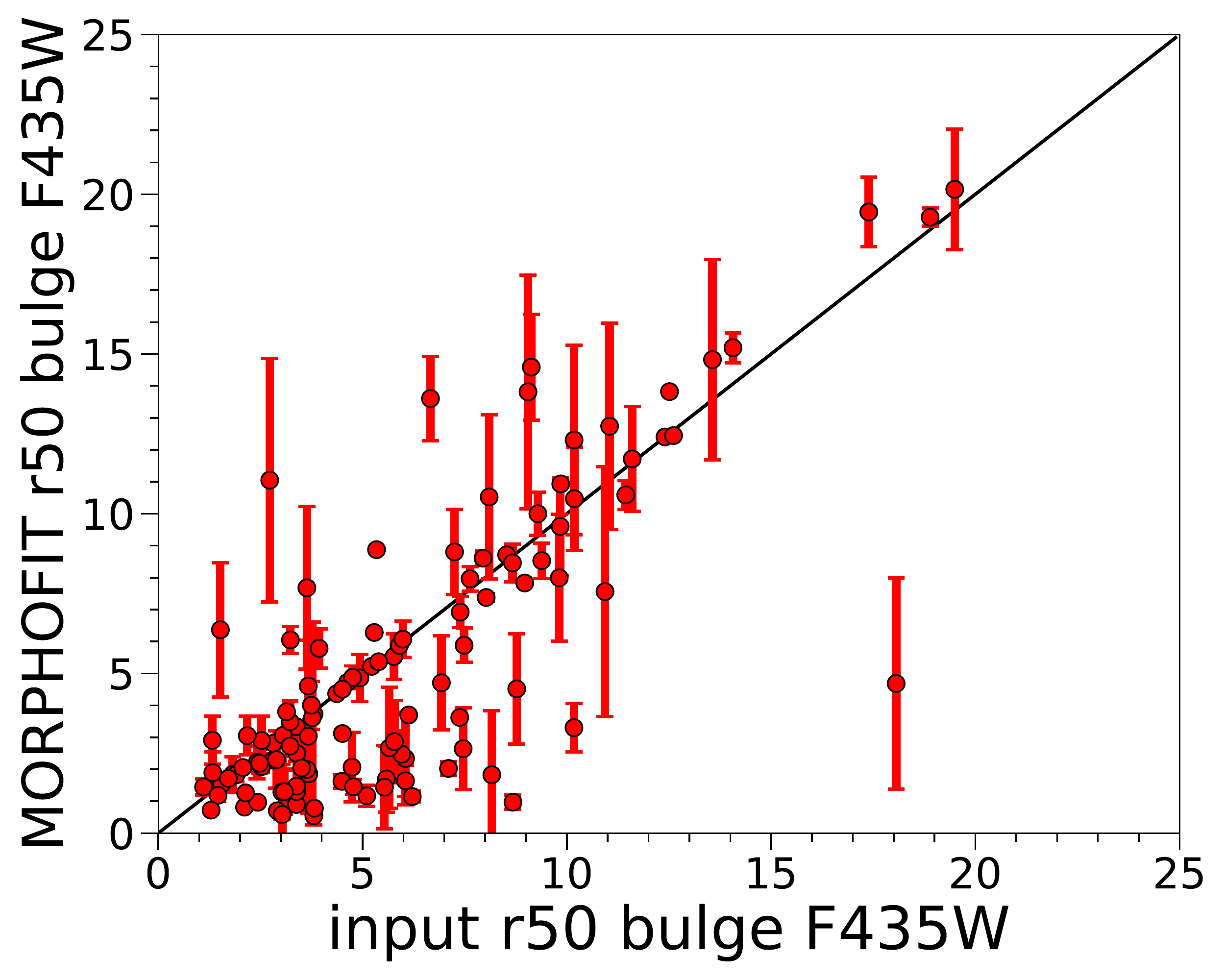}
\includegraphics[scale=0.2]{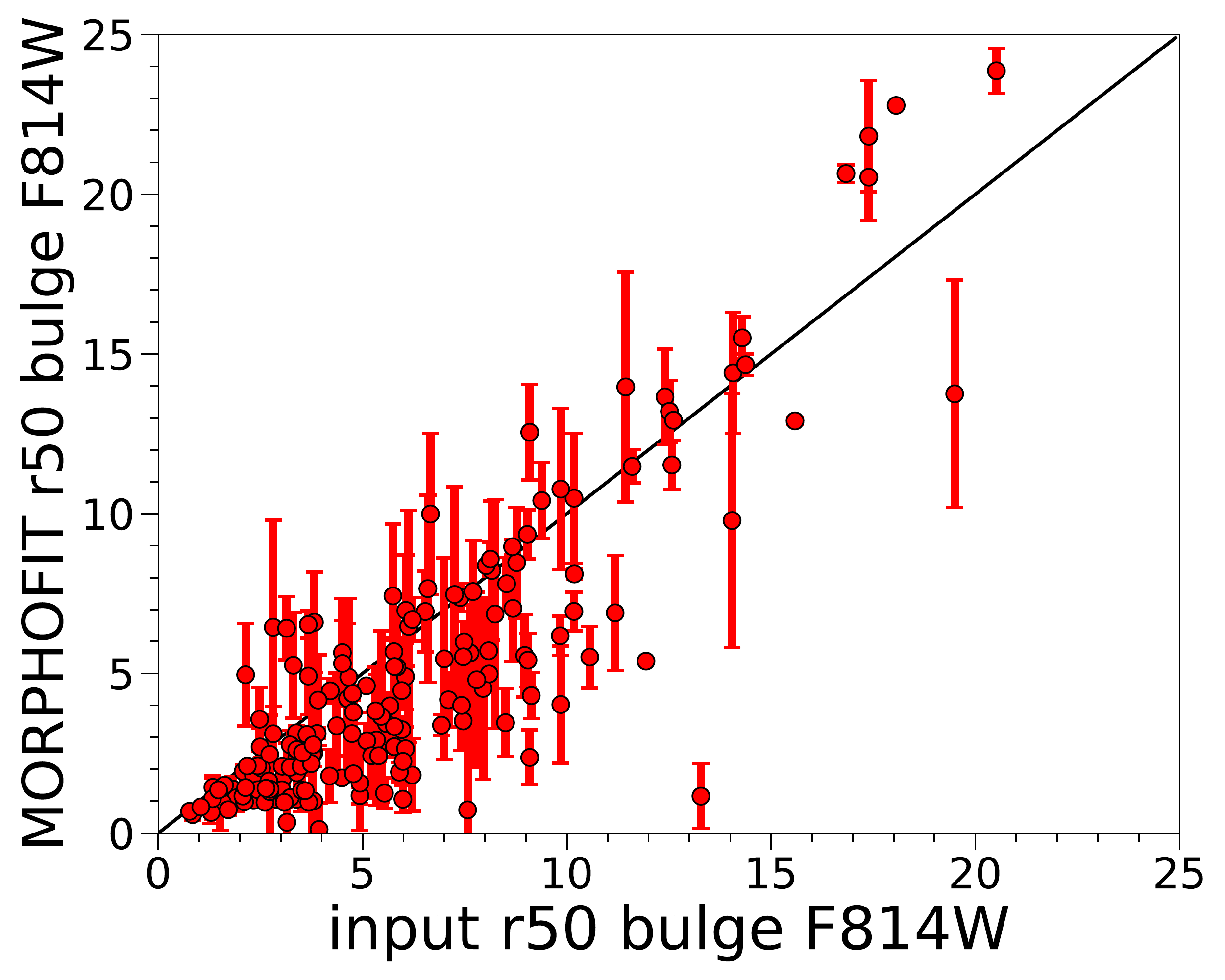}
\includegraphics[scale=0.2]{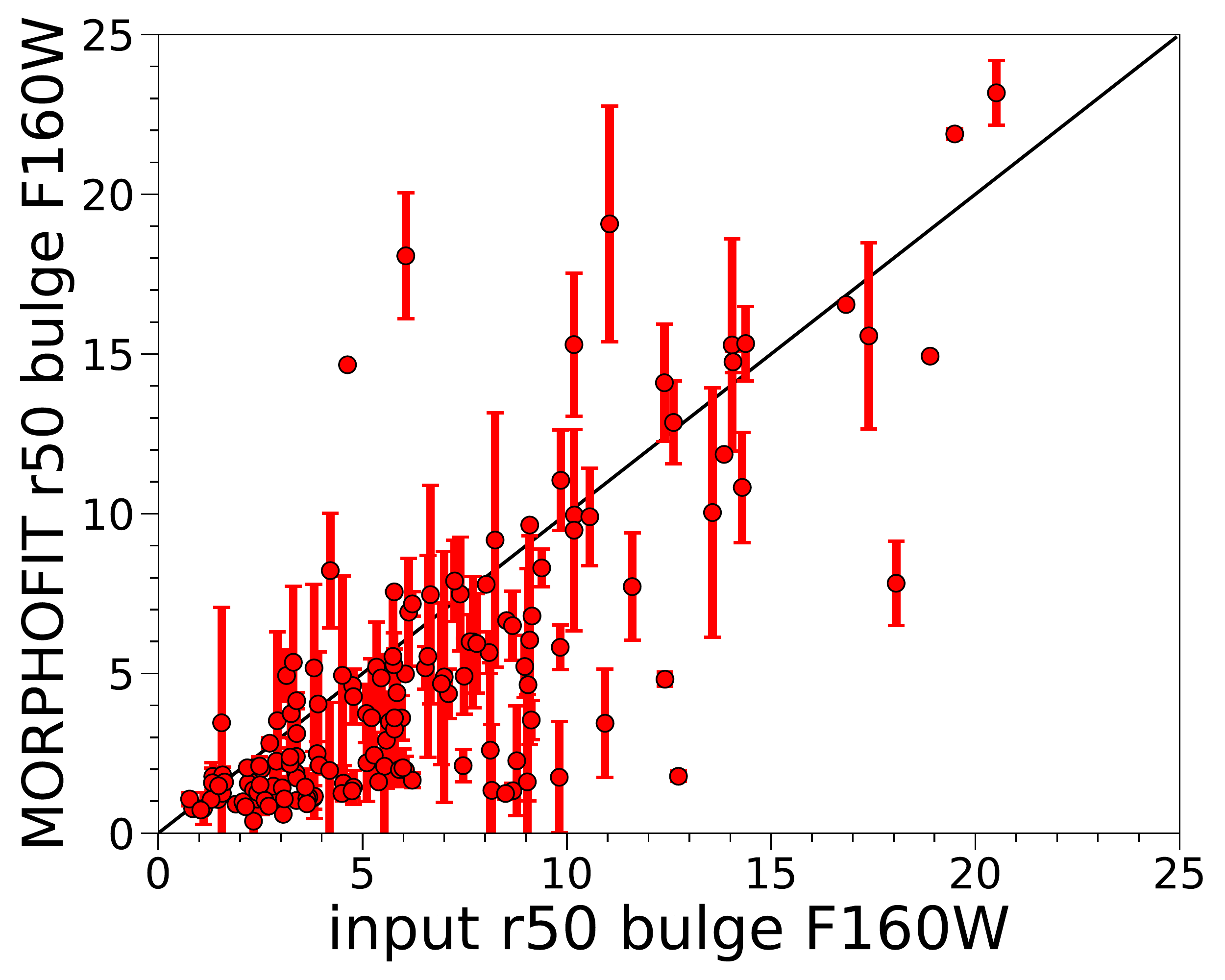}
\includegraphics[scale=0.2]{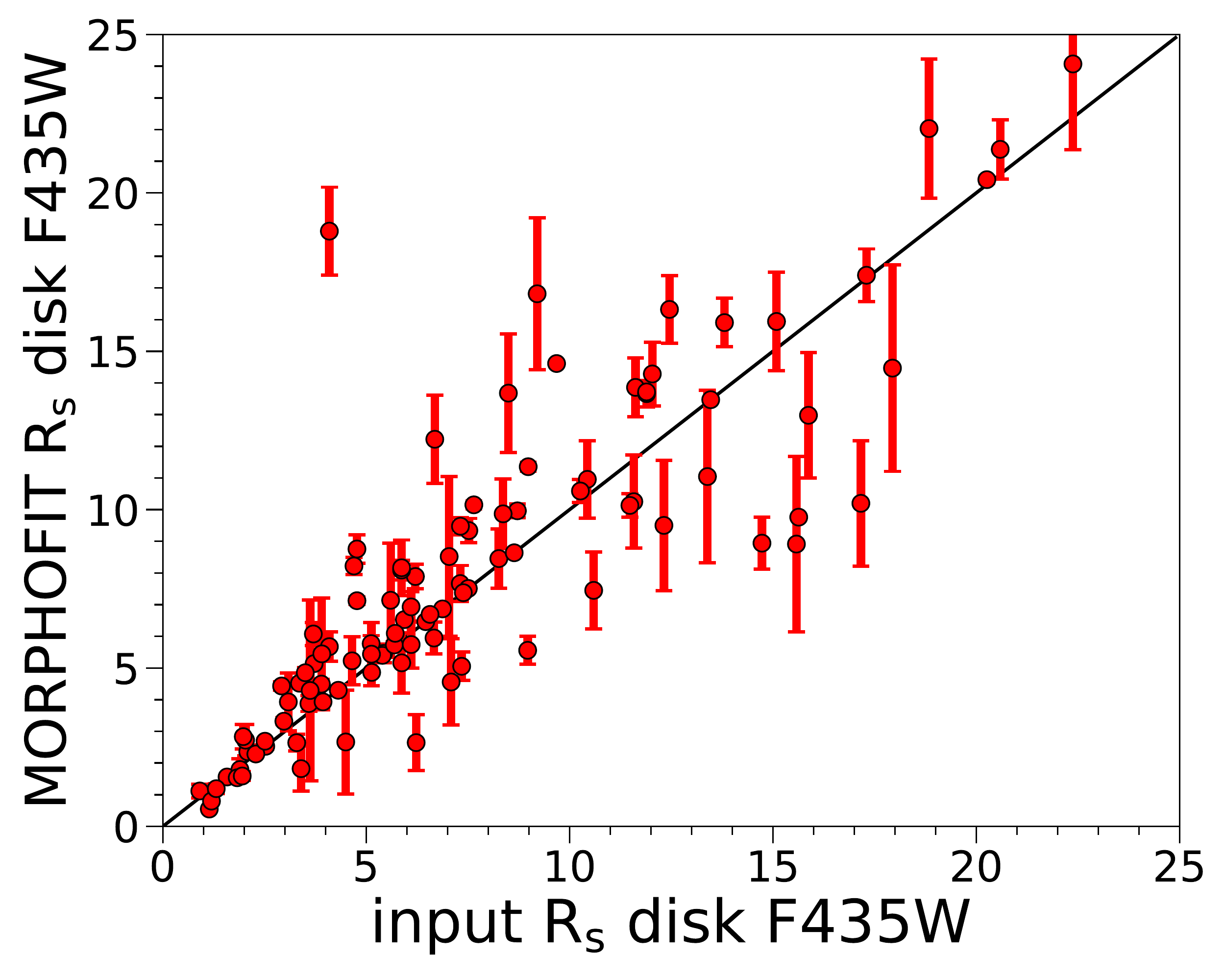}
\includegraphics[scale=0.2]{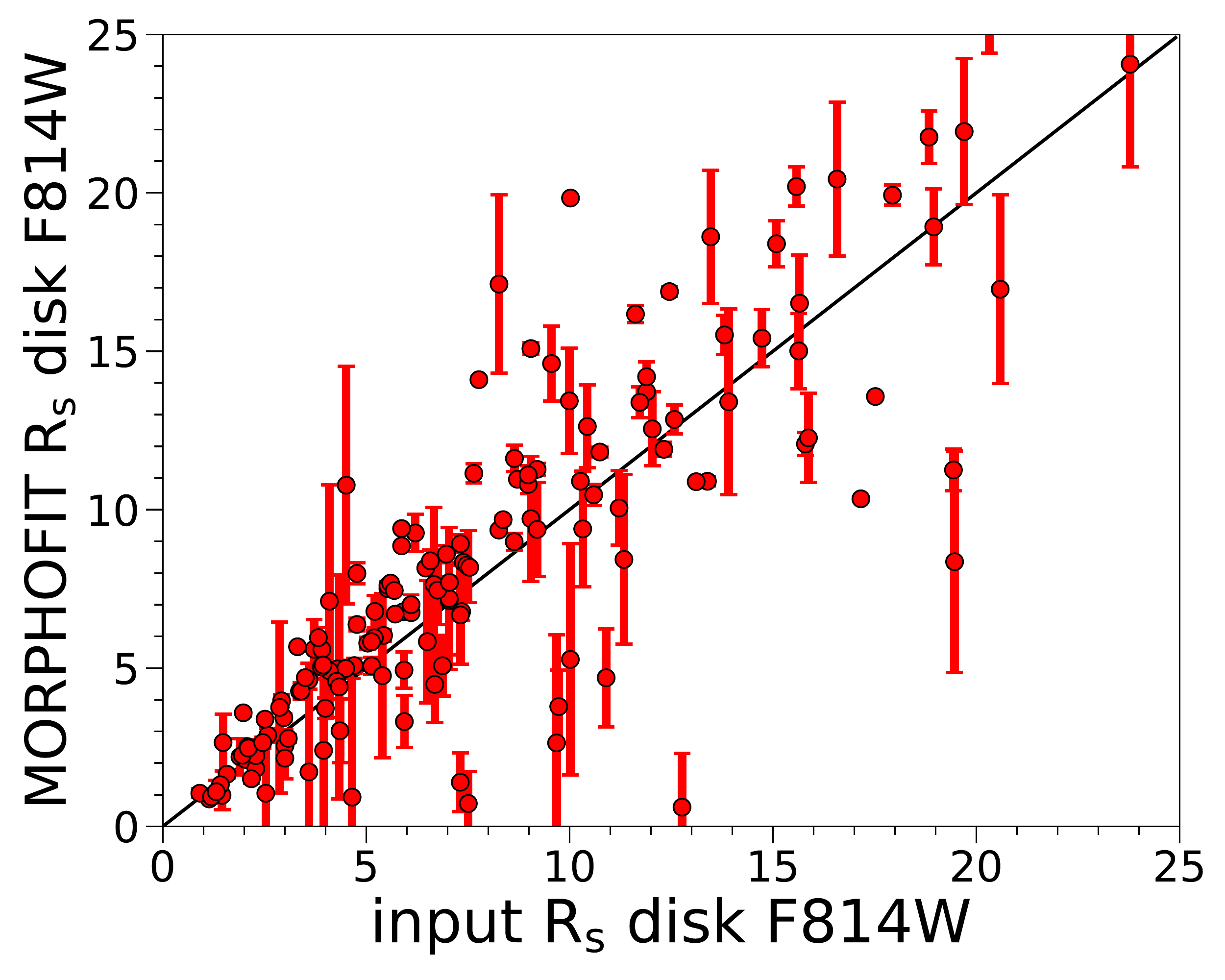}
\includegraphics[scale=0.2]{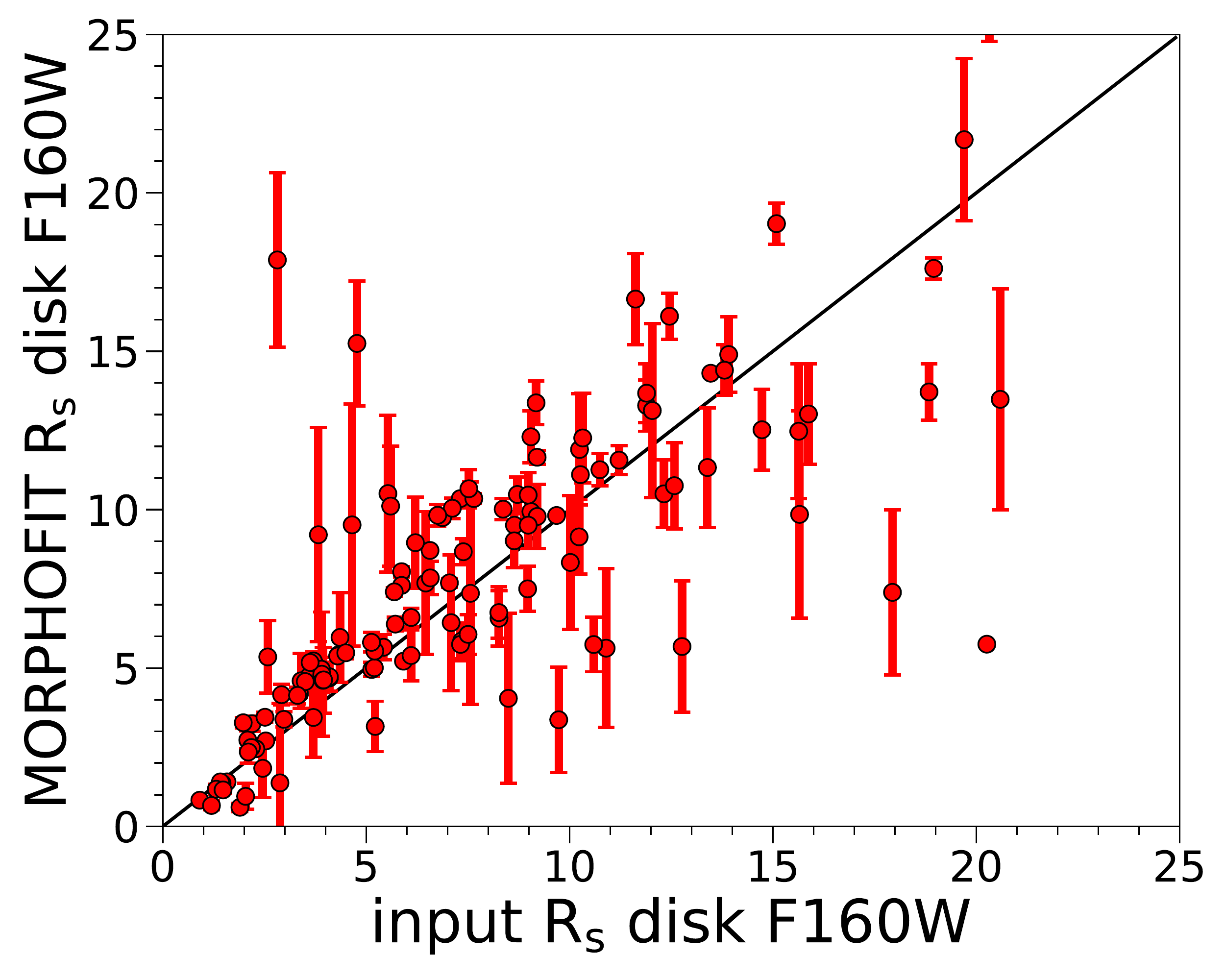}
\includegraphics[scale=0.2]{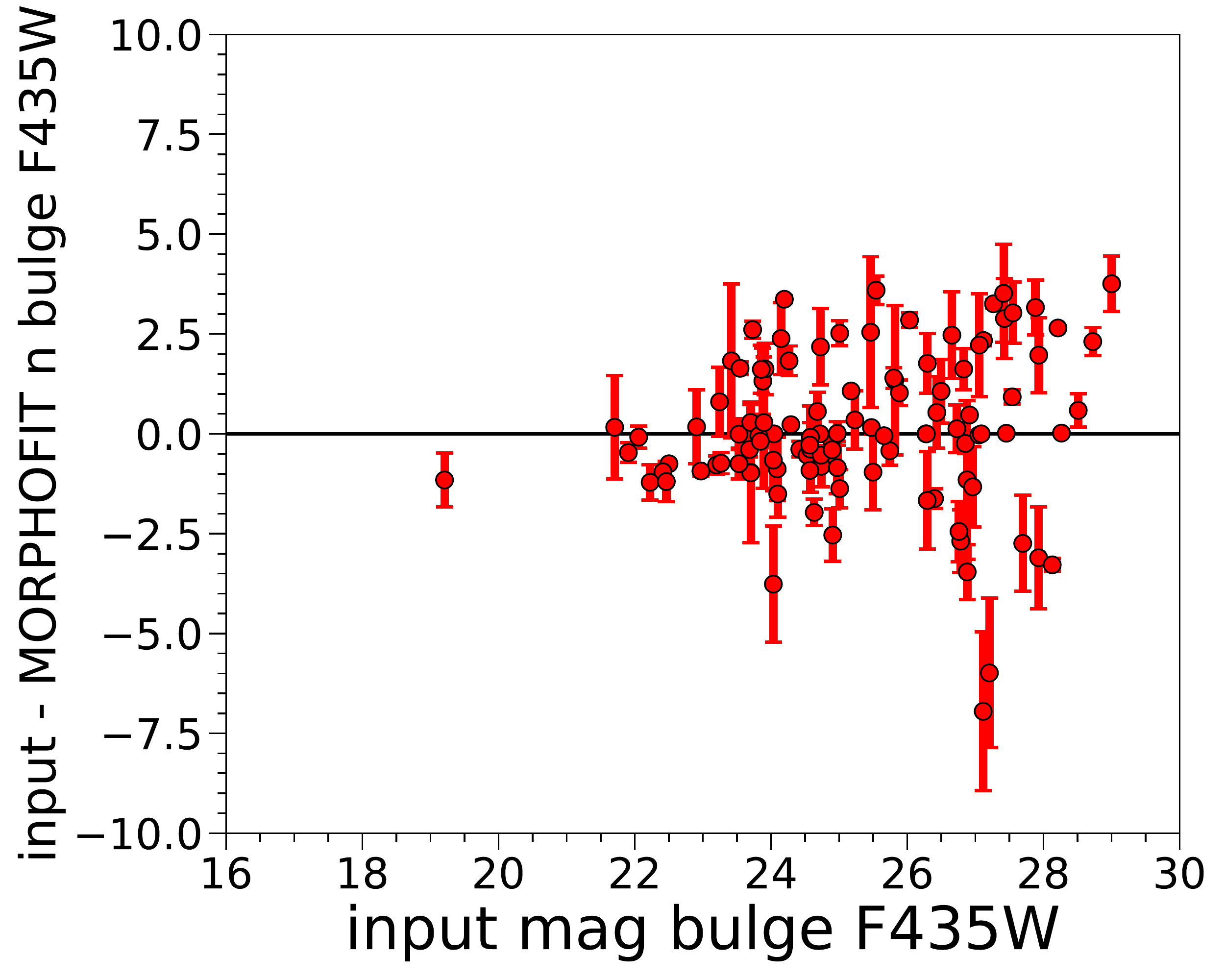}
\includegraphics[scale=0.2]{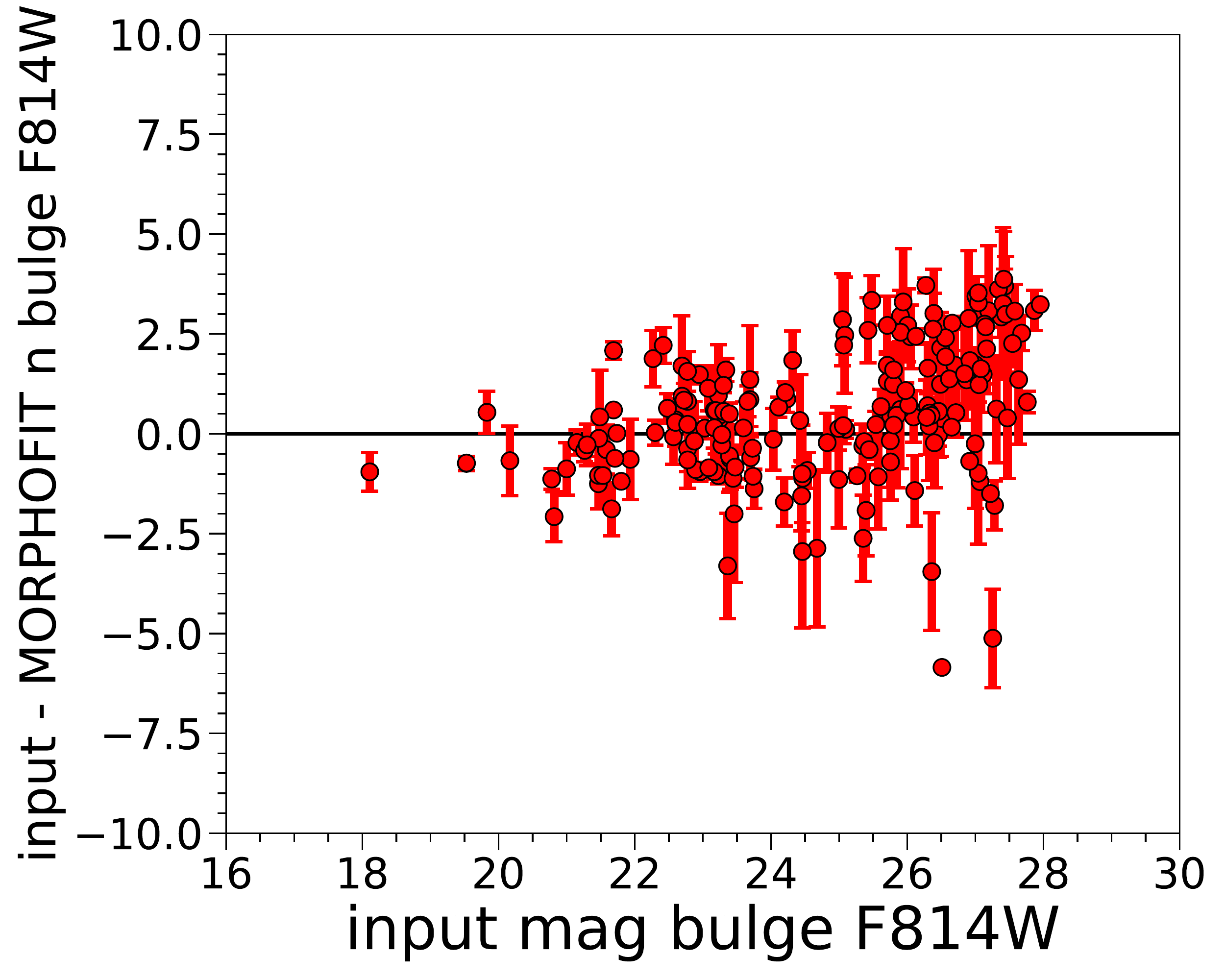}
\includegraphics[scale=0.2]{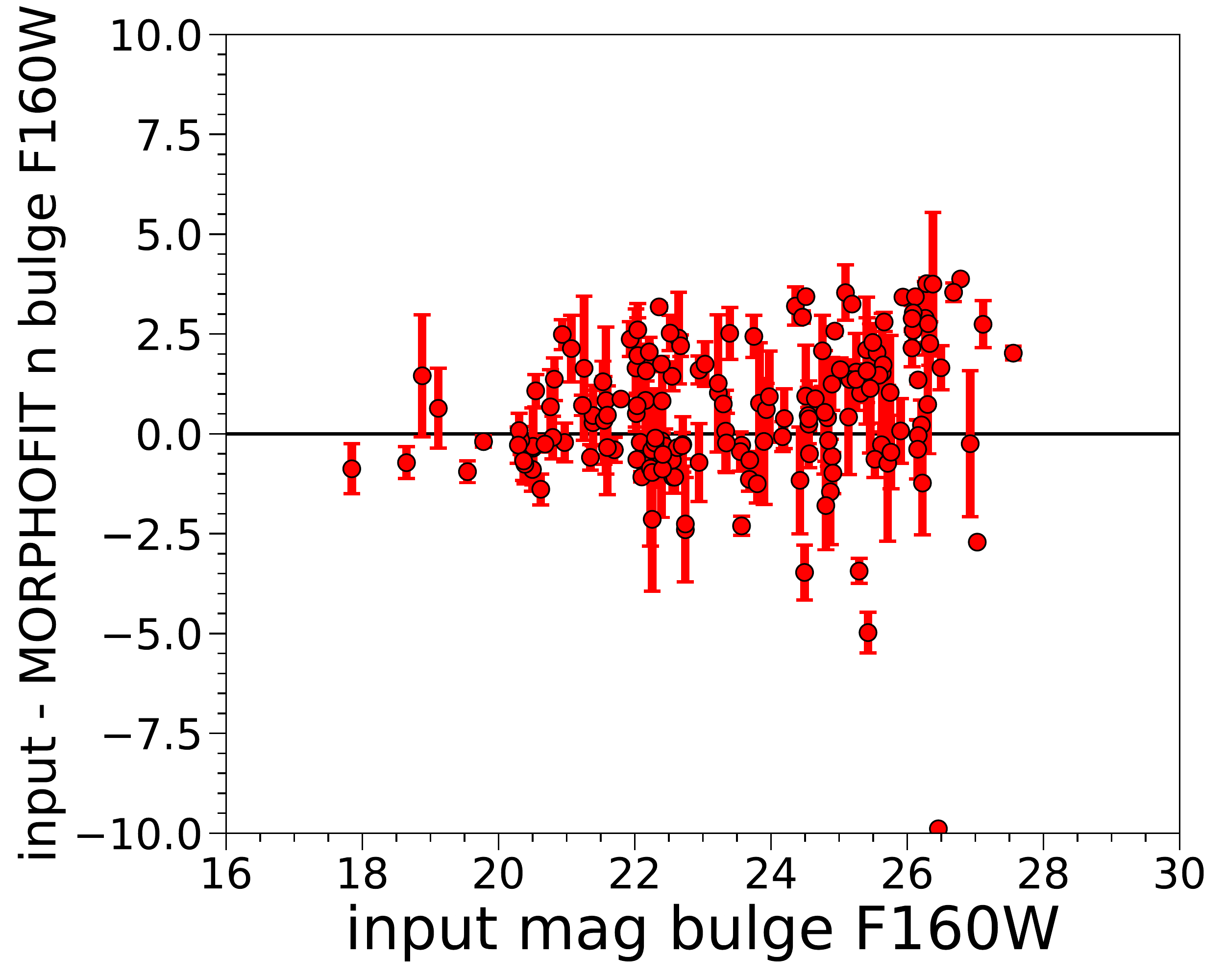}
\caption{The first and second rows of panels show the comparison between the input simulated galaxy magnitudes of bulge (first row) and disk (second row),  and the corresponding best-fitting magnitudes from \textsc{morphofit}. The third and fourth rows of panels show the comparison for the effective radii and the scale lengths.  The fifth row shows the deviation of the recovered bulge S\'ersic index with respect to the input one.  For each row,  we only select a sub-sample of the seven wavebands for plot clarity, comparing similar bands for the bulge and disk components.  Red points refer to the simulated target galaxies and their neighbouring objects. The black solid line represents the one-to-one relation between the input and the best-fitting quantities.}
\label{input_galfit_bulgedisk_comparison}
\end{figure*}

We draw new properties from the empirical galaxy population model for the bulge plus disk simulations.  Galaxies have B/T ratios that are drawn from a uniform distribution in the range $0.1 \le \mathrm{B/T} \le 0.9$.  The bulge and disk profiles of an individual galaxy have the same centers,  while different S\'ersic indices ($n=4$ for bulge,  $n=1$ for disk) and different effective radii.  The disk effective radius is $r_{\mathrm{e},\mathrm{disk}} = N \times r_{\mathrm{e},\mathrm{bulge}}$,  where $N$ is drawn from a uniform distribution in the range $[1, 4]$.  When fitting an exponential disk profile with \textsc{galfit}, the package requires as input the scale radius of the disk,  $R_{\mathrm{s}}$.  Therefore,  we input it by dividing the $r_{\mathrm{e},\mathrm{disk}}$ by 1.678 as explained in \citealt{Peng2010}.

The surface brightness fitting of these galaxies follows the pipeline in Figure \ref{flowchart} as well.  We use the same selection of target galaxies and number of combinations for AS1063 and M1149 fit on stamps as in Section \ref{single_sersic_sims}.  We use \textsc{morphofit} to fit the surface brightness profile of each target galaxy using a S\'ersic and exponential disk profiles for the bulge and the disk,  respectively.  The two profiles share the same initial values for the centroid (\textsc{sextractor} $\tt{ALPHAWIN\_J2000}$ and $\tt{DELTAWIN\_J2000}$),  magnitude (\textsc{sextractor} $\tt{MAG\_AUTO}$),  axis ratio (\textsc{sextractor} $\tt{BWIN\_IMAGE/AWIN\_IMAGE}$),  and position angle (\textsc{sextractor} $\tt{THETAWIN\_SKY}$).  The initial values differ for the S\'ersic index,  which is let free to vary for the bulge, while fixed to $n=1$ for the disk,  and for the effective radius.  The initial value for the bulge effective radius is given by the \textsc{sextractor} $\tt{FLUX\_RADIUS}$ parameter, which is the radius that encloses $50\%$ of the total light from the galaxy.  The initial value for the disk scale length is given by the \textsc{sextractor} $\tt{FLUX\_RADIUS}$ parameter divided by $1.678$, since this is the relation between the scale length and the effective radius quoted in \citealt{Peng2010}. The best-fitting estimates from the fit on stamps are then used as initial values for the fit on regions and on the full images.

Figure \ref{input_galfit_bulgedisk_comparison} shows the comparison between the structural parameters estimated with \textsc{morphofit} and the input properties of simulated bulge plus disk galaxies.  We report the results for the bulge and disk magnitudes and sizes,  and for the bulge S\'ersic indices,  since the disk ones are kept fixed at the values used to generate the simulations.  The results in this case show a larger scatter with respect to the single S\'ersic profile. This is expected since we are trying to simultaneously minimise a larger number of parameters per galaxy,  where the slightest change in light contribution of one of the components is immediately reflected on the parameters of the other one.  Additionally,  the disk component is more susceptible to the estimate of the image background, therefore caution must be applied when fitting the sky background value. 

Bright galaxy magnitudes keep being close to the one-to-one relation.  The percentage of objects consistent with the input magnitudes is around $50\%$ if we consider the $1\sigma$ level,  while it is around $70\%$ at the $3\sigma$ level.  If we consider all galaxies, these percentages go to the $25\%$ at the $1\sigma$ level and $55\%$ at the $3\sigma$ level.  The degeneracy arising in the simultaneous minimisation of two components is particularly evident in the larger errors and scatter of the effective radii and S\'ersic indices.  Bright galaxy bulges and disk effective radii have percentages of agreement with the input values within $1\sigma$ that are slightly lower than those obtained for the magnitudes ($\sim 30\%$),  but similar for the $3\sigma$ case.  The percentages of galaxy bulges S\'ersic indices that are consistent with the input values are $30\%$ for the $1\sigma$ case and $60\%$ for the $3\sigma$ case.  If we consider the bright galaxies,  the percentages rise to $35\%$ and $70\%$ for the $1\sigma$ and $3\sigma$ cases, respectively.

The simulations of single S\'ersic profile galaxies show that \textsc{morphofit} is able to recover with good accuracy and within the errors the input parameters of the simulations.  In the case of the bulge plus disk profile simulations,  the simultaneous minimisation of two components per galaxy causes an increase in the scatter and in the uncertainties of the parameters.  Despite that,  \textsc{morphofit} provides good performances in the recovery of the input parameters for bright galaxies. It also provides a roughly $70\%$ consistency within $3\sigma$ with the input values if we consider galaxies spanning the whole range of simulated magnitudes.

\section{Comparison with published catalogues}
\label{section:comparison_with_catalogues}

The tests on the simulated data demonstrate the good performances of \textsc{morphofit} in estimating the structural parameters,  especially in the case of single S\'ersic profiles.  \textsc{morphofit} is used in LT23 to measure the structural parameters of intermediate redshift cluster galaxies in order to build the Kormendy relation as a function of wavelength. 

\begin{figure*}
\centering
\includegraphics[scale=0.22]{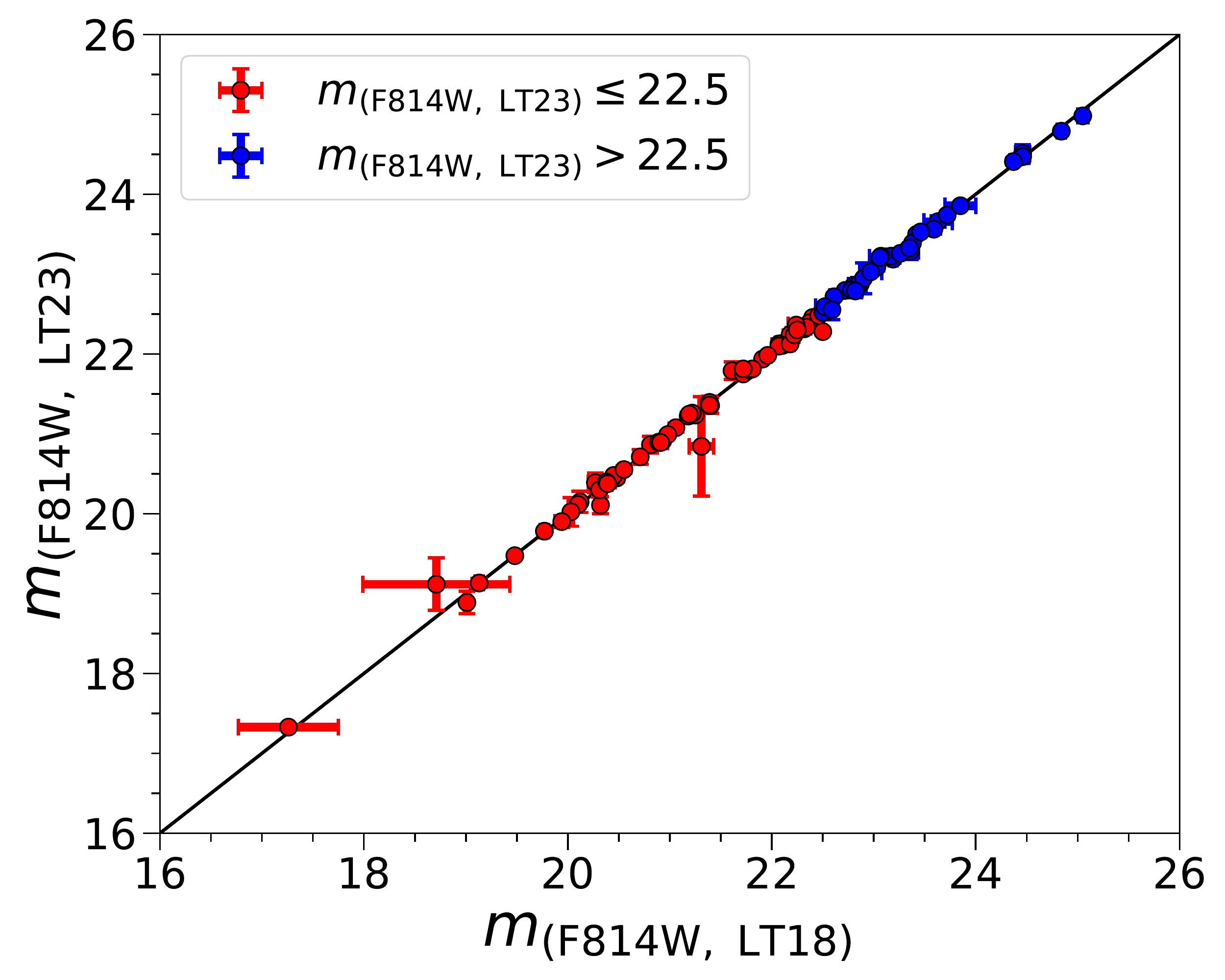}
\includegraphics[scale=0.22]{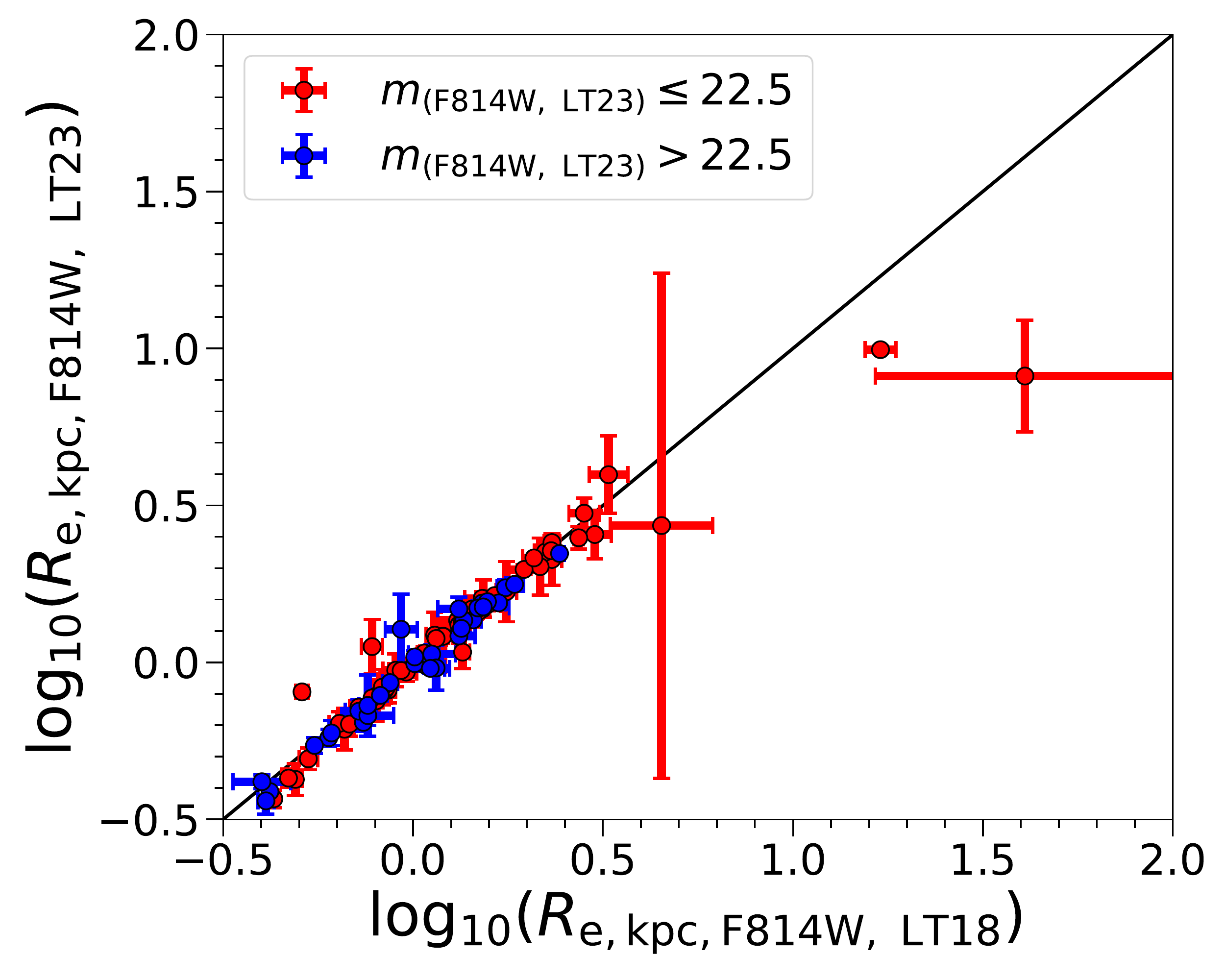}
\includegraphics[scale=0.22]{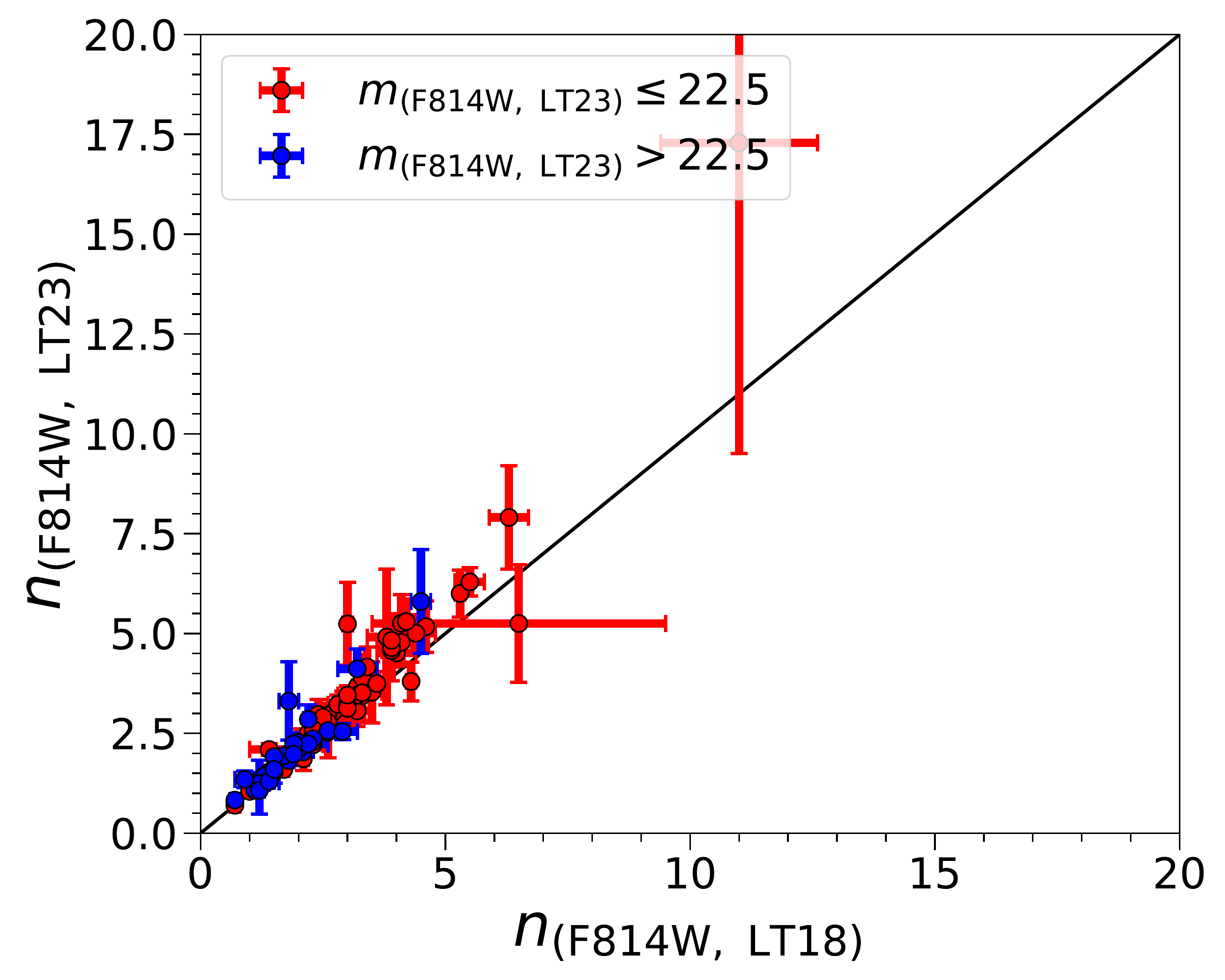}
\includegraphics[scale=0.22]{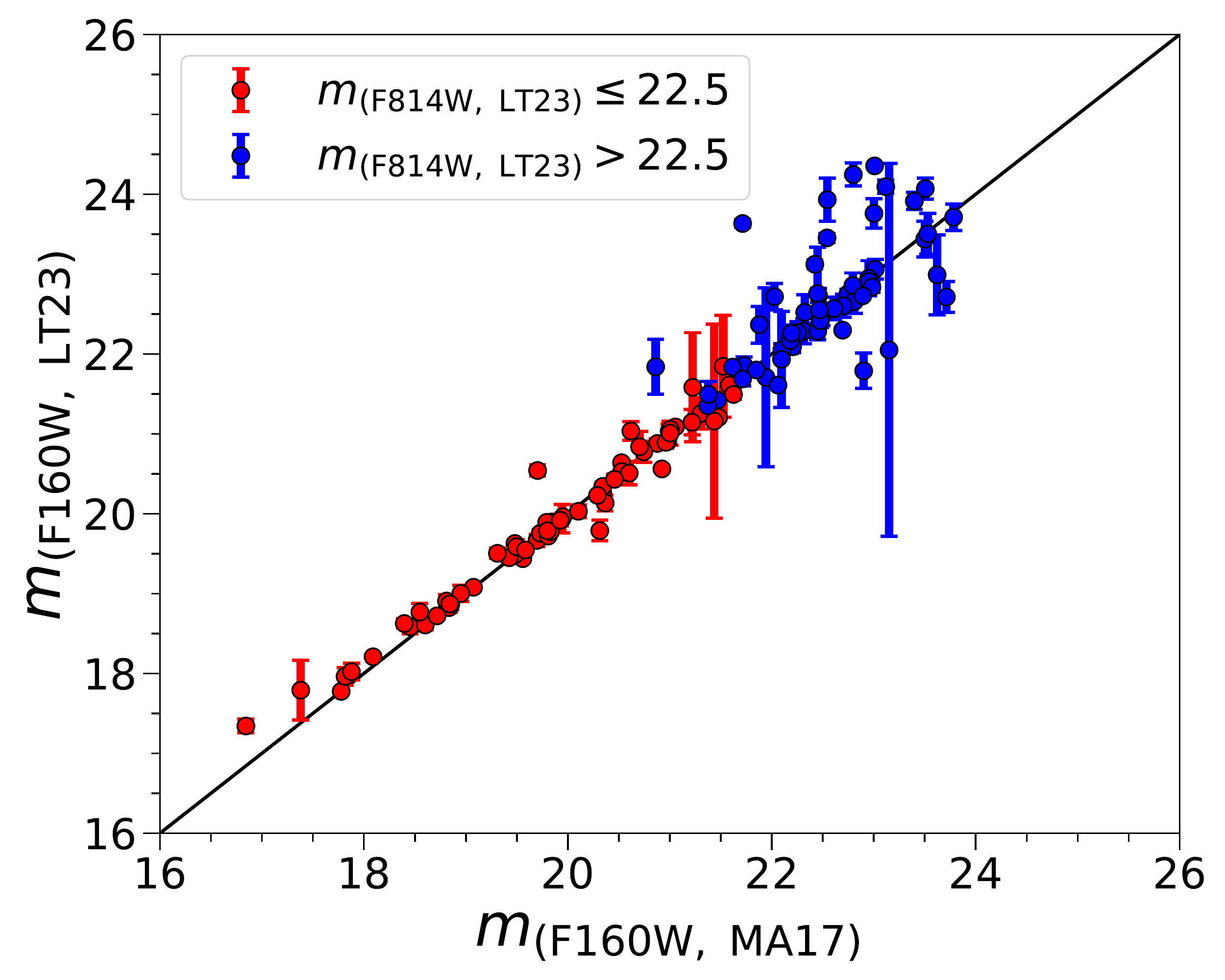}
\includegraphics[scale=0.22]{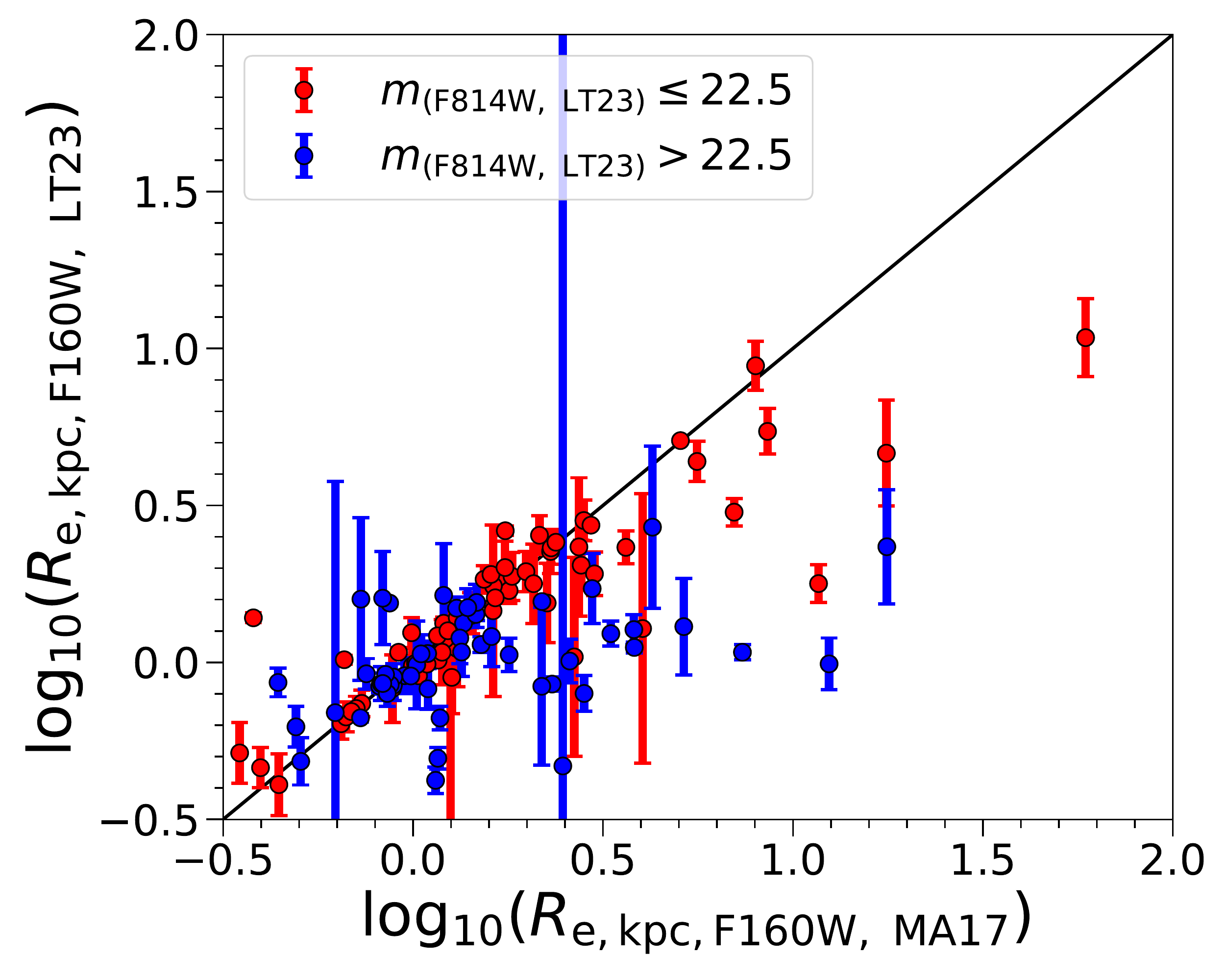}
\includegraphics[scale=0.22]{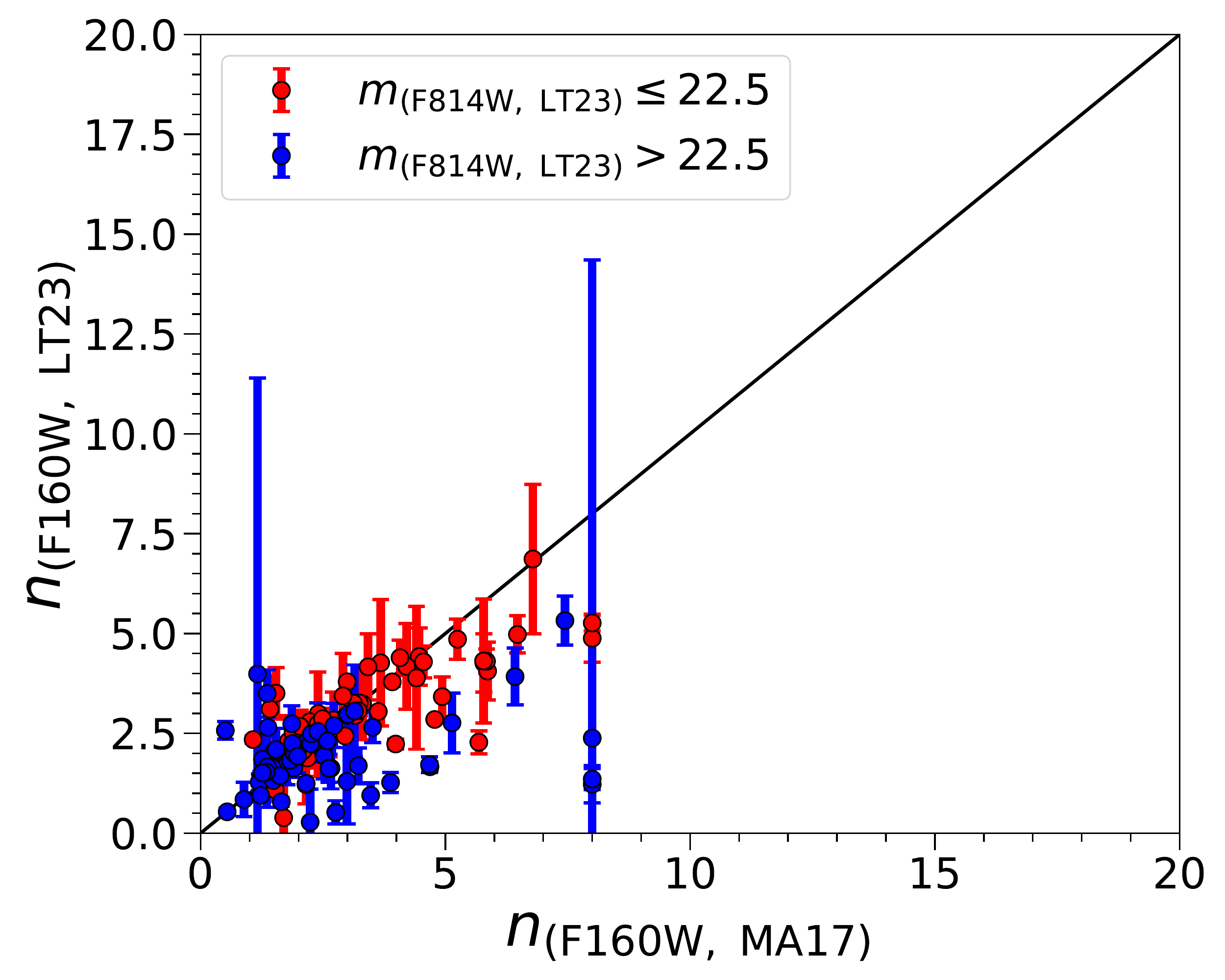}
\includegraphics[scale=0.22]{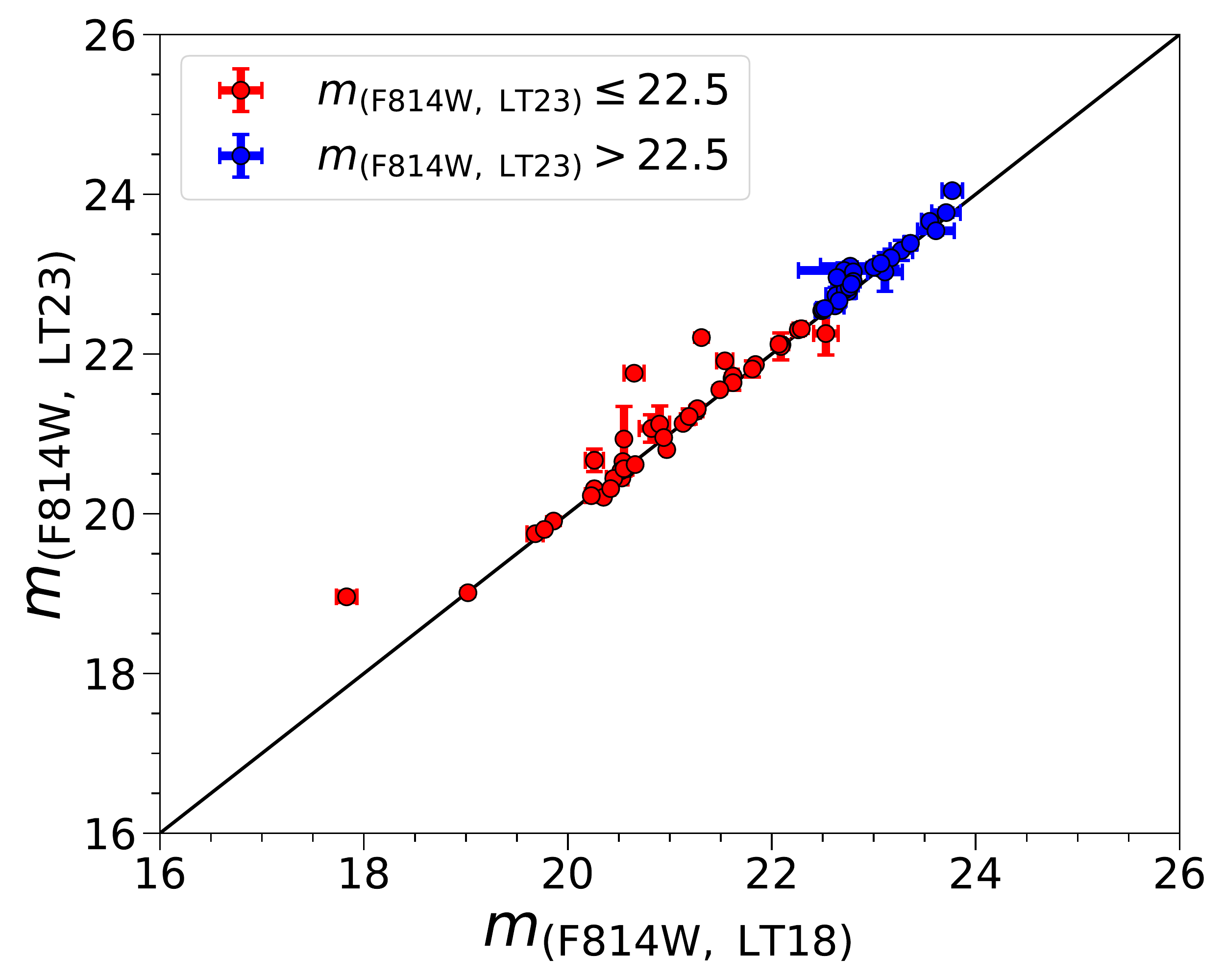}
\includegraphics[scale=0.22]{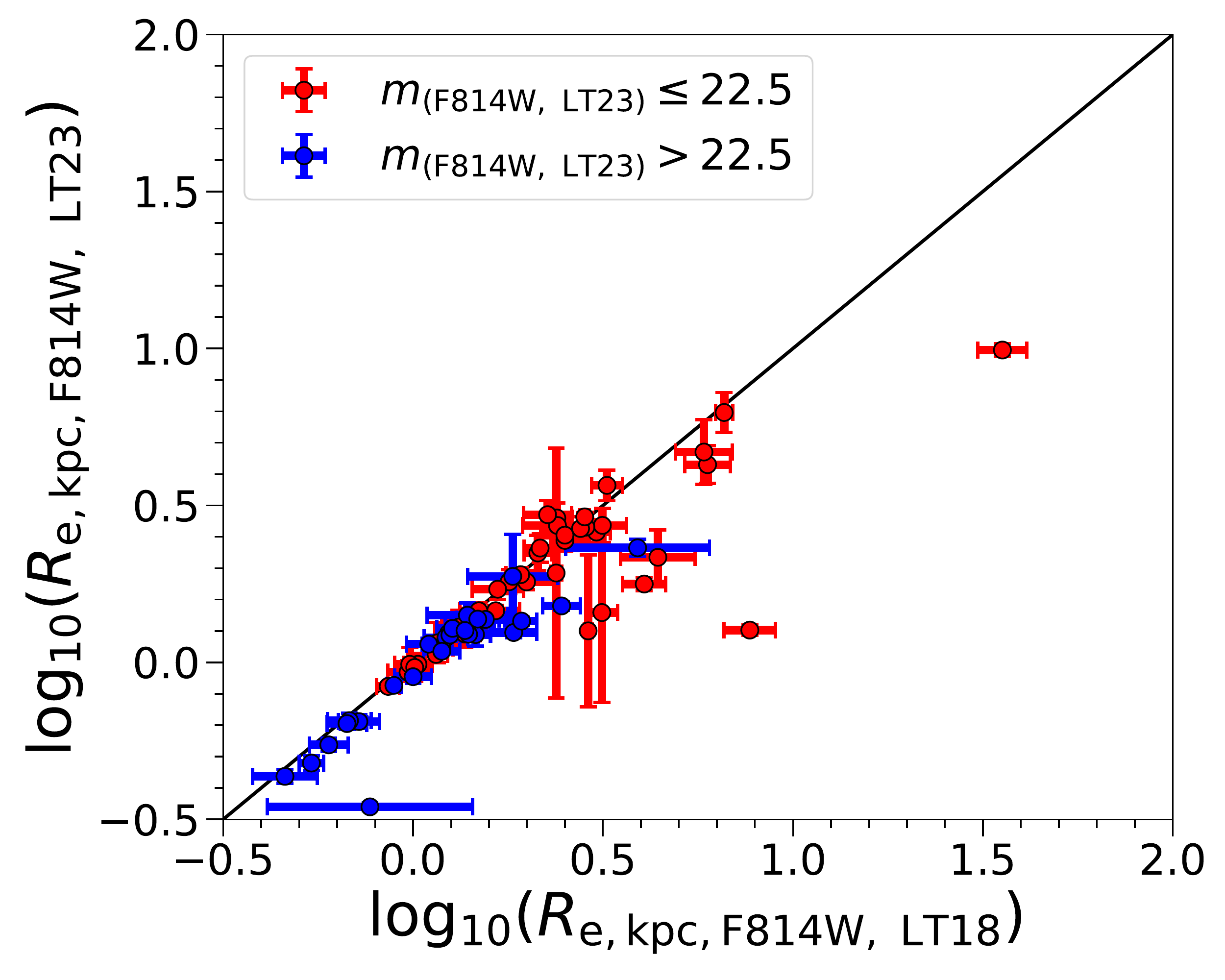}
\includegraphics[scale=0.22]{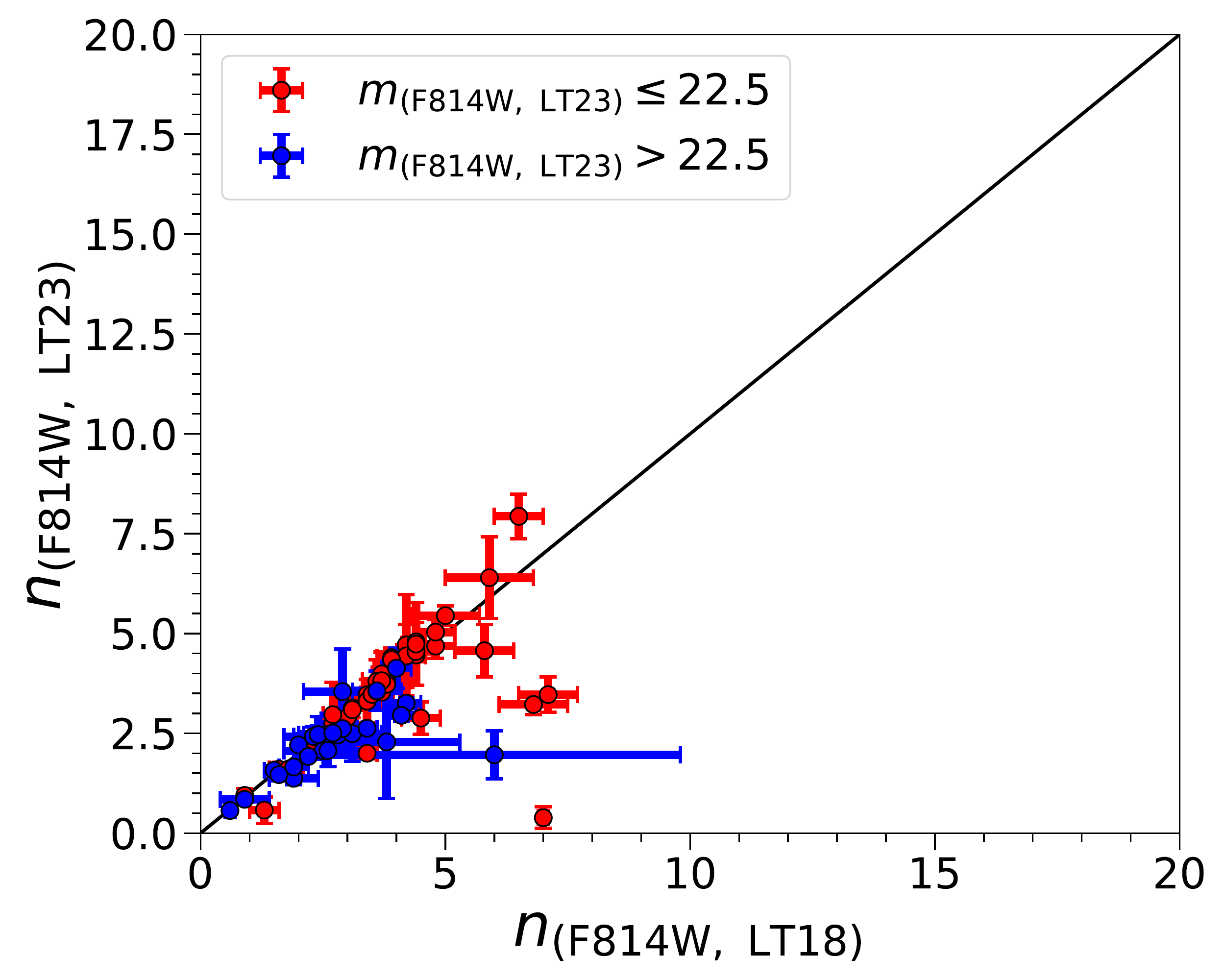}
\caption{The panels show the comparison between the magnitudes,  the $\log_{10}{R_{\mathrm{e}}}$,  with $R_{\mathrm{e}}$ in $\mathrm{kpc}$, and the S\'ersic indices in LT23 and those of LT18 (top and bottom panels) and MA17 (middle panels).  LT18 structural parameters are measured for AS1063 and M1149,  while MA17 for M0416. The comparisons with LT18 are in the \textit{F814W} waveband,  while the comparison with MA17 in the \textit{F160W} waveband. In all panels,  red points represent galaxies with magnitude $m \le 22.5$ in the \textit{F814W} waveband,  while blue points represent galaxies fainter than this limit in the same waveband. The black solid lines represent the one-to-one relations.}
\label{comparison_with_literature}
\end{figure*}

In Figure \ref{comparison_with_literature}, we compare the structural parameters estimated with \textsc{morphofit} in LT23 with already existing results in the literature for the same clusters.  In particular,  we fit galaxies with single S\'ersic profiles and we compare the parameters of the \textit{F814W} waveband with those measured in LT18 for AS1063 and M1149.  We also reviewed papers with independent and publicly available catalogues. We compare the \textit{F160W} waveband structural parameters with those measured in \citealt{Annunziatella2017} (hereafter,  MA17) for MACS J0416.1-2403 (M0416).  This allows us to check whether the \textsc{morphofit} structural parameters estimation and the methodology behind it perform similarly to already existing results in the literature.  The studies of \citealt{Merlin2016,DiCriscienzo2017,Shipley2018,Bradac2019,Pagul2021} have published photometric catalogues for AS1063, M1149,  and M0416.  However, either they do not use \textsc{galfit} at all,  or they use \textsc{galapagos} \citep{Barden2012} solely to account for the contamination of the intracluster light. The photometry that is released in these catalogues has been mainly obtained by running \textsc{sextractor} (or a similar software), therefore it cannot be compared to our results and used to demonstrate that automatising the process yields similar results as those obtained from manual adjustments to the fit (see Appendix \ref{appendix} for a comparison between the \textsc{sextractor} structural parameters estimate and the \textsc{morphofit} one).

The magnitude comparison for AS1063 (upper left panel) shows that all galaxies are consistent within the errors with the one-to-one relation.  For M0416 (central left panel),  the consistency holds for bright galaxies.  Faint galaxies show a large scatter, particularly those with $m_{\mathrm{F160W}} > 22.5$, and some of them are not consistent within the errors with the one-to-one relation.  This is expected since fainter objects are increasingly harder to fit and different methodologies adopted in different works may lead to very different estimates.  For M1149 (bottom left panel),  the results are similar to those of AS1063, except for three objects (one of them being the BCG) that resides in the centre of the cluster where there is strong contamination from lensed objects.

The comparison of $\log_{10}{R_{\mathrm{e}}}$, with $R_{\mathrm{e}}$ in $\mathrm{kpc}$, for AS1063 (top central panel) shows that both faint and bright galaxies are consistent within the errors with the one-to-one relation at $\log_{10}{R_{\mathrm{e}}} < 1.0$.  The two largest galaxies of the sample (one of them being the BCG) are not consistent.  For such large galaxies,  the different treatments of the background noise among different studies have a tremendous impact on the final value of the size.  The trend for M0416 shows a more pronounced scatter with respect to the other two clusters.  The majority of the bright objects at $\log_{10}{R_{\mathrm{e}}} < 0.5$ are consistent with the one-to-one relation, but with errors larger than the other two clusters.  At $\log_{10}{R_{\mathrm{e}}} > 0.5$,  MA17 sizes tend to be larger than the results from \textsc{morphofit}.  For faint galaxies,  only the smallest ones are consistent within the errors with the one-to-one relation, otherwise MA17 sizes are overestimated with respect to the \textsc{morphofit} estimates. M1149 shows a similar behaviour as AS1063,  with the two largest galaxies (BCG included) not consistent with the one-to-one relation.

The AS1063 S\'ersic indices (top right panel) estimated  with \textsc{morphofit} are all consistent within the errors with the one-to-one relation.  The larger the input S\'ersic index, the larger is the error on the estimate.  The M0416 S\'ersic indices of bright galaxies below $n < 5$ are consistent within the errors with the one-to-one relation,  while, above that threshold,  MA17 S\'ersic indices tend to be overestimated with respect to the \textsc{morphofit} estimates.  For faint galaxies,  only the ones with $n < 2.5$ are consistent with the one-to-one relation, while, at higher S\'ersic indices,  the \textsc{morphofit} estimate tends to be smaller than that from MA17.  The M1149 S\'ersic indices follow a similar trend as in AS1063,  albeit with five sources at $n > 5$ that are not consistent with the one-to-one relation. 

Overall, the comparison shows that there is a very good agreement of the \textsc{morphofit} properties with the literature ones, especially for AS1063 and M1149,  while for M0416 the scatter is larger,  but the consistency is still good.  Faint galaxies in MA17 show a larger discrepancy with respect to the \textsc{morphofit} estimate.  This is due to the larger impact that different treatments of PSF,  sigma image and background estimation have on the morphological analysis of faint galaxies.  The tests on simulations and literature data confirm the robustness of the parameters estimated with \textsc{morphofit} and,  consequently,  the analysis of the Kormendy relation carried out in LT23.

\section{Conclusions}
\label{section:conclusions}

In this paper we introduce \textsc{morphofit},  a package for the estimate of galaxy structural parameters.  The package is designed to answer to the necessity of modern wide-field surveys for surface brightness decomposition codes that are accurate,  highly parallelisable and require a small degree of human intervention.  The package is written in \textsc{python} and it is constituted by modules that can be run independently or in a pipeline.  The latter allows the user to estimate the galaxy structural parameters in stamps around galaxies or using all the galaxies present in an image and it contains routines to create PSF images from imaged stars.  \textsc{morphofit} allows the user to  fit multiple components to each galaxy among those currently implemented in the code, namely the S\'ersic,  the deVaucouleurs and the exponential disk profiles.  It makes use of wide-spread and reliable codes, such as \textsc{sextractor} and \textsc{galfit}.  \textsc{morphofit} is also designed to take advantage of the pipeline first introduced in \citealt{Tortorelli2018},  refined in this work and used in \citealt{Tortorelli2023} to measure the Kormendy relation dependence on wavelength.  Its parallelisation capabilities relies on the use of the \textsc{python} package \textsc{esub-epipe}.

We test the accuracy of the results by creating simulated images mimicking the observational and instrumental conditions of the HST Frontier Fields survey.  We create two sets of simulations using \textsc{ufig}: one where galaxies are rendered as single S\'ersic profiles and another one where galaxies are rendered as bulge plus disk components with a variable bulge-to-total luminosity ratio.  This second set of simulations is of great importance because,  although galaxies may have a lot of structures (like rings, pseudorings, lenses, bars, envelops, and spiral arms),  a parametric bulge-disk decomposition is one of the first possible quantitative steps to tag a galaxy as an early or late-type system.  We use \textsc{morphofit} and the pipeline described in Figure \ref{flowchart} to estimate the structural parameters of simulated galaxies and check whether we are able to recover the input parameters of the simulations within the errors.  We find that the structural parameters measured by \textsc{morphofit} on the single S\'ersic simulated galaxies are all consistent within the errors with the input ones.  For the bulge plus disk components,  the simultaneous minimisation of two components increases the degeneracies in the recovery of the parameters.  This leads to roughly a $50\%$ ($70\%$) of bright objects that show recovered magnitudes consistent with the input ones at the $1\sigma$ ($3\sigma$) level,  for both bulges and disks.  A smaller degree of consistency is present for bright galaxies effective radii and S\'ersic indices at the $1\sigma$ level ($35\%$), while at the $3\sigma$ level the consistency is similar to the magnitude case.

We also compare the structural parameters estimated with \textsc{morphofit} on observed data with those from existing literature studies.  We use the data of the clusters Abell S1063,  MACS J1149.5+2223 and MACS J0416.1-2403 from the HST Frontier Fields survey.  We compare the parameters of the \textit{F814W} waveband for Abell S1063 and MACS J1149.5+2223 with the results from \citealt{Tortorelli2018} and of the \textit{F160W} waveband for MACS J0416.1-2403 with the results from \citealt{Annunziatella2017}.  We find a very good consistency of the structural parameters for Abell S1063 and MACS J1149.5+2223 with \citealt{Tortorelli2018}. The consistency with \citealt{Annunziatella2017} is also very good,  especially for bright galaxies.  Faint galaxies have a more sparse behaviour, but this is expected given the larger impact the different choice of PSF,  sigma images and background estimations have on faint galaxies morphological properties.

Overall,  \textsc{morphofit} is able to correctly recover the input simulated parameters,  as well as being consistent with existing literature results.  This shows that this package constitutes a promising tool that can be exploited with modern wide-field surveys to measure the structural parameters of a large number of galaxies in a parallel,  accurate,  and multi-component way, with a very small degree of human intervention.  Future prospects include the addition of all the available \textsc{galfit} surface brightness profiles and the possibility to choose between the use of \textsc{galfit} and \textsc{galfitm} \citep{Haussler2013} for the simultaneous fit of all the wavebands.

\section*{Acknowledgments}
The authors are grateful to Mario Nonino, Claudio Grillo and Giovanni Granata for helpful discussions and feedbacks on the package.  LT acknowledges support by Swiss National Science Foundation (SNF) grant 200021\_169130.  The numerical simulations were performed on the Euler cluster operated by the High Performance Computing group at ETH Z\"urich.  The CLASH Multi-Cycle Treasury Program is based on observations made with the NASA/ESA Hubble Space Telescope.  The Space Telescope Science Institute is operated by the Association of Universities for Research in Astronomy, Inc., under NASA contract NAS 5-26555. Based on observations made with the European Southern Observatory Very Large Telescope (ESO/VLT) at Cerro Paranal, under programme IDs 60.A-9345(A), 095.A-0653(A), 294.A-5032(A) and 186.A-0798(A).

\section*{Data Availability Statement}
The raw data supporting the conclusions of this article will be made available by the authors in the github page https://github.com/torluca/morphofit, without undue reservation.

\newpage

\appendix
\section*{Appendix A: Comparison between \textsc{morphofit} and \textsc{sextractor}}
\label{appendix}

\begin{figure*}
\centering
\includegraphics[scale=0.2]{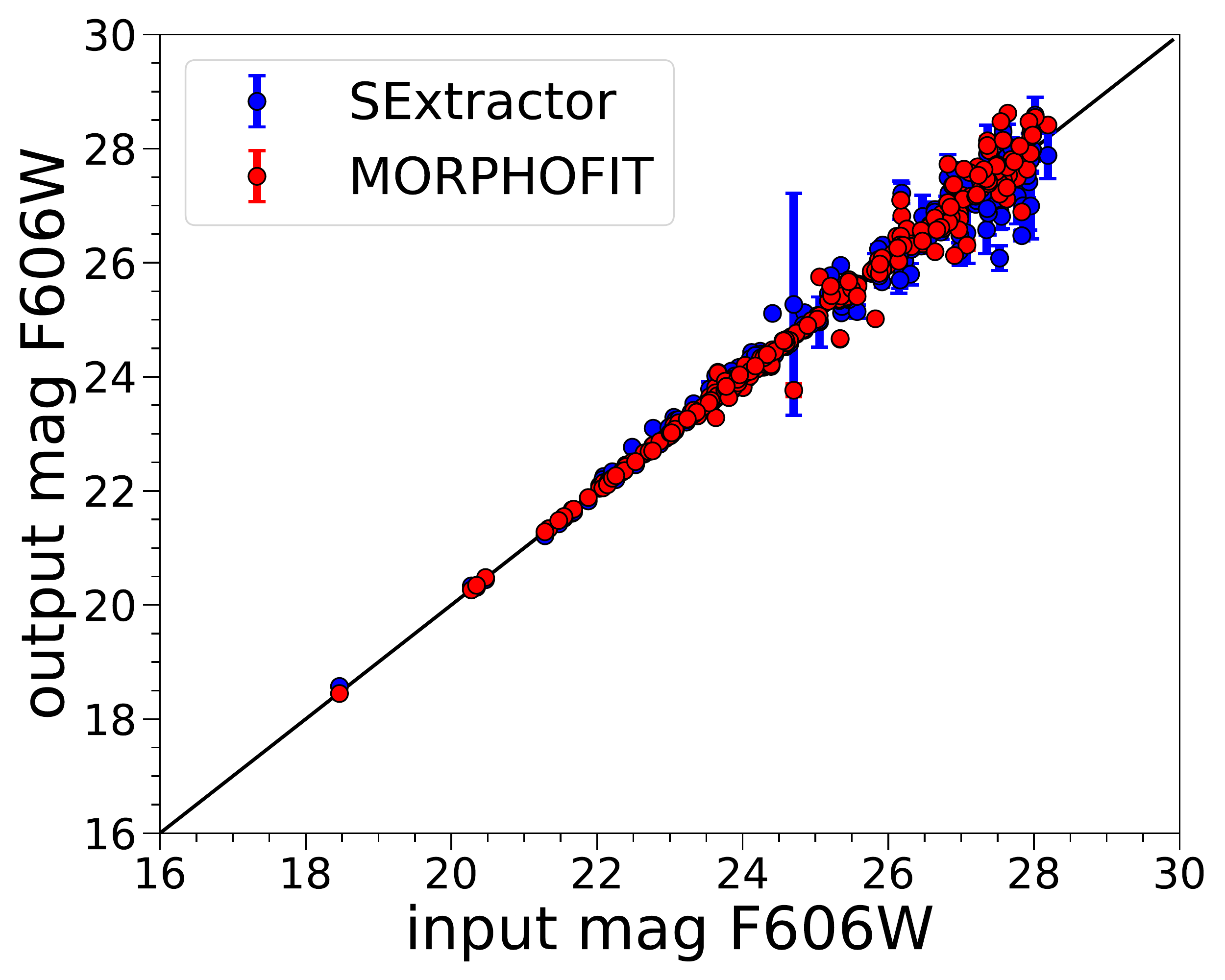}
\includegraphics[scale=0.2]{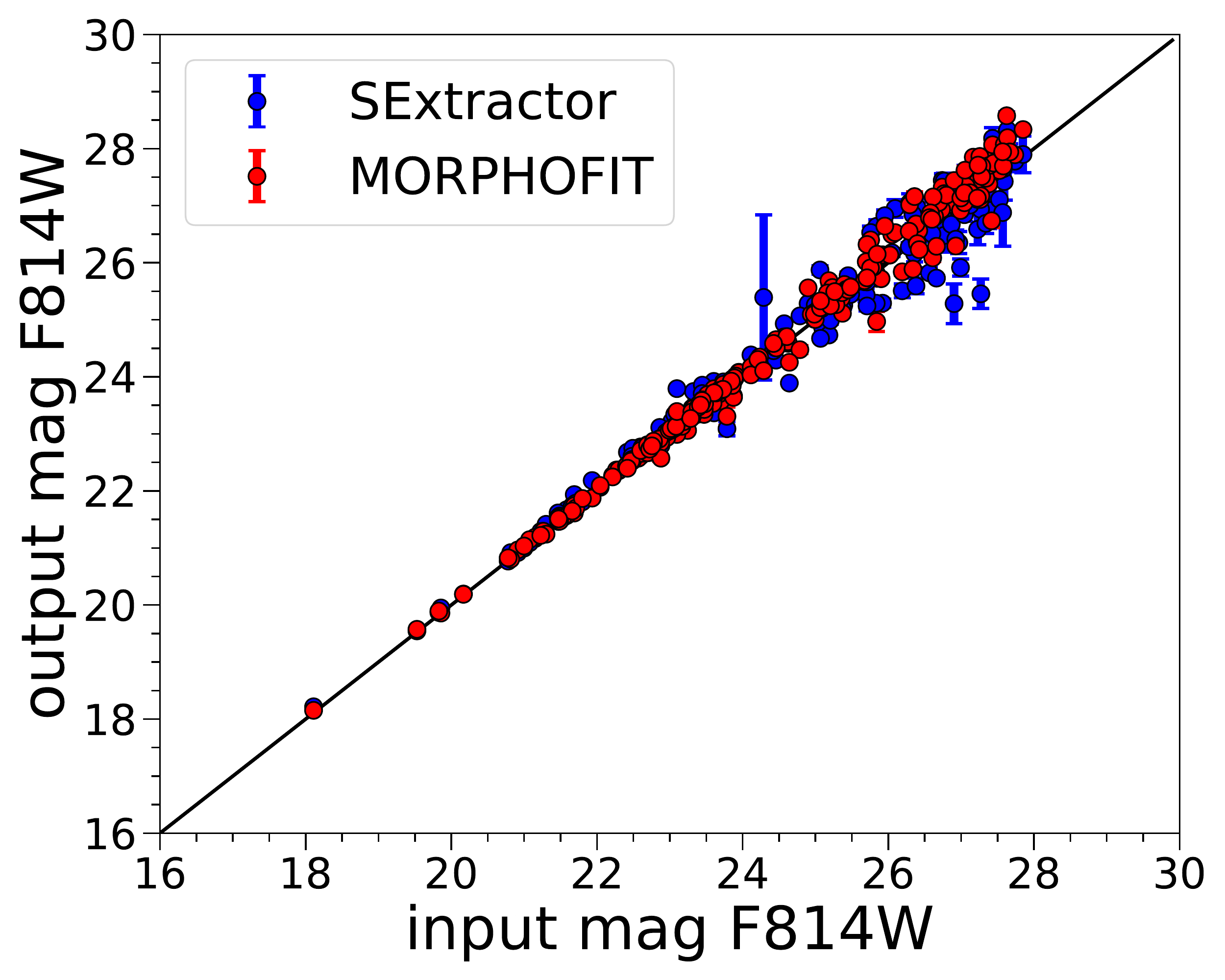}
\includegraphics[scale=0.2]{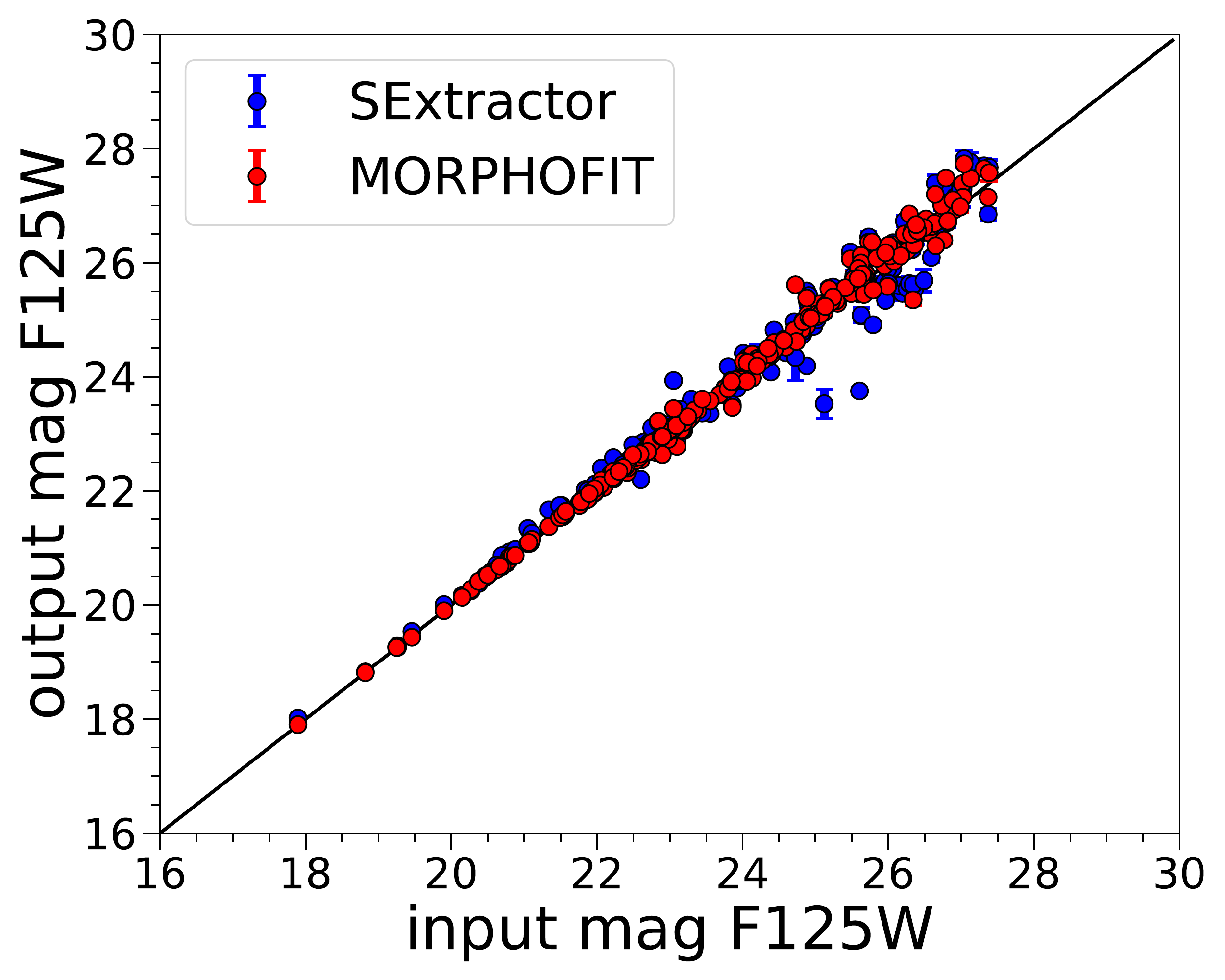}
\includegraphics[scale=0.2]{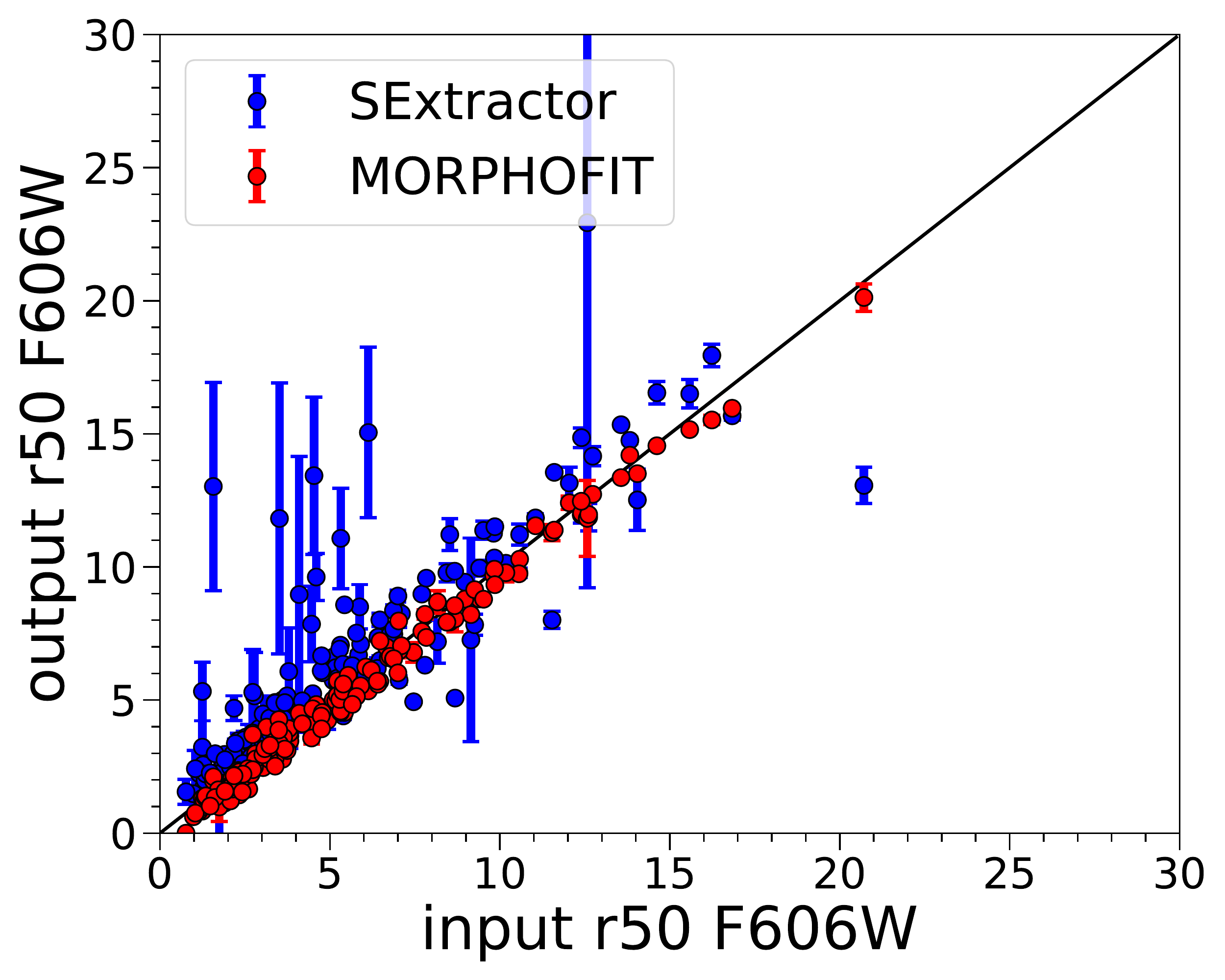}
\includegraphics[scale=0.2]{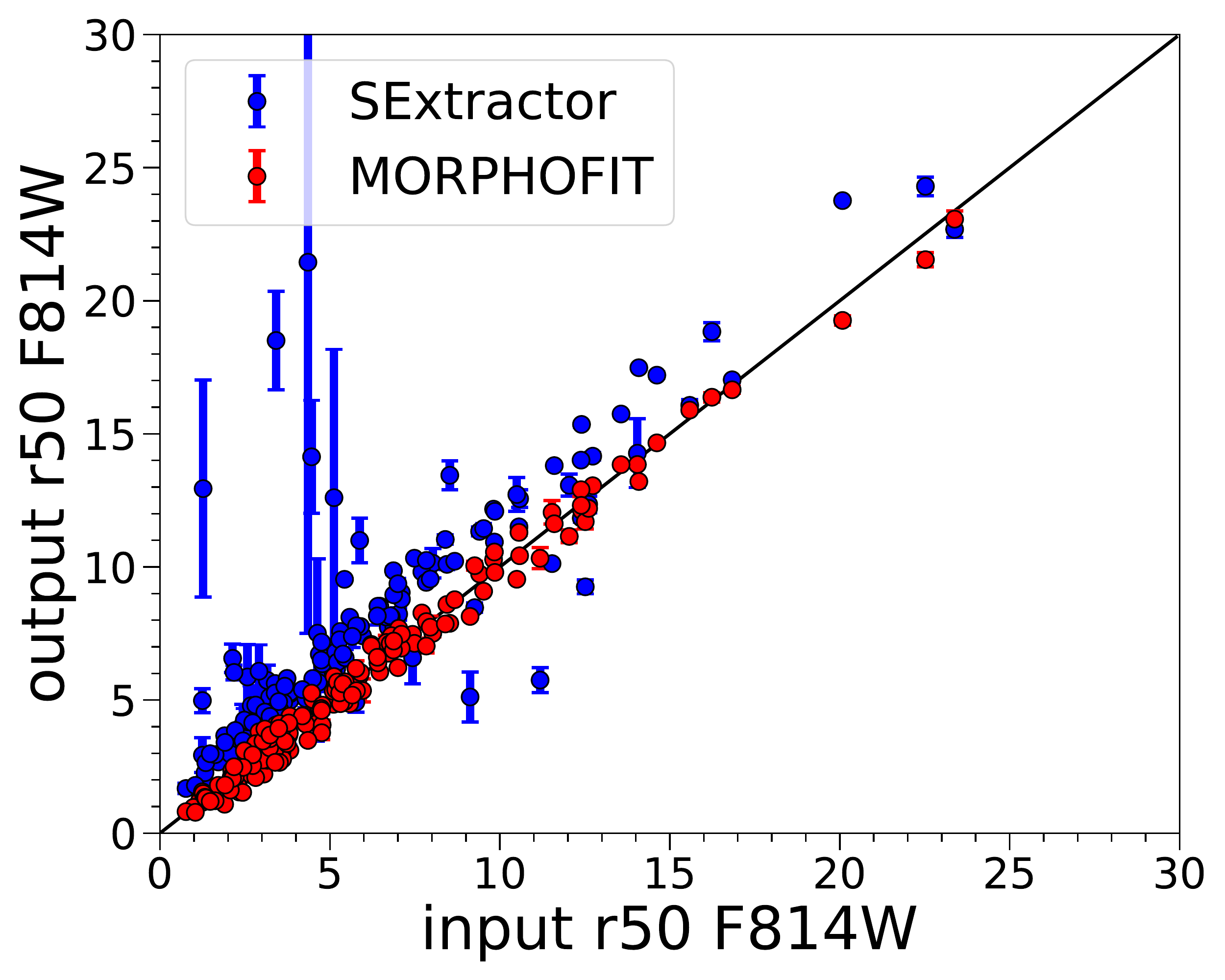}
\includegraphics[scale=0.2]{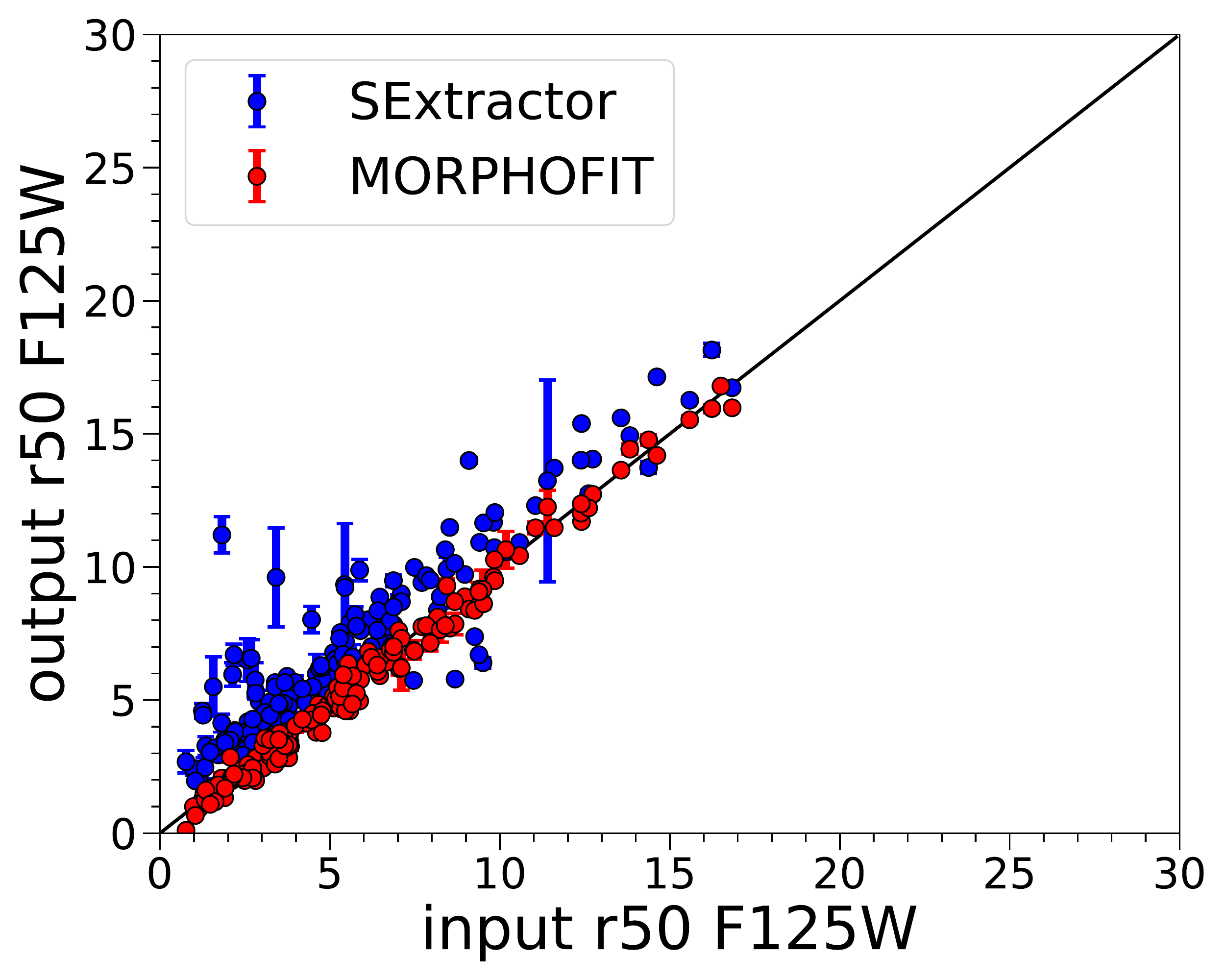}
\includegraphics[scale=0.2]{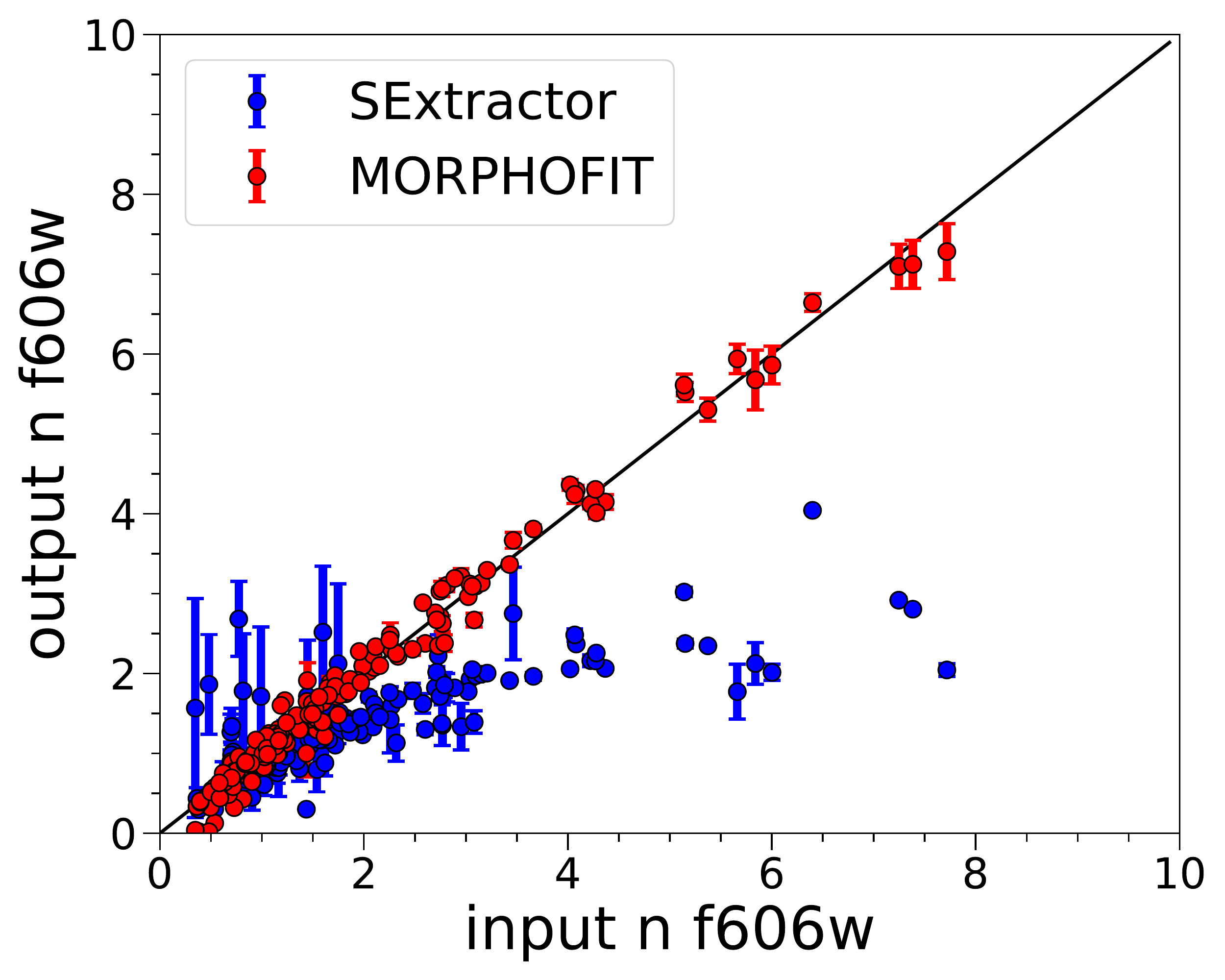}
\includegraphics[scale=0.2]{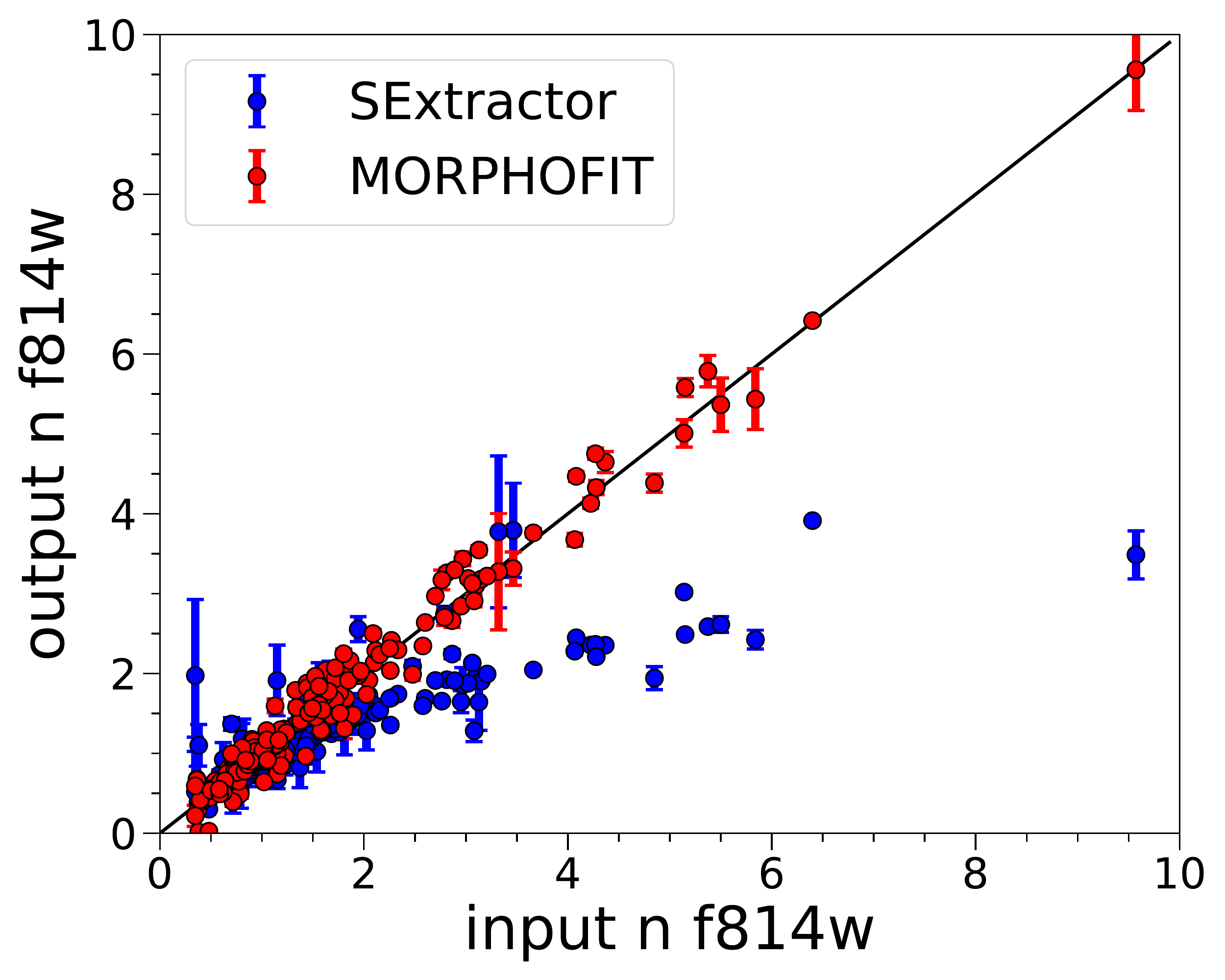}
\includegraphics[scale=0.2]{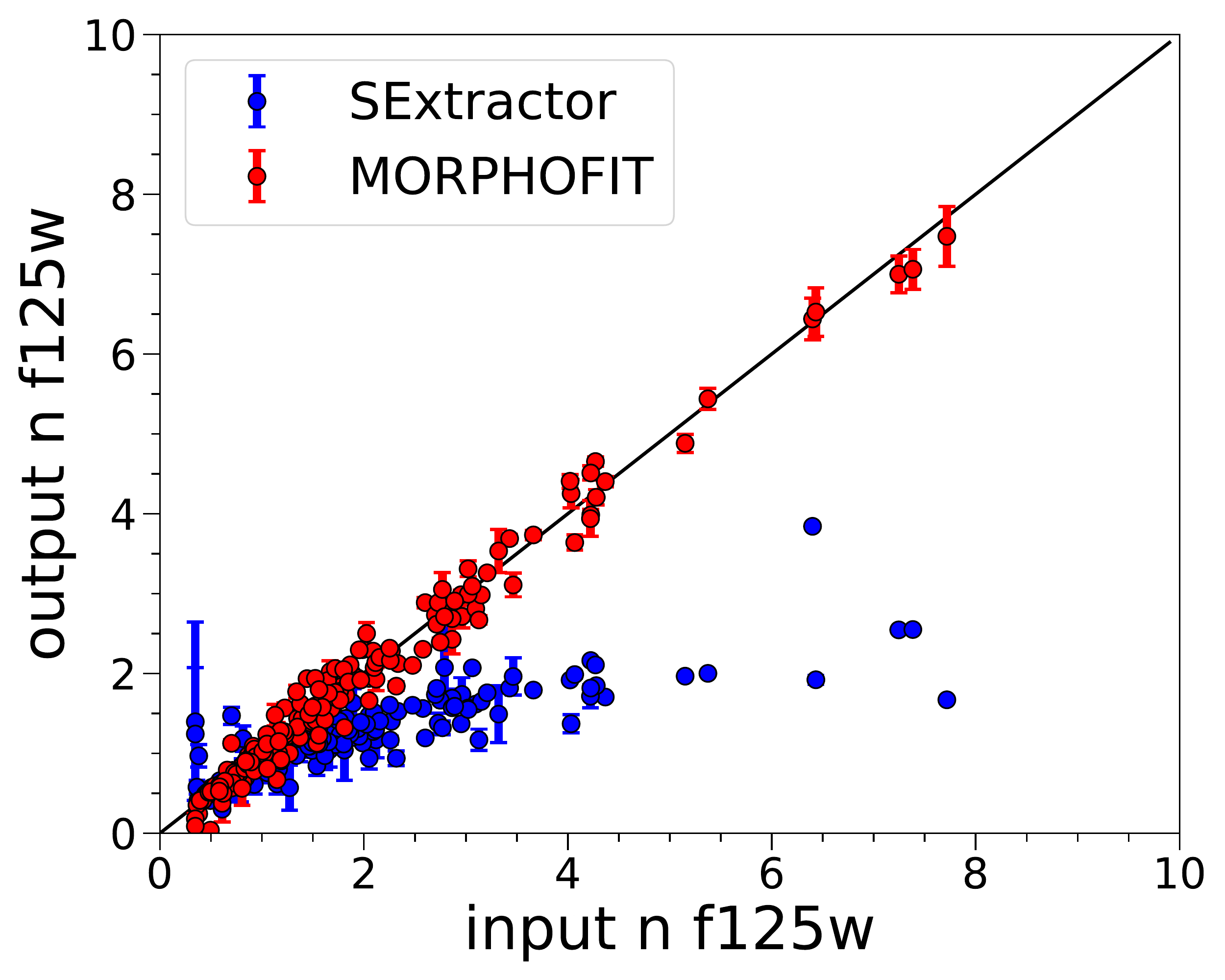}
\caption{The top, middle and bottom panels show the comparison between the input simulated galaxy magnitudes, the input simulated galaxy effective radii, the input simulated galaxy S\'ersic indices and the best-fitting magnitudes,  effective radii, S\'ersic indices from \textsc{morphofit} (in red) and \textsc{sextractor} (in blue), respectively.  Magnitudes are in the AB system,  while the effective radii are expressed in units of pixels.  $\mathrm{mag}$, $\mathrm{r_{\mathrm{e}}}$ and $\mathrm{n}$ refer to the total magnitude,  the effective radius and the S\'ersic index of the input and fitted S\'ersic profile. For each row,  we only select a sub-sample of the seven wavebands for plot clarity.  The black solid line represents the one-to-one relation between the input and the best-fitting quantities}
\label{morphofit_sextractor_comparison}
\end{figure*}

\textsc{sextractor} contains routines that allow the user to perform parametric surface brightness fitting. We compare the results of running \textsc{morphofit} on simulated single S\'ersic profile galaxies (see section \ref{single_sersic_sims}) with the parametric fit performed by \textsc{sextractor} on the same galaxies.  Figure \ref{morphofit_sextractor_comparison} shows that the magnitudes predicted by \textsc{sextractor} lie close to the one-to-one relation with the input parameters, albeit with a larger scatter with respect to the \textsc{morphofit} estimates and larger errors at faint magnitudes.  The figure also shows that \textsc{sextractor} predicted S\'ersic indices have larger errors with respect to the \textsc{morphofit} ones and are systematically under-estimated with respect to the input values for $n > 2$.  Consequently,  the effective radii are systematically over-estimated with respect to the input ones.  We implement in \textsc{morphofit} the possibility for the user to access the \textsc{sextractor} parametric fit feature and use those values as initial estimates for the fit with \textsc{galfit}. 

\bibliographystyle{Frontiers-Harvard} 
\bibliography{tortorelli_2023_bibliography_frontiers}

\begin{thebibliography}{63}
\providecommand{\natexlab}[1]{#1}
\expandafter\ifx\csname urlstyle\endcsname\relax
  \providecommand{\doi}[1]{doi:\discretionary{}{}{}#1}\else
  \providecommand{\doi}{doi:\discretionary{}{}{}\begingroup
  \urlstyle{rm}\Url}\fi
\providecommand{\selectlanguage}[1]{\relax}
\providecommand{\bibAnnoteFile}[1]{%
  \IfFileExists{#1}{\begin{quotation}\noindent\textsc{Key:} #1\\
  \textsc{Annotation:}\ \input{#1}\end{quotation}}{}}
\providecommand{\bibAnnote}[2]{%
  \begin{quotation}\noindent\textsc{Key:} #1\\
  \textsc{Annotation:}\ #2\end{quotation}}

\bibitem[{{Amara} and {Quanz}(2012)}]{Amara2012}
{Amara}, A. and {Quanz}, S.~P. (2012).
\newblock {PYNPOINT: an image processing package for finding exoplanets}.
\newblock \emph{Monthly Notices of the Royal Astronomical Society} 427,
  948--955.
\newblock \doi{10.1111/j.1365-2966.2012.21918.x}
\bibAnnoteFile{Amara2012}

\bibitem[{{Anderson}(2016)}]{Anderson2016}
[Dataset] {Anderson}, J. (2016).
\newblock {Empirical Models for the WFC3/IR PSF}.
\newblock Space Telescope WFC Instrument Science Report
\bibAnnoteFile{Anderson2016}

\bibitem[{{Anderson} and {King}(2000)}]{Anderson2000}
{Anderson}, J. and {King}, I.~R. (2000).
\newblock {Toward High-Precision Astrometry with WFPC2. I. Deriving an Accurate
  Point-Spread Function}.
\newblock \emph{Publications of the ASP} 112, 1360--1382.
\newblock \doi{10.1086/316632}
\bibAnnoteFile{Anderson2000}

\bibitem[{{Annunziatella} et~al.(2017){Annunziatella}, {Bonamigo}, {Grillo},
  {Mercurio}, {Rosati}, {Caminha} et~al.}]{Annunziatella2017}
{Annunziatella}, M., {Bonamigo}, M., {Grillo}, C., {Mercurio}, A., {Rosati},
  P., {Caminha}, G., et~al. (2017).
\newblock {Mass Profile Decomposition of the Frontier Fields Cluster MACS
  J0416-2403: Insights on the Dark-matter Inner Profile}.
\newblock \emph{Astrophysical Journal} 851, 81.
\newblock \doi{10.3847/1538-4357/aa9845}
\bibAnnoteFile{Annunziatella2017}

\bibitem[{{Barden} et~al.(2012){Barden}, {H{\"a}u{\ss}ler}, {Peng}, {McIntosh},
  and {Guo}}]{Barden2012}
{Barden}, M., {H{\"a}u{\ss}ler}, B., {Peng}, C.~Y., {McIntosh}, D.~H., and
  {Guo}, Y. (2012).
\newblock {GALAPAGOS: from pixels to parameters}.
\newblock \emph{Monthly Notices of the Royal Astronomical Society} 422,
  449--468.
\newblock \doi{10.1111/j.1365-2966.2012.20619.x}
\bibAnnoteFile{Barden2012}

\bibitem[{{Berg{\'e}} et~al.(2013){Berg{\'e}}, {Gamper}, {R{\'e}fr{\'e}gier},
  and {Amara}}]{Berge2013}
{Berg{\'e}}, J., {Gamper}, L., {R{\'e}fr{\'e}gier}, A.~r., and {Amara}, A.
  (2013).
\newblock {An Ultra Fast Image Generator (UFIG) for wide-field astronomy}.
\newblock \emph{Astronomy and Computing} 1, 23--32.
\newblock \doi{10.1016/j.ascom.2013.01.001}
\bibAnnoteFile{Berge2013}

\bibitem[{{Bertin} and {Arnouts}(1996)}]{Bertin1996}
{Bertin}, E. and {Arnouts}, S. (1996).
\newblock {SExtractor: Software for source extraction.}
\newblock \emph{Astronomy and Astrophysicss} 117, 393--404.
\newblock \doi{10.1051/aas:1996164}
\bibAnnoteFile{Bertin1996}

\bibitem[{{Binney} et~al.(1990){Binney}, {Davies}, and
  {Illingworth}}]{Binney1990}
{Binney}, J.~J., {Davies}, R.~L., and {Illingworth}, G.~D. (1990).
\newblock {Velocity Mapping and Models of the Elliptical Galaxies NGC 720, NGC
  1052, and NGC 4697}.
\newblock \emph{The Astrophysical Journal} 361, 78.
\newblock \doi{10.1086/169169}
\bibAnnoteFile{Binney1990}

\bibitem[{{Bonfini}(2014)}]{Bonfini2014}
{Bonfini}, P. (2014).
\newblock {GALFIT-CORSAIR: Implementing the Core-S{\'e}rsic Model Into GALFIT}.
\newblock \emph{Publications of the ASP} 126, 935.
\newblock \doi{10.1086/678566}
\bibAnnoteFile{Bonfini2014}

\bibitem[{{Brada{\v{c}}} et~al.(2019){Brada{\v{c}}}, {Huang}, {Fontana},
  {Castellano}, {Merlin}, {Amor{\'\i}n} et~al.}]{Bradac2019}
{Brada{\v{c}}}, M., {Huang}, K.-H., {Fontana}, A., {Castellano}, M., {Merlin},
  E., {Amor{\'\i}n}, R., et~al. (2019).
\newblock {Hubble Frontier Field photometric catalogues of Abell 370 and RXC
  J2248.7-4431: multiwavelength photometry, photometric redshifts, and stellar
  properties}.
\newblock \emph{Monthly Notices of the Royal Astronomical Society} 489,
  99--107.
\newblock \doi{10.1093/mnras/stz2119}
\bibAnnoteFile{Bradac2019}

\bibitem[{Bradley et~al.(2020)Bradley, Sipőcz, Robitaille, Tollerud,
  Vinícius, Deil et~al.}]{photutils}
[Dataset] Bradley, L., Sipőcz, B., Robitaille, T., Tollerud, E., Vinícius,
  Z., Deil, C., et~al. (2020).
\newblock astropy/photutils: 1.0.0.
\newblock \doi{10.5281/zenodo.4044744}
\bibAnnoteFile{photutils}

\bibitem[{{Bruderer} et~al.(2016){Bruderer}, {Chang}, {Refregier}, {Amara},
  {Berg{\'e}}, and {Gamper}}]{Bruderer2016}
{Bruderer}, C., {Chang}, C., {Refregier}, A., {Amara}, A., {Berg{\'e}}, J., and
  {Gamper}, L. (2016).
\newblock {Calibrated Ultra Fast Image Simulations for the Dark Energy Survey}.
\newblock \emph{Astrophysical Journal} 817, 25.
\newblock \doi{10.3847/0004-637X/817/1/25}
\bibAnnoteFile{Bruderer2016}

\bibitem[{{Cappellari} et~al.(2013){Cappellari}, {Scott}, {Alatalo}, {Blitz},
  {Bois}, {Bournaud} et~al.}]{Cappellari2013}
{Cappellari}, M., {Scott}, N., {Alatalo}, K., {Blitz}, L., {Bois}, M.,
  {Bournaud}, F., et~al. (2013).
\newblock {The ATLAS$^{3D}$ project - XV. Benchmark for early-type galaxies
  scaling relations from 260 dynamical models: mass-to-light ratio, dark
  matter, Fundamental Plane and Mass Plane}.
\newblock \emph{Monthly Notices of the Royal Astronomical Society} 432,
  1709--1741.
\newblock \doi{10.1093/mnras/stt562}
\bibAnnoteFile{Cappellari2013}

\bibitem[{{Coe} et~al.(2006){Coe}, {Ben{\'\i}tez}, {S{\'a}nchez}, {Jee},
  {Bouwens}, and {Ford}}]{Coe2006}
{Coe}, D., {Ben{\'\i}tez}, N., {S{\'a}nchez}, S.~F., {Jee}, M., {Bouwens}, R.,
  and {Ford}, H. (2006).
\newblock {Galaxies in the Hubble Ultra Deep Field. I. Detection, Multiband
  Photometry, Photometric Redshifts, and Morphology}.
\newblock \emph{Astronomical Journal} 132, 926--959.
\newblock \doi{10.1086/505530}
\bibAnnoteFile{Coe2006}

\bibitem[{{Conselice}(2003)}]{Conselice2003}
{Conselice}, C.~J. (2003).
\newblock {The Relationship between Stellar Light Distributions of Galaxies and
  Their Formation Histories}.
\newblock \emph{The Astrophysical Journal Supplement series} 147, 1--28.
\newblock \doi{10.1086/375001}
\bibAnnoteFile{Conselice2003}

\bibitem[{{Costantin} et~al.(2018){Costantin}, {Corsini}, {M{\'e}ndez-Abreu},
  {Morelli}, {Dalla Bont{\`a}}, and {Pizzella}}]{Costantin2018}
{Costantin}, L., {Corsini}, E.~M., {M{\'e}ndez-Abreu}, J., {Morelli}, L.,
  {Dalla Bont{\`a}}, E., and {Pizzella}, A. (2018).
\newblock {On the observational diagnostics to separate classical and disk-like
  bulges}.
\newblock \emph{Monthly Notices of the Royal Astronomical Society} 481,
  3623--3642.
\newblock \doi{10.1093/mnras/sty1754}
\bibAnnoteFile{Costantin2018}

\bibitem[{{de Souza} et~al.(2004){de Souza}, {Gadotti}, and {dos
  Anjos}}]{DeSouza2004}
{de Souza}, R.~E., {Gadotti}, D.~A., and {dos Anjos}, S. (2004).
\newblock {BUDDA: A New Two-dimensional Bulge/Disk Decomposition Code for
  Detailed Structural Analysis of Galaxies}.
\newblock \emph{The Astrophysical Journal Supplement Series} 153, 411--427.
\newblock \doi{10.1086/421554}
\bibAnnoteFile{DeSouza2004}

\bibitem[{{Di Criscienzo} et~al.(2017){Di Criscienzo}, {Merlin}, {Castellano},
  {Santini}, {Fontana}, {Amorin} et~al.}]{DiCriscienzo2017}
{Di Criscienzo}, M., {Merlin}, E., {Castellano}, M., {Santini}, P., {Fontana},
  A., {Amorin}, R., et~al. (2017).
\newblock {The ASTRODEEP Frontier Fields catalogues. III. Multiwavelength
  photometry and rest-frame properties of MACS-J0717 and MACS-J1149}.
\newblock \emph{Astronomy and Astrophysics} 607, A30.
\newblock \doi{10.1051/0004-6361/201731172}
\bibAnnoteFile{DiCriscienzo2017}

\bibitem[{{Ding} et~al.(2021){Ding}, {Birrer}, {Treu}, and
  {Silverman}}]{Ding2021}
{Ding}, X., {Birrer}, S., {Treu}, T., and {Silverman}, J.~D. (2021).
\newblock {Galaxy shapes of Light (GaLight): a 2D modeling of galaxy images}.
\newblock \emph{arXiv e-prints} , arXiv:2111.08721
\bibAnnoteFile{Ding2021}

\bibitem[{{Erwin}(2015)}]{Erwin2015}
{Erwin}, P. (2015).
\newblock {IMFIT: A Fast, Flexible New Program for Astronomical Image Fitting}.
\newblock \emph{The Astrophysical Journal} 799, 226.
\newblock \doi{10.1088/0004-637X/799/2/226}
\bibAnnoteFile{Erwin2015}

\bibitem[{{Fagioli} et~al.(2020){Fagioli}, {Tortorelli}, {Herbel},
  {Z{\"u}rcher}, {Refregier}, and {Amara}}]{Fagioli2020}
{Fagioli}, M., {Tortorelli}, L., {Herbel}, J., {Z{\"u}rcher}, D., {Refregier},
  A., and {Amara}, A. (2020).
\newblock {Spectro-imaging forward model of red and blue galaxies}.
\newblock \emph{Journal of Cosmology and Astroparticle Physics} 2020, 050.
\newblock \doi{10.1088/1475-7516/2020/06/050}
\bibAnnoteFile{Fagioli2020}

\bibitem[{{Ferrari} et~al.(2015){Ferrari}, {de Carvalho}, and
  {Trevisan}}]{Ferrari2015}
{Ferrari}, F., {de Carvalho}, R.~R., and {Trevisan}, M. (2015).
\newblock {Morfometryka{\textemdash}A New Way of Establishing Morphological
  Classification of Galaxies}.
\newblock \emph{The Astrophysical Journal} 814, 55.
\newblock \doi{10.1088/0004-637X/814/1/55}
\bibAnnoteFile{Ferrari2015}

\bibitem[{{Fisher} and {Drory}(2008)}]{Fisher2008}
{Fisher}, D.~B. and {Drory}, N. (2008).
\newblock {The Structure of Classical Bulges and Pseudobulges: the Link Between
  Pseudobulges and S{\'E}RSIC Index}.
\newblock \emph{The Astronomical Journal} 136, 773--839.
\newblock \doi{10.1088/0004-6256/136/2/773}
\bibAnnoteFile{Fisher2008}

\bibitem[{{Gaia Collaboration} et~al.(2021){Gaia Collaboration}, {Brown},
  {Vallenari}, {Prusti}, {de Bruijne}, {Babusiaux} et~al.}]{Gaia2021}
{Gaia Collaboration}, {Brown}, A.~G.~A., {Vallenari}, A., {Prusti}, T., {de
  Bruijne}, J.~H.~J., {Babusiaux}, C., et~al. (2021).
\newblock {Gaia Early Data Release 3. Summary of the contents and survey
  properties}.
\newblock \emph{Astronomy and Astrophysics} 649, A1.
\newblock \doi{10.1051/0004-6361/202039657}
\bibAnnoteFile{Gaia2021}

\bibitem[{{Gao} and {Ho}(2017)}]{Gao2017}
{Gao}, H. and {Ho}, L.~C. (2017).
\newblock {An Optimal Strategy for Accurate Bulge-to-disk Decomposition of Disk
  Galaxies}.
\newblock \emph{The Astrophysical Journal} 845, 114.
\newblock \doi{10.3847/1538-4357/aa7da4}
\bibAnnoteFile{Gao2017}

\bibitem[{{Ghosh} et~al.(2020){Ghosh}, {Urry}, {Wang}, {Schawinski}, {Turp},
  and {Powell}}]{Ghosh2020}
{Ghosh}, A., {Urry}, C.~M., {Wang}, Z., {Schawinski}, K., {Turp}, D., and
  {Powell}, M.~C. (2020).
\newblock {Galaxy Morphology Network: A Convolutional Neural Network Used to
  Study Morphology and Quenching in {\ensuremath{\sim}}100,000 SDSS and
  {\ensuremath{\sim}}20,000 CANDELS Galaxies}.
\newblock \emph{The Astrophysical Journal} 895, 112.
\newblock \doi{10.3847/1538-4357/ab8a47}
\bibAnnoteFile{Ghosh2020}

\bibitem[{{H{\"a}u{\ss}ler} et~al.(2013){H{\"a}u{\ss}ler}, {Bamford}, {Vika},
  {Rojas}, {Barden}, {Kelvin} et~al.}]{Haussler2013}
{H{\"a}u{\ss}ler}, B., {Bamford}, S.~P., {Vika}, M., {Rojas}, A.~L., {Barden},
  M., {Kelvin}, L.~S., et~al. (2013).
\newblock {MegaMorph - multiwavelength measurement of galaxy structure:
  complete S{\'e}rsic profile information from modern surveys}.
\newblock \emph{Monthly Notices of the Royal Astronomical Society} 430,
  330--369.
\newblock \doi{10.1093/mnras/sts633}
\bibAnnoteFile{Haussler2013}

\bibitem[{{H{\"a}u{\ss}ler} et~al.(2011){H{\"a}u{\ss}ler}, {Barden}, {Bamford},
  and {Rojas}}]{Haussler2011}
{H{\"a}u{\ss}ler}, B., {Barden}, M., {Bamford}, S.~P., and {Rojas}, A. (2011).
\newblock {Galapagos: A Semi-Automated Tool for Galaxy Profile Fitting}.
\newblock In \emph{Astronomical Data Analysis Software and Systems XX}, eds.
  I.~N. {Evans}, A.~{Accomazzi}, D.~J. {Mink}, and A.~H. {Rots}. vol. 442 of
  \emph{Astronomical Society of the Pacific Conference Series}, 155
\bibAnnoteFile{Haussler2011}

\bibitem[{{H{\"a}u{\ss}ler} et~al.(2022){H{\"a}u{\ss}ler}, {Vika}, {Bamford},
  {Johnston}, {Brough}, {Casura} et~al.}]{Haussler2022}
{H{\"a}u{\ss}ler}, B., {Vika}, M., {Bamford}, S.~P., {Johnston}, E.~J.,
  {Brough}, S., {Casura}, S., et~al. (2022).
\newblock {Galapagos-2/Galfitm/GAMA -- multi-wavelength measurement of galaxy
  structure: separating the properties of spheroid and disk components in
  modern surveys}.
\newblock \emph{arXiv e-prints} , arXiv:2204.05907
\bibAnnoteFile{Haussler2022}

\bibitem[{{Herbel} et~al.(2017){Herbel}, {Kacprzak}, {Amara}, {Refregier},
  {Bruderer}, and {Nicola}}]{Herbel2017}
{Herbel}, J., {Kacprzak}, T., {Amara}, A., {Refregier}, A., {Bruderer}, C., and
  {Nicola}, A. (2017).
\newblock {The redshift distribution of cosmological samples: a forward
  modeling approach}.
\newblock \emph{Journal of Cosmology and Astroparticle Physics} 2017, 035.
\newblock \doi{10.1088/1475-7516/2017/08/035}
\bibAnnoteFile{Herbel2017}

\bibitem[{{Herbel} et~al.(2018){Herbel}, {Kacprzak}, {Amara}, {Refregier}, and
  {Lucchi}}]{Herbel2018}
{Herbel}, J., {Kacprzak}, T., {Amara}, A., {Refregier}, A., and {Lucchi}, A.
  (2018).
\newblock {Fast point spread function modeling with deep learning}.
\newblock \emph{Journal of Cosmology and Astroparticle Physics} 2018, 054.
\newblock \doi{10.1088/1475-7516/2018/07/054}
\bibAnnoteFile{Herbel2018}

\bibitem[{{Kannawadi} et~al.(2019){Kannawadi}, {Hoekstra}, {Miller}, {Viola},
  {Fenech Conti}, {Herbonnet} et~al.}]{Kannawadi2019}
{Kannawadi}, A., {Hoekstra}, H., {Miller}, L., {Viola}, M., {Fenech Conti}, I.,
  {Herbonnet}, R., et~al. (2019).
\newblock {Towards emulating cosmic shear data: revisiting the calibration of
  the shear measurements for the Kilo-Degree Survey}.
\newblock \emph{Astronomy and Astrophysics} 624, A92.
\newblock \doi{10.1051/0004-6361/201834819}
\bibAnnoteFile{Kannawadi2019}

\bibitem[{{Kawinwanichakij} et~al.(2021){Kawinwanichakij}, {Silverman}, {Ding},
  {George}, {Damjanov}, {Sawicki} et~al.}]{Kawinwanichakij2021}
{Kawinwanichakij}, L., {Silverman}, J.~D., {Ding}, X., {George}, A.,
  {Damjanov}, I., {Sawicki}, M., et~al. (2021).
\newblock {Hyper Suprime-Cam Subaru Strategic Program: A Mass-dependent Slope
  of the Galaxy Size-Mass Relation at z < 1}.
\newblock \emph{The Astrophysical Journal} 921, 38.
\newblock \doi{10.3847/1538-4357/ac1f21}
\bibAnnoteFile{Kawinwanichakij2021}

\bibitem[{{Kelvin} et~al.(2012){Kelvin}, {Driver}, {Robotham}, {Hill},
  {Alpaslan}, {Baldry} et~al.}]{Kelvin2012}
{Kelvin}, L.~S., {Driver}, S.~P., {Robotham}, A. S.~G., {Hill}, D.~T.,
  {Alpaslan}, M., {Baldry}, I.~K., et~al. (2012).
\newblock {Galaxy And Mass Assembly (GAMA): Structural Investigation of
  Galaxies via Model Analysis}.
\newblock \emph{Monthly Notices of the Royal Astronomical Society} 421,
  1007--1039.
\newblock \doi{10.1111/j.1365-2966.2012.20355.x}
\bibAnnoteFile{Kelvin2012}

\bibitem[{{Kluge} et~al.(2021){Kluge}, {Bender}, {Riffeser}, {Goessl}, {Hopp},
  {Schmidt} et~al.}]{Kluge2021}
{Kluge}, M., {Bender}, R., {Riffeser}, A., {Goessl}, C., {Hopp}, U., {Schmidt},
  M., et~al. (2021).
\newblock {Photometric Dissection of Intracluster Light and Its Correlations
  with Host Cluster Properties}.
\newblock \emph{The Astrophysical Journal Supplement Series} 252, 27.
\newblock \doi{10.3847/1538-4365/abcda6}
\bibAnnoteFile{Kluge2021}

\bibitem[{{Kormendy}(1977)}]{Kormendy1977}
{Kormendy}, J. (1977).
\newblock {Brightness distributions in compact and normal galaxies. II -
  Structure parameters of the spheroidal component}.
\newblock \emph{Astrophysical Journal} 218, 333--346.
\newblock \doi{10.1086/155687}
\bibAnnoteFile{Kormendy1977}

\bibitem[{{Krist} et~al.(2011){Krist}, {Hook}, and {Stoehr}}]{Krist2011}
{Krist}, J.~E., {Hook}, R.~N., and {Stoehr}, F. (2011).
\newblock {20 years of Hubble Space Telescope optical modeling using Tiny Tim}.
\newblock In \emph{Optical Modeling and Performance Predictions V}, ed. M.~A.
  {Kahan}. vol. 8127 of \emph{Society of Photo-Optical Instrumentation
  Engineers (SPIE) Conference Series}, 81270J.
\newblock \doi{10.1117/12.892762}
\bibAnnoteFile{Krist2011}

\bibitem[{{La Barbera} et~al.(2010){La Barbera}, {de Carvalho}, {de La Rosa},
  and {Lopes}}]{LaBarbera2010}
{La Barbera}, F., {de Carvalho}, R.~R., {de La Rosa}, I.~G., and {Lopes},
  P.~A.~A. (2010).
\newblock {SPIDER - II. The Fundamental Plane of early-type galaxies in
  grizYJHK}.
\newblock \emph{Monthly Notices of the Royal Astronomical Society} 408,
  1335--1360.
\newblock \doi{10.1111/j.1365-2966.2010.17091.x}
\bibAnnoteFile{LaBarbera2010}

\bibitem[{{Lee} et~al.(2018){Lee}, {Chary}, and {Wright}}]{Lee2018}
{Lee}, B., {Chary}, R.-R., and {Wright}, E.~L. (2018).
\newblock {Galaxy Ellipticity Measurements in the Near-infrared for Weak
  Lensing}.
\newblock \emph{The Astrophysical Journal} 866, 157.
\newblock \doi{10.3847/1538-4357/aadfd7}
\bibAnnoteFile{Lee2018}

\bibitem[{{Li} et~al.(2022){Li}, {Napolitano}, {Roy}, {Tortora}, {La Barbera},
  {Sonnenfeld} et~al.}]{Li2022}
{Li}, R., {Napolitano}, N.~R., {Roy}, N., {Tortora}, C., {La Barbera}, F.,
  {Sonnenfeld}, A., et~al. (2022).
\newblock {Galaxy Light Profile Convolutional Neural Networks (GaLNets). I.
  Fast and Accurate Structural Parameters for Billion-galaxy Samples}.
\newblock \emph{The Astrophysical Journal} 929, 152.
\newblock \doi{10.3847/1538-4357/ac5ea0}
\bibAnnoteFile{Li2022}

\bibitem[{{Lotz} et~al.(2017){Lotz}, {Koekemoer}, {Coe}, {Grogin}, {Capak},
  {Mack} et~al.}]{Lotz2017}
{Lotz}, J.~M., {Koekemoer}, A., {Coe}, D., {Grogin}, N., {Capak}, P., {Mack},
  J., et~al. (2017).
\newblock {The Frontier Fields: Survey Design and Initial Results}.
\newblock \emph{Astrophysical Journal} 837, 97.
\newblock \doi{10.3847/1538-4357/837/1/97}
\bibAnnoteFile{Lotz2017}

\bibitem[{{Lotz} et~al.(2004){Lotz}, {Primack}, and {Madau}}]{Lotz2004}
{Lotz}, J.~M., {Primack}, J., and {Madau}, P. (2004).
\newblock {A New Nonparametric Approach to Galaxy Morphological
  Classification}.
\newblock \emph{The Astronomical Journal} 128, 163--182.
\newblock \doi{10.1086/421849}
\bibAnnoteFile{Lotz2004}

\bibitem[{{M{\'e}ndez-Abreu} et~al.(2008){M{\'e}ndez-Abreu}, {Aguerri},
  {Corsini}, and {Simonneau}}]{Mendez-Abreu2008}
{M{\'e}ndez-Abreu}, J., {Aguerri}, J.~A.~L., {Corsini}, E.~M., and {Simonneau},
  E. (2008).
\newblock {Structural properties of disk galaxies. I. The intrinsic equatorial
  ellipticity of bulges}.
\newblock \emph{Astronomy and Astrophysics} 478, 353--369.
\newblock \doi{10.1051/0004-6361:20078089}
\bibAnnoteFile{Mendez-Abreu2008}

\bibitem[{{Merlin} et~al.(2016){Merlin}, {Amor{\'\i}n}, {Castellano},
  {Fontana}, {Buitrago}, {Dunlop} et~al.}]{Merlin2016}
{Merlin}, E., {Amor{\'\i}n}, R., {Castellano}, M., {Fontana}, A., {Buitrago},
  F., {Dunlop}, J.~S., et~al. (2016).
\newblock {The ASTRODEEP Frontier Fields catalogues. I. Multiwavelength
  photometry of Abell-2744 and MACS-J0416}.
\newblock \emph{Astronomy and Astrophysics} 590, A30.
\newblock \doi{10.1051/0004-6361/201527513}
\bibAnnoteFile{Merlin2016}

\bibitem[{{Pagul} et~al.(2021){Pagul}, {S{\'a}nchez}, {Davidzon}, and
  {Mobasher}}]{Pagul2021}
{Pagul}, A., {S{\'a}nchez}, F.~J., {Davidzon}, I., and {Mobasher}, B. (2021).
\newblock {Hubble Frontier Field Clusters and Their Parallel Fields:
  Photometric and Photometric Redshift Catalogs}.
\newblock \emph{The Astrophysical Journal Supplement Series} 256, 27.
\newblock \doi{10.3847/1538-4365/abea9d}
\bibAnnoteFile{Pagul2021}

\bibitem[{{Peng} et~al.(2010){Peng}, {Ho}, {Impey}, and {Rix}}]{Peng2010}
{Peng}, C.~Y., {Ho}, L.~C., {Impey}, C.~D., and {Rix}, H.-W. (2010).
\newblock {Detailed Decomposition of Galaxy Images. II. Beyond Axisymmetric
  Models}.
\newblock \emph{The Astronomical Journal} 139, 2097--2129.
\newblock \doi{10.1088/0004-6256/139/6/2097}
\bibAnnoteFile{Peng2010}

\bibitem[{{Peng} et~al.(2011){Peng}, {Ho}, {Impey}, and {Rix}}]{Peng2011}
[Dataset] {Peng}, C.~Y., {Ho}, L.~C., {Impey}, C.~D., and {Rix}, H.~W. (2011).
\newblock {GALFIT: Detailed Structural Decomposition of Galaxy Images}.
\newblock Astrophysics Source Code Library
\bibAnnoteFile{Peng2011}

\bibitem[{{S{\'e}rsic}(1963)}]{Sersic1963}
{S{\'e}rsic}, J.~L. (1963).
\newblock {Influence of the atmospheric and instrumental dispersion on the
  brightness distribution in a galaxy}.
\newblock \emph{Boletin de la Asociacion Argentina de Astronomia La Plata
  Argentina} 6, 41--43
\bibAnnoteFile{Sersic1963}

\bibitem[{{Shipley} et~al.(2018){Shipley}, {Lange-Vagle}, {Marchesini},
  {Brammer}, {Ferrarese}, {Stefanon} et~al.}]{Shipley2018}
{Shipley}, H.~V., {Lange-Vagle}, D., {Marchesini}, D., {Brammer}, G.~B.,
  {Ferrarese}, L., {Stefanon}, M., et~al. (2018).
\newblock {HFF-DeepSpace Photometric Catalogs of the 12 Hubble Frontier Fields,
  Clusters, and Parallels: Photometry, Photometric Redshifts, and Stellar
  Masses}.
\newblock \emph{The Astrophysical Journal Supplement Series} 235, 14.
\newblock \doi{10.3847/1538-4365/aaacce}
\bibAnnoteFile{Shipley2018}

\bibitem[{{Simard} et~al.(2011){Simard}, {Mendel}, {Patton}, {Ellison}, and
  {McConnachie}}]{Simard2011}
{Simard}, L., {Mendel}, J.~T., {Patton}, D.~R., {Ellison}, S.~L., and
  {McConnachie}, A.~W. (2011).
\newblock {A Catalog of Bulge+disk Decompositions and Updated Photometry for
  1.12 Million Galaxies in the Sloan Digital Sky Survey}.
\newblock \emph{The Astrophysical Journal Supplement} 196, 11.
\newblock \doi{10.1088/0067-0049/196/1/11}
\bibAnnoteFile{Simard2011}

\bibitem[{{Simard} et~al.(2002){Simard}, {Willmer}, {Vogt}, {Sarajedini},
  {Phillips}, {Weiner} et~al.}]{Simard2002}
{Simard}, L., {Willmer}, C. N.~A., {Vogt}, N.~P., {Sarajedini}, V.~L.,
  {Phillips}, A.~C., {Weiner}, B.~J., et~al. (2002).
\newblock {The DEEP Groth Strip Survey. II. Hubble Space Telescope Structural
  Parameters of Galaxies in the Groth Strip}.
\newblock \emph{The Astrophysical Journal Supplement Series} 142, 1--33.
\newblock \doi{10.1086/341399}
\bibAnnoteFile{Simard2002}

\bibitem[{{Sonnenfeld}(2022)}]{Sonnenfeld2022}
{Sonnenfeld}, A. (2022).
\newblock {The effect of spiral arms on the S{\'e}rsic photometry of galaxies}.
\newblock \emph{Astronomy and Astrophysics} 659, A141.
\newblock \doi{10.1051/0004-6361/202142786}
\bibAnnoteFile{Sonnenfeld2022}

\bibitem[{{Tortorelli} et~al.(2018{\natexlab{a}}){Tortorelli}, {Della Bruna},
  {Herbel}, {Amara}, {Refregier}, {Alarcon} et~al.}]{Tortorelli2018b}
{Tortorelli}, L., {Della Bruna}, L., {Herbel}, J., {Amara}, A., {Refregier},
  A., {Alarcon}, A., et~al. (2018{\natexlab{a}}).
\newblock {The PAU Survey: a forward modeling approach for narrow-band
  imaging}.
\newblock \emph{Journal of Cosmology and Astroparticle Physics} 2018, 035.
\newblock \doi{10.1088/1475-7516/2018/11/035}
\bibAnnoteFile{Tortorelli2018b}

\bibitem[{{Tortorelli} et~al.(2020){Tortorelli}, {Fagioli}, {Herbel}, {Amara},
  {Kacprzak}, and {Refregier}}]{Tortorelli2020}
{Tortorelli}, L., {Fagioli}, M., {Herbel}, J., {Amara}, A., {Kacprzak}, T., and
  {Refregier}, A. (2020).
\newblock {Measurement of the B-band galaxy Luminosity Function with
  Approximate Bayesian Computation}.
\newblock \emph{Journal of Cosmology and Astroparticle physics} 2020, 048.
\newblock \doi{10.1088/1475-7516/2020/09/048}
\bibAnnoteFile{Tortorelli2020}

\bibitem[{{Tortorelli} et~al.(2023){Tortorelli}, {Mercurio}, {Granata},
  {Rosati}, {Grillo}, {Nonino} et~al.}]{Tortorelli2023}
{Tortorelli}, L., {Mercurio}, A., {Granata}, G., {Rosati}, P., {Grillo}, C.,
  {Nonino}, M., et~al. (2023).
\newblock {The Kormendy relation of early-type galaxies as a function of
  wavelength in Abell S1063, MACS J0416.1-2403 and MACS J1149.5+2223}.
\newblock \emph{arXiv e-prints} ,
  arXiv:2302.07896\doi{10.48550/arXiv.2302.07896}
\bibAnnoteFile{Tortorelli2023}

\bibitem[{{Tortorelli} et~al.(2018{\natexlab{b}}){Tortorelli}, {Mercurio},
  {Paolillo}, {Rosati}, {Gargiulo}, {Gobat} et~al.}]{Tortorelli2018}
{Tortorelli}, L., {Mercurio}, A., {Paolillo}, M., {Rosati}, P., {Gargiulo}, A.,
  {Gobat}, R., et~al. (2018{\natexlab{b}}).
\newblock {The Kormendy relation of galaxies in the Frontier Fields clusters:
  Abell S1063 and MACS J1149.5+2223}.
\newblock \emph{Monthly Notices of the Royal Astronomical Society} 477,
  648--668.
\newblock \doi{10.1093/mnras/sty617}
\bibAnnoteFile{Tortorelli2018}

\bibitem[{{Tortorelli} et~al.(2021){Tortorelli}, {Siudek}, {Moser}, {Kacprzak},
  {Berner}, {Refregier} et~al.}]{Tortorelli2021}
{Tortorelli}, L., {Siudek}, M., {Moser}, B., {Kacprzak}, T., {Berner}, P.,
  {Refregier}, A., et~al. (2021).
\newblock {The PAU survey: measurement of narrow-band galaxy properties with
  approximate bayesian computation}.
\newblock \emph{Journal of Cosmology and Astroparticle physics} 2021, 013.
\newblock \doi{10.1088/1475-7516/2021/12/013}
\bibAnnoteFile{Tortorelli2021}

\bibitem[{{Trujillo} et~al.(2001){Trujillo}, {Aguerri}, {Cepa}, and
  {Guti{\'e}rrez}}]{Trujillo2001}
{Trujillo}, I., {Aguerri}, J.~A.~L., {Cepa}, J., and {Guti{\'e}rrez}, C.~M.
  (2001).
\newblock {The effects of seeing on S{\'e}rsic profiles - II. The Moffat PSF}.
\newblock \emph{Monthly Notices of the Royal Astronomical Society} 328,
  977--985.
\newblock \doi{10.1046/j.1365-8711.2001.04937.x}
\bibAnnoteFile{Trujillo2001}

\bibitem[{{Tuccillo} et~al.(2018){Tuccillo}, {Huertas-Company},
  {Decenci{\`e}re}, {Velasco-Forero}, {Dom{\'\i}nguez S{\'a}nchez}, and
  {Dimauro}}]{Tuccillo2018}
{Tuccillo}, D., {Huertas-Company}, M., {Decenci{\`e}re}, E., {Velasco-Forero},
  S., {Dom{\'\i}nguez S{\'a}nchez}, H., and {Dimauro}, P. (2018).
\newblock {Deep learning for galaxy surface brightness profile fitting}.
\newblock \emph{Monthly Notices of the Royal Astronomical Society} 475,
  894--909.
\newblock \doi{10.1093/mnras/stx3186}
\bibAnnoteFile{Tuccillo2018}

\bibitem[{{van der Marel}(1991)}]{vanderMarel1991}
{van der Marel}, R.~P. (1991).
\newblock {The velocity dispersion anisotropy and mass-to-light ratio of
  elliptical galaxies.}
\newblock \emph{Monthly Notices of the Royal Astronomical Society} 253,
  710--726.
\newblock \doi{10.1093/mnras/253.4.710}
\bibAnnoteFile{vanderMarel1991}

\bibitem[{{Vikram} et~al.(2010){Vikram}, {Wadadekar}, {Kembhavi}, and
  {Vijayagovindan}}]{Vikram2010}
{Vikram}, V., {Wadadekar}, Y., {Kembhavi}, A.~K., and {Vijayagovindan}, G.~V.
  (2010).
\newblock {PYMORPH: automated galaxy structural parameter estimation using
  PYTHON}.
\newblock \emph{Monthly Notices of the Royal Astronomical Society} 409,
  1379--1392.
\newblock \doi{10.1111/j.1365-2966.2010.17426.x}
\bibAnnoteFile{Vikram2010}

\bibitem[{{Z{\"u}rcher} et~al.(2022){Z{\"u}rcher}, {Fluri}, {Sgier},
  {Kacprzak}, {Gatti}, {Doux} et~al.}]{Zuercher2022}
{Z{\"u}rcher}, D., {Fluri}, J., {Sgier}, R., {Kacprzak}, T., {Gatti}, M.,
  {Doux}, C., et~al. (2022).
\newblock {Dark energy survey year 3 results: Cosmology with peaks using an
  emulator approach}.
\newblock \emph{Monthly Notices of the Royal Astronomical Society} 511,
  2075--2104.
\newblock \doi{10.1093/mnras/stac078}
\bibAnnoteFile{Zuercher2022}

\bibitem[{{Z{\"u}rcher} et~al.(2021){Z{\"u}rcher}, {Fluri}, {Sgier},
  {Kacprzak}, and {Refregier}}]{Zuercher2021}
{Z{\"u}rcher}, D., {Fluri}, J., {Sgier}, R., {Kacprzak}, T., and {Refregier},
  A. (2021).
\newblock {Cosmological forecast for non-Gaussian statistics in large-scale
  weak lensing surveys}.
\newblock \emph{Journal of Cosmology and Astroparticle physics} 2021, 028.
\newblock \doi{10.1088/1475-7516/2021/01/028}
\bibAnnoteFile{Zuercher2021}

\end{thebibliography}


\end{document}